\documentclass{statsoc}
\usepackage[a4paper]{geometry}
\usepackage{amsmath, amssymb}
\usepackage{amsfonts, multirow, epsfig, floatrow}
\usepackage{graphicx, pdflscape, verbatim, enumerate, colortbl, setspace}
\usepackage{setspace, color,bm}
\usepackage[normalem]{ulem}
\usepackage{cite}
\usepackage{multirow} 
\usepackage{booktabs,array}
\usepackage{hyperref}
\usepackage{float}
\usepackage{caption}
\usepackage{subcaption} 
\usepackage{tikz}
 \usepackage[linesnumbered,ruled]{algorithm2e}

\newtheorem{Corollary}{Corollary}
\newtheorem{Proposition}{Proposition}
\newtheorem{Lemma}{Lemma}

\newtheorem{Theorem}{Theorem}
 
\newtheorem{Remark}{Remark}

\newcommand{\argmin}{\mathop{\rm arg\min}}

\newcommand{\R}{\mathbb{R}}

\newcommand{\E}{{\mathbb{E}}}
\newcommand{\PP}{{\mathbb{P}}}
\newcommand{\RR}{{\mathbf{R}}}

\newcommand{\FF}{{\mathcal{F}}}

\newcommand{\HH}{{\mathcal{H}}}

\newcommand{\TV}{L_1}
\newcommand{\T}{{\rm T}}

\newcommand{\supp}{{\rm supp}}
\newcommand{\dd}{{\bf d}}

\newcommand{\suppx}{{\rm supp}(\bx)}

\newcommand{\nb}{n_2}
\newcommand{\uu}{{u}}

\newcommand{\Sigmab}{{{\Sigma}}}

\newcommand{\bX}{X}

\newcommand{\bw}{w}

\newcommand{\cip}{\overset{p}{\to}}

\def\limsup{\mathop{\overline{\rm lim}}}
\def\liminf{\mathop{\underline{\rm lim}}}

\usepackage{enumitem}

\newcommand{\alphab}{\boldsymbol{\alpha}}
\newcommand{\omegab}{\boldsymbol{\omega}}
\newcommand{\taumini}{\tau_{\rm mini}({\kt,\bx_{\rm new}})}
\newcommand{\tauadpt}{\tau_{\rm adap}({\ku,\kt,\bx_{\rm new}})}

\newcommand{\kt}{s}
\newcommand{\ku}{s_u}

\newcommand{\esta}{\widehat{\bx_{\rm new}^{\intercal}{\bbeta}_1}}
\newcommand{\estb}{\widehat{\bx_{\rm new}^{\intercal}{\bbeta}_2}}
\def\xnew{\bx_{\rm new}}
\def\trans{^{\intercal}}
\def\bY{\mathbf{Y}}
\def\bX{\mathbf{X}}

\def\Xbb{\mathbb{X}}
\def\bbeta{\boldsymbol{\beta}}
\def\bgamma{\boldsymbol{\eta}}
\def\bepsilon{\boldsymbol{\epsilon}}

\def\bx{\mathbf{x}}
\def\subnew{_{\sf\scriptscriptstyle new}}
\def\bxnew{\bx\subnew}
\def\Deltanew{\Delta\subnew}
\def\bXi{\mathbf{X}_{1,i}}
\def\bXib{\mathbf{X}_{2,i}}
\def\btheta{\boldsymbol{\theta}}

\def\bbetahat{\widehat{\bbeta}}
\def\be{\mathbf{e}}
\def\bu{\mathbf{u}}
\def\butilde{\widetilde{\bu}}
\def\supth{^{\rm \scriptscriptstyle th}}

\def\bE{\mathbf{E}}
\def\bEhat{\widehat{\bE}}
\def\buhat{\widehat{\bu}}
\def\bSigma{\boldsymbol{\Sigma}}
\def\bSigmahat{\widehat{\bSigma}}
\def\bmu{\boldsymbol{\mu}}
\def\bB{\mathbf{B}}

\def\Deltanewtilde{\widetilde{\Delta\subnew}}
\def\bv{\mathbf{v}}
\def\bvhat{\widehat{\bv}}

\def\Hbb{\mathbb{H}}
\def\Ssc{\mathcal{S}}

\title[Individualized Treatment Selection]{{Optimal Statistical Inference for Individualized Treatment Effects in High-dimensional Models}}

\author{Tianxi Cai}
\address{Harvard University, Boston, USA}
\author{T. Tony Cai}
\address{University of Pennsylvania, Philadelphia, USA}
\author[Cai, Cai $\&$ Guo]{Zijian Guo}
\address{Rutgers University, Piscataway, USA}

\begin{document}
\maketitle 

\begin{abstract}
The ability to predict individualized treatment effects (ITEs) based on a given patient's profile is essential for personalized medicine.
We propose a hypothesis testing approach to choosing between two potential treatments for a given individual in the framework of high-dimensional linear models.  The methodological novelty lies in the construction of a debiased estimator of the ITE and establishment of its asymptotic normality uniformly for an arbitrary  future high-dimensional observation, while the existing methods can only handle certain specific forms of observations. We introduce a testing procedure with the type-I error controlled and establish its asymptotic power. The proposed method can be extended to making inference for general linear contrasts, including both the average treatment effect and outcome prediction.  We introduce the optimality framework for hypothesis testing from both the minimaxity and adaptivity perspectives and establish the optimality of the proposed procedure. An extension to  high-dimensional approximate linear models is also considered. The finite sample performance of the procedure is demonstrated in simulation studies and further illustrated through an analysis of electronic health records data from patients with rheumatoid arthritis.
\end{abstract}
 

\newcommand{\Abb}{\mathbb{A}}
\section{Introduction}\label{sec-intro}
It has been well recognized that the effectiveness and potential risk of a treatment often vary significantly by patient subgroups. The ability to predict individualized treatment effects (ITEs) based on a given covariate profile is essential for precision medicine.  Although trial-and-error and one-size-fits-all approaches  remain a common practice, much recent focus has been placed on predicting treatment effects at a more individual level \citep{la2011predictive,ong2012personalized}.  Genetic mutations and gene-expression profiles are increasingly used to guide treatment selection for cancer patients  \citep{albain2010prognostic,eberhard2005mutations}. Large scale clinical trials are being conducted to evaluate individualized treatment strategies \citep{chantrill2015precision,evans2004moving,simon2007feasibility}. The increasing availability of electronic health records (EHR) systems with detailed patient data promises a new paradigm for translational precision medicine research. 
Models for predicting ITE can be estimated using real world data and can potentially be deployed more efficiently to clinical practice. 
 
Motivated by the ITE estimation using EHR data with high-dimensional covariates, we consider in this paper efficient estimation and inference procedures for predicting a future patient's ITE given his/her $p$ dimensional covariates when $p$ is potentially much larger than the sample size $n$. Specifically, we consider high-dimensional linear regression models for the outcomes in the two treatment groups:
\begin{equation}
\bY_{k}=\Xbb_{k}\bbeta_{k}+\bepsilon_{k} , \quad k = 1, 2,
\label{eq: linear models}
\end{equation}
where $\bY_k = (y_{k,1}, ..., y_{k,n_k})^{\intercal}$ and $\Xbb_k = (\bX_{k,1}, ..., \bX_{k,n_k})\trans$ are the response and covariates observed independently for the $n_k$ subjects in the treatment group $k$ respectively, 
$\bepsilon_k = (\epsilon_{k,1}, ..., \epsilon_{k,n_k})\trans$ is the error vector with constant variance $\sigma_k^2 = \mbox{var}(\epsilon_{k,i})$ and $\bbeta_{k}\in \R^{p}$ is the regression vector for the $k\supth$ treatment group. For a given patient with covariate vector $\bxnew$, we construct point and interval estimators for the ITE $\Deltanew = \bxnew\trans(\bbeta_1-\bbeta_2)$ and  
consider the hypothesis testing 
\begin{equation}
H_0: \bxnew\trans(\bbeta_1-\bbeta_2)\leq 0 \quad \text{vs.} \quad H_1: \bxnew\trans(\bbeta_1-\bbeta_2)>0.
\label{eq: testing problem}
\end{equation}



\subsection{Individualized Treatment Selection}
\label{sec: ITS motivation}
While clinical trials and traditional cohort studies remain critical sources for precision medicine research, they have limitations including the generalizability of study findings and the limited ability to test broader hypotheses. 
In recent years, due to the increasing adoption of EHR and the linkage of EHR with bio-repositories and other research registries, integrated large datasets now exist as a new source for precision medicine studies. 
For example, the Partner's Healthcare System (PHS) biobank contains both a wealth of clinical (e.g. diagnoses, treatments, laboratory values) and biological measurements including 
genomic data \citep{gainer2016biobank}. These integrated datasets open opportunities for developing EHR-based individualized  treatment selection models, which can potentially be fed back to the EHR system for guiding clinical decision making. 

To enable EHR for such precision medicine research, different patients cohorts with specific diseases of interest have been constructed at PHS via the efforts of the Informatics for Integrating Biology and the Bedside (i2b2) \citep{kohane2012translational}. An example of such disease cohort is rheumatoid arthritis (RA), consisting of 4453 patients identified as having RA using a machine learning algorithm \citep{liao2010electronic}. A small subset of these patients have their genetic and biological markers measured. The biomarker data integrated with EHR data can be used to derive ITE models for guiding treatment strategies for RA patients. 
A range of disease modifying treatment options are now available for RA patients, including methotrexate,  tumor necrosis factor inhibitors often referred to as anti-TNF, and the combination of the two  \citep{calabrese20162015}. The superiority of the combination therapy over monotherapy has been well established  \citep{emery2008comparison,breedveld2006premier,van2006comparison}. Despite its superiority, a significant proportion of patients do not respond to the combination therapy with reported response rates ranging from about 30\% to 60\%. Due to the high cost and significant side effects including serious infection and malignancy associated with anti-TNF therapy \citep{bongartz2006anti}, there is a pressing need to develop ITE models to guide RA treatment selection. We address this need by deriving an ITE  model for RA using the biomarker linked EHR data at PHS. The proposed procedures are desirable tools for application since the number of potential predictors is large in this setting. 

\vspace{-5mm}

\subsection{Statistical Framework and Contributions} 
\label{sec: contribution}

Many statistical and machine learning algorithms have been proposed for estimating the ITEs \citep{zhou2017residual,zhao2012estimating,imai2013estimating,qian2011performance}. However, existing methods largely focused on the low-dimensional settings. In the presence of high dimensional predictors, inference for the ITEs becomes significantly more challenging. Several regularized procedures can be used to obtain point estimators for $\Deltanew = \bxnew\trans(\bbeta_1-\bbeta_2)$ 
\citep{chen2001atomic,tibshirani1996regression,fan2001variable,candes2007dantzig,sun2012scaled,zhang2010nearly,belloni2011square,moon2007ensemble,Song_2015,belloni2014high}. However, when the goal is to construct confidence intervals (CIs) for $\Deltanew$, it is problematic to estimate  $\Deltanew$ by simply plugging in the regularized estimators due to their inherent biases. 
These biases can accumulate when projecting along the direction of $\bxnew$ and result in a significant bias in $\Deltanew$. 

In this paper, we develop the High-dimensional Individualized Treatment Selection (HITS) method that aims to choose between two treatments for a given individual based on the observed high-dimensional covariates. We propose a novel bias-corrected estimator for $\bxnew\trans(\bbeta_1-\bbeta_2)$ and establish its asymptotic normality for any given $\bxnew$. This is achieved by imposing an additional novel constraint in the construction of the projection direction, which is used to correct the bias of the plug-in estimator. This additional constraint guarantees that the variance of the HITS estimator dominates its bias for any $\bxnew$. 
With this bias-corrected estimator, we construct CIs and carry out hypothesis test for $\Deltanew$ under the challenging setting where  $\bx_{\rm new}$ is of high-dimension and of no special structure. Rigorous justifications are given for the coverage and length properties of the resulting CIs and also for Type I error control and power of the proposed testing procedure. More generally, the HITS method can be adapted for making inference about any linear contrasts $\bxnew\trans \bbeta_k$ for $k=1,2$, which are crucial to inference for average treatment effect (ATE) and inference related to prediction; see Sections \ref{sec: ATE} and \ref{sec: prediction} for details.  We have also extended the asymptotic normality results to high-dimensional approximate linear models.
We further introduce an optimality framework for hypothesis testing in the high-dimensional sparse linear model and establish the optimality of HITS from two perspectives, minimaxity and adaptivity, where minimaxity captures the difficulty of the testing problem with true sparsity level known a priori and adaptivity is for the more challenging setting with unknown sparsity. 

We summarize two key contributions of the current paper below and then  compare the present work to existing high dimensional inference literature in section \ref{sec: liter comparison}.
 
 \begin{itemize}
\item To the best of our knowledge, the method proposed in the current paper is the first unified inference procedure with theoretical guarantees for general linear contrasts $ \bxnew\trans(\bbeta_1-\bbeta_2)$ and $\bxnew\trans \bbeta_k$ for $k=1,2$, where no structural assumptions are made on the high-dimensional loading $\bxnew$. This is a challenging task as noted in prior literature on
inference for linear contrast in high dimensional regression \citep{cai2015regci,athey2018approximate,zhu2018linear}.

\item Optimal detection boundary without knowledge of the exact sparsity level, noted as an open question in \citet{zhu2017projection}, is addressed in the current paper.  It is shown that HITS is adaptively optimal  for testing the hypotheses \eqref{eq: testing problem} over a large class of loadings $\bxnew$ with unknown and unconstrained sparsity level.
\end{itemize}

\vspace{-7mm}
\subsection{Comparisons with High-dimensional Inference Literature}
\label{sec: liter comparison}

For a single regression coefficient under sparse linear models, \citet{zhang2014confidence,van2014asymptotically,javanmard2014confidence} introduced  debiasing methods for CI construction. Inference for more general linear contrasts has been investigated recently in \citet{cai2015regci,athey2018approximate,zhu2018linear}. These all require special structure on the loading $\bxnew$.
Our work is the first to provide valid inference procedures for general contrasts with arbitrary high-dimensional loading $\bxnew$ without special structures. More specifically, in the context of constructing CIs, \citet{cai2015regci} showed a significance difference between sparse and dense $\bxnew$. The methods developed for a single regression coefficient can be extended to a sparse $\bxnew$ but the construction of a dense $\bxnew$ relies on a conservative upper bound and requires the information on sparsity level.  \citet{athey2018approximate} constructed CI for the general linear contrasts for $\bxnew$ only if the loading $\bxnew$ has a bounded weighted $\ell_2$ norm and constructed CI for ATE under the {\em overlap} assumption; See Section \ref{sec: ATE} for detailed discussion. \citet{zhu2018linear} constructed a CI for the linear contrast under the condition that the conditional expectation $\E[\bxnew^{\intercal} \bX_{1,i}\mid \bv_{1}^{\intercal}\bX_{1,i},\cdots, \bv_{p-1}^{\intercal}\bX_{1,i}]$ is a sparse linear combination of $\bv_{1}^{\intercal}\bX_{1,i},\cdots, \bv_{p-1}^{\intercal}\bX_{1,i}$, where $\{\bv_{j}\}_{1\leq j \leq p-1}$ span the space orthogonal to $\bxnew$. The most significant distinction of the proposed HITS method from the aforementioned literature is a unified uncertainty quantification method for all high-dimensional loadings $\bxnew$. 
 
{Hypothesis testing for more general functionals has been recently considered in \citet{javanmard2017flexible} and \citet{zhu2017projection}. 
\citet{javanmard2017flexible}  reduces the testing problem for a general functional to that for the projection of the functional of interest to a given orthogonal basis and then construct a debiased estimator of the corresponding basis expansion. The test statistic is constructed by comparing this debiased estimator and its projection to the null parameter space. This strategy can also be used to construct CIs for a linear contrast, but is only valid if $\bxnew$ is sparse.
\citet{zhu2017projection} proposed a general testing procedure by first constructing an estimator by $\ell_1$ projection of the penalized estimator to the null parameter space and then debiasing both the penalized and projected estimators. The test is based on the difference between these two debiased estimators and  a critical value computed via bootstrap. Although this test, in principle, controls  the type I error of \eqref{eq: testing problem}, the asymptotic power is established only when the true parameter is well separated from the null under the $\ell_{\infty}$ norm by $n^{-\frac{1}{4}}$. There are no results on the power if the true parameters in the alternative outside this region. Our approach is distinct; we establish the asymptotic normality of the proposed estimator of $\Deltanew$, uniformly over all loadings $\bxnew$ and the whole parameter spaces of approximately sparse regression vectors. As a consequence, 1) we have an asymptotic expression of the power of the proposed test for all $\Deltanew$; 2) since the asymptotic power in \citet{zhu2017projection} is established by inequalities instead of the limiting distribution, the CIs for $\Deltanew$  by inverting the testing procedure in \citet{zhu2017projection} are more conservative than the CI constructed in the present paper. Additionally, we resolved the open question raised in \citet{zhu2017projection} ``the minimax detection rate for this problem without knowledge of the sparsity level is also an open question" in Corollary \ref{cor: optimality} in the present paper.  We provide more technical comparsions to these two approaches in Remark \ref{rem: tech}. } 

Another intuitive inference method for a general linear contrast is to plug-in the debiased estimator for individual regression coefficients developed in \citet{zhang2014confidence,van2014asymptotically,javanmard2014confidence}. A numerical comparison of this estimator with the proposed HITS procedure is given in Section \ref{sec: simulation}. The results show that  HITS  not only is computationally more efficient but also has uniformly better coverage properties than the plug-in estimator.

From another perspective, we compare the optimality results for hypothesis testing established here with those for CIs given in \citet{cai2015regci}. The adaptivity framework for hypothesis testing is different from that for CI construction. 
In addition, the current paper considers a broader classes of loadings than those in \citet{cai2015regci}, including the case of decaying loadings and a more general class of sparse exact loadings.

\subsection{Organization of the Paper}
The rest of the paper is organized as follows. Section \ref{sec: method} presents the proposed testing and CI procedures for $\Deltanew$. Theoretical properties are given in Section \ref{sec: theory};  Optimality of the testing procedure is discussed in Section \ref{sec: optimality}; The proposed method is extended in Section \ref{sec: prediction} to quantify uncertainty for prediction in high dimensional linear regression; The numerical performance is investigated in Section \ref{sec: simulation}.  In Section \ref{sec: real data}, we apply the proposed method to infer about ITE of the aforementioned combination therapy over methotrexate alone for treating RA using EHR data from PHS. Discussions are provided in Section \ref{sec: discussion} and proofs of the main results are given in Section \ref{sec:proof}. Additional discussions, simulations and proofs are presented in the supplementary materials.

\subsection{Notations}
For a matrix $\bX\in \R^{n\times p}$, $\bX_{i\cdot}$,  $\bX_{\cdot j}$, and $\bX_{ij}$ denote respectively its $i\supth$ row,  $j\supth$ column, and  $(i,j)$ entry. 
For a vector $\bx\in \R^{p}$, $\bx_{-j}$ denotes the subvector of $\bx$ excluding the $j\supth$ element, $\suppx$ denotes the support of $\bx$ and
the $\ell_q$ norm of $\bx$ is defined as $\|\bx\|_{q}=(\sum_{j=1}^{p}|x_j|^q)^{\frac{1}{q}}$ for $q \geq 0$ with $\|\bx\|_0=|\suppx|$ and $\|\bx\|_{\infty}=\max_{1\leq j \leq p}|x_j|$. 
For a matrix $\Abb$, we define the spectral norm $\|\Abb\|_{2}=\sup_{\|\bx\|_2 = 1} \|\Abb\bx\|_2$; 
For a symmetric matrix $\Abb$, $\lambda_{\min}\left(\Abb\right)$ and $\lambda_{\max}\left(\Abb\right)$  denote respectively the smallest and largest eigenvalue of $\Abb$. We use $c$ and $C$ to denote generic positive constants that may vary from place to place. For two positive sequences $a_n$ and $b_n$,  $a_n \lesssim b_n$ means $a_n \leq C b_n$ for all $n$ and $a_n \gtrsim b_n $ if $b_n\lesssim  a_n$ and $a_n \asymp b_n $ if $a_n \lesssim b_n$ and $b_n \lesssim a_n$, and $a_n \ll b_n$ if $\limsup_{n\rightarrow\infty} \frac{a_n}{b_n}=0$ and $a_n \gg b_n$ if $b_n \ll a_n$. 

\section{Methodology} 
\label{sec: method} 
In this section, we detail proposed inference procedures for the ITE $\Deltanew = \bxnew\trans(\bbeta_1-\bbeta_2)$. We first discuss existing bias correction methods in high-dimensional regression in Section \ref{sec: bias correction} and introduce a novel construction of projection direction which adapts to any given loading $\bx_{\rm new}$ in Section \ref{sec: construction}, where throughout we use subscript $k \in \{1,2\}$ to index the treatment group.  
Then in Section \ref{sec: statistical inference}, we propose point and interval estimators as well as a hypothesis testing procedure for $\Deltanew$. In Section \ref{sec: ATE}, we extend the proposed method to inference for average treatment effect.
\subsection{Existing Method of Bias Correction: Minimize Variance with Bias Constrained} 
\label{sec: bias correction}
Given the observations $\bY_k\in \R^{n_k}$ and $\Xbb_k\in \R^{n_k\times p}$,  $\bbeta_k$ can be estimated by the Lasso estimator,
\begin{equation}
\widehat{\bbeta}_k=\argmin_{\bbeta_k\in \R^{p},\in \R^{+}}\frac{\|\bY_k-\Xbb_k\bbeta_k\|_2^2}{2n_k}+A\sqrt{\frac{\log p}{n_k}} \sum_{j=1}^{p} {W_{k,j}} |\bbeta_{k,j}|,  \quad \text{for} \quad k=1,2,
\label{eq: Lasso estimator a}
\end{equation}
with a pre-specified positive constant $A>0$ and $W_{k,j}=\sqrt{\frac{1}{n_k}\sum_{i=1}^{n_k}X^2_{k,ij}}$ denoting the penalization weight for the $j\supth$ variable in the $k\supth$ treatment group. The variance $\sigma_k^2$ is then estimated by $\widehat{\sigma}_k^2=\tfrac{1}{n_k}\|\bY_k-\Xbb_k\widehat{\bbeta}_k\|^2_2$ for $k=1,2.$  
We note that, other initial estimators can also be used, including the Dantzig Selector \citep{candes2007dantzig} and tuning-free penalized estimators, such as the scaled Lasso \citep{sun2012scaled}, square-root Lasso \citep{belloni2011square}, and iterated Lasso \citep{belloni2012sparse}, as long as the initial estimators $\widehat{\beta}_k$ and $\widehat{\sigma}_k^2$ satisfy certain consistency properties as stated in Section \ref{sec: asyn}.




We discuss the bias correction idea for estimating $\bxnew\trans\bbeta_1$ and the same approach can be extended to $k=2$. 
 A natural and simple way to estimate $\bx_{\rm new}\trans \bbeta_1$ is to plug in the Lasso estimator $\widehat{\bbeta}_1$ in \eqref{eq: Lasso estimator a}.
However, this plug-in estimator $\bx_{\rm new}\trans \widehat{\bbeta}_1$  is known to suffer from the bias induced by the penalty in \eqref{eq: Lasso estimator a}. For the special case $\bx_{\rm new}=\be_j$, where $\be_j$ is the $j\supth$ Euclidean basis vector, various forms of debiased estimators have been introduced in \citet{zhang2014confidence,van2014asymptotically,javanmard2014confidence} to correct the bias of the plug-in estimator $\widehat{\bbeta}_{1,j}$ and then construct CIs centered at the debiased estimators. 
The idea 
can be extended to general linear contrasts $\bx_{\rm new}^{\intercal}\bbeta_1$ for certain class of $\bxnew$, where a key step is to estimate the bias $\bx_{\rm new}\trans (\widehat{\bbeta}_1-\bbeta_1)$. To this end, we aim to identify an effective projection direction $\bu \in \RR^{p}$ to construct a debiased estimator for $\bx\subnew\trans\bbeta_1$ as
\begin{equation}
\bx\subnew\trans\bbetahat_1 + \bu \trans \bEhat_1, \quad \mbox{where}\quad
\bEhat_k = \frac{1}{n_k}\sum_{i=1}^{n_k} \bX_{k,i}\left(Y_{k,i}-\bX_{k,i}\trans \widehat{\bbeta}_k\right) \; \text{for}\; k=1,2.
\label{eq: previous est}
\end{equation}
The error decomposition of the above bias-corrected estimator is 
\begin{align}
(\bx\subnew\trans\bbetahat_1 + \bu \trans \bEhat_1)-\bxnew\trans\bbeta_1= \bu\trans \frac{1}{n_1}\sum_{i=1}^{n_1} \bXi\epsilon_{1,i}+(\bSigmahat_1 \bu-\bx_{\rm new})\trans  (\widehat{\bbeta}_1-\bbeta_1)
\label{eq: projection decomposition}
\end{align}
{where} 
$\bSigmahat_k =  \frac{1}{n_k}\sum_{i=1}^{n_k} \bX_{k,i}\bX_{k,i}\trans$. 

To correct the bias of each individual regression coefficient (that is $\xnew=\be_j$), \citet{zhang2014confidence,javanmard2014confidence} proposed the foundational idea of constructing a projection direction for bias correction via minimization of variance with the bias constrained. Specifically, in \eqref{eq: projection decomposition}, the projection direction is identified such that the variance of $\bu\trans \frac{1}{n_1}\sum_{i=1}^{n_1} \bXi\epsilon_{1,i}$ is minimized while the bias component $(\bSigmahat_1 \bu-\bx_{\rm new})\trans  (\widehat{\bbeta}_1-\bbeta_1)$ is constrained.  This idea is generalized in \citet{cai2015regci} to deal with sparse $\xnew$ via the following algorithm 
\begin{equation}
\butilde_1 = \argmin_{\bu\in \R^{p}} \left\{\bu\trans \bSigmahat_1  \bu:\;
\| \bSigmahat_1 \bu-\bx_{\rm new}\|_{\infty}\leq  \|\bx_{\rm new}\|_2 \lambda_{1}\right\}
\label{eq: initial correction}
\end{equation}
where $\lambda_{1}\asymp \sqrt{{\log p}/{n_1}}$. Here, $\bu\trans \bSigmahat_1  \bu$ measures the variance of $\bu\trans \frac{1}{n_1}\sum_{i=1}^{n_1} \bXi\epsilon_{1,i}$ and the constraint on $\| \bSigmahat_1 \bu-\bx_{\rm new}\|_{\infty}$ further controls the bias term $(\bSigmahat_1 \bu-\bx_{\rm new})\trans  (\widehat{\bbeta}_1-\bbeta_1)$ in \eqref{eq: projection decomposition} via the inequality 
\begin{equation}
|(\bSigmahat_1 \bu-\bx_{\rm new})\trans  (\widehat{\bbeta}_1-\bbeta_1)|\leq \| \bSigmahat_1 \bu-\bx_{\rm new}\|_{\infty}\|\widehat{\bbeta}_1-\bbeta_1\|_1.
\label{eq: holder}
\end{equation}
The bias corrected estimator $\bx\subnew\trans\bbetahat_1 + \butilde_1 \trans \bEhat_1$  and its variant  have been studied in the literature
\citep{cai2015regci,Nilesh2019prediction,athey2018approximate}. \citet{cai2015regci}  and \citet{athey2018approximate} considered inference for $\bxnew$ of specific structures. It was also shown that $\butilde_1$ is effective for bias-correction when $\bx_{\rm new}$ is of certain sparse structure \citep{cai2015regci}  and when $\bxnew$ is of a bounded weighted $\ell_2$ norm
 \citep{athey2018approximate}. \citet{Nilesh2019prediction} focused exclusively on its estimation 
error  instead of confidence interval construction for $\bxnew^{\intercal}\beta$. In fact, this type of projection direction \eqref{eq: initial correction}, used in \citet{cai2015regci,athey2018approximate} and \citet{Nilesh2019prediction}, is not always effective for bias correction or the subsequent confidence interval construction. Proposition \ref{prop: challenge} of Section \ref{sec: discussion} shows that if the loading $\bx_{\rm new}$ is of certain dense structure, then the projection direction $\butilde_1$  is zero and hence the ``bias-corrected" estimator using $\butilde_1$ is reduced to the plug-in estimator $\bx\subnew\trans\bbetahat_1$.  As shown in Table \ref{tab: Gaussian setting 2}, the plug-in estimator can have reasonable estimation error but is not suitable for confidence interval construction due to large bias. These existing inference procedures cannot automatically adapt to arbitrary structure of $\bx_{\rm new}$. 
\subsection{New Projection Direction: Minimize variance and Constrain Bias and Variance}
\label{sec: construction}
To effectively debias for an arbitrary $\bxnew$, we propose to identify the projection direction $\widehat{\bu}_k$ for the $k\supth$ treatment as 
\begin{align}
\buhat_k=\;\argmin_{\bu\in \R^{p}} \bu\trans \bSigmahat_k\bu \quad \text{subject to}\;
&\|\bSigmahat_k\bu-\bx_{\rm new}\|_{\infty}\leq  \|\bx_{\rm new}\|_2 \lambda_{k} \label{eq: constraint 1}\\
&\;|\bx_{\rm new}\trans \bSigmahat_k\bu-\|\bx_{\rm new}\|_2^2|\leq \|\bx_{\rm new}\|_2^2\lambda_{k} ,\label{eq: constraint 2}
\end{align}
where $\lambda_{k}\asymp \sqrt{{\log p}/{n_k}}$. Our proposed bias corrected estimator for $\bx\subnew\trans\bbeta_k$ is then
\begin{equation}
\widehat{\bx_{\rm new}\trans \bbeta_k}=\bx_{\rm new}\trans \widehat{\bbeta}_k+\buhat_k\trans  \bEhat_k \quad \text{for}\quad k=1,2.
\label{eq: correction single}
\end{equation}



The main difference between $\buhat_1$ and $\butilde_1$ in \eqref{eq: initial correction} is the additional constraint \eqref{eq: constraint 2}. 
As mentioned earlier, a general strategy for bias correction  is to minimize the variance component while constraining the bias \citep{zhang2014confidence,javanmard2014confidence}. However, to develop a unified inference method for all $\bxnew$, the optimization strategy developed in the current paper is a triplet, minimizing the variance, constraining the bias and constraining the variance. In particular, the additional constraint \eqref{eq: constraint 2} is imposed to constrain the variance such that it is dominating the bias component, which is essential for CI construction. We refer to the construction defined in \eqref{eq: constraint 1} and \eqref{eq: constraint 2} as ``Variance-enhancement Projection Direction". 
 Such a general triplet optimization strategy can be of independent interest and applied to other inference problems.

We shall remark that, the further constraint \eqref{eq: constraint 2}  on the variance  is not as intuitive as the other constraints, in the sense that the error decomposition of the bias-corrected estimator in \eqref{eq: projection decomposition} shows that it is sufficient to obtain an accurate estimator of $\xnew^{\intercal}\bbeta_1$ as long as we adopt the existing idea of minimizing variance under the bias constraint; see the detailed discussion between \eqref{eq: projection decomposition} and \eqref{eq: holder}. In the error decomposition \eqref{eq: projection decomposition}, it seems that the additional constraint \eqref{eq: constraint 2} is not needed. However, \eqref{eq: constraint 2} is the key component to construct a valid inference procedure uniformly over all $\bxnew$. For statistical inference, one not only needs an accurate estimator, but also an accurate assessment of the uncertainty of the estimator. This is the main reason for adding the additional constraint.

\def\bw{\mathbf{w}}
An equivalent way of constructing the projection direction defined in \eqref{eq: constraint 1} and \eqref{eq: constraint 2} is,
\begin{equation}
\buhat_k=\;\argmin_{\bu\in \R^{p}} \bu\trans \bSigmahat_k\bu \quad \text{subject to}\; \sup_{\bf w \in \mathcal{C}}|\langle {\bf w}, \bSigmahat_k\bu-\bx_{\rm new}\rangle|\leq  \|\bx_{\rm new}\|_2 \lambda_{k} \label{eq: max constraint}
\end{equation}
where 
$
\mathcal{C}=\{{\bf e}_1,\cdots,{\bf e}_p,{\bx_{\rm new}}/{\|\bx_{\rm new}\|_2}\}
$
with ${\bf e}_{i}$ denoting the $i$th standard Euclidean basis vector. The feasible set in \eqref{eq: max constraint} ensures that the projected values $\langle \bw, \widehat{\bSigma}_k\bu -\bx_{\rm new}\rangle$ of $\bSigmahat_k\bu-\bx_{\rm new}$ to any of the $p+1$ vectors in $\mathcal{C}$ are small. In contrast, as motivated in \eqref{eq: holder}, the existing debiased procedures only constrain that the projected values to all the standard Euclidean basis vectors, $\max_{1\leq j\leq p}|\langle {\bf e}_{j}, \widehat{\bSigma}_k\bu -\bx_{\rm new}\rangle|$, are small. In the case where $\bx_{\rm new}=e_i$, these two constraints are the same; however, in the case where $\bx_{\rm new}$ is of complicated structures, the additional direction $\bx_{\rm new}/\|\bx_{\rm new}\|_2$ contained in $\mathcal{C}$ is the key component to conduct the bias correction; see more discussion in Section \ref{sec: discussion}. 

\subsection{Statistical Inference for Individualized Treatment Effect}
\label{sec: statistical inference}

\smallskip
Combining $\esta$ and $\estb$, we propose  to estimate  $\Deltanew$ as 
\begin{equation}
\widehat{\Deltanew}=  \esta-\estb.
\label{eq: correction}
\end{equation}
Compared to the plug-in estimator $\bx_{\rm new}\trans (\widehat{\bbeta}_1-\widehat{\bbeta}_2)$, the main advantage of $\widehat{\Deltanew}$ is that the variance of $\widehat{\Deltanew}$ is dominating the bias of $\widehat{\Deltanew}$ while the bias of the plug-in estimator is usually as large as its variance. (See Table \ref{tab: Gaussian setting 1} for the numerical illustration.) Such a rebalance of bias and variance is useful for inference as the variance component is much easier to characterize than the bias component. 

To conduct HITS, it remains to quantify the variability of $\widehat{\Deltanew}$. The asymptotic  variance of $\widehat{\Deltanew}$ is
\begin{equation}
{\rm V}=\frac{\sigma_1^2}{n_1}\buhat_1\trans  \bSigmahat_1\buhat_1+\frac{\sigma_2^2}{n_2}\buhat_2\trans  \bSigmahat_2\buhat_2, 
\label{eq: variance}
\end{equation}
which can be estimated by
$
\widehat{\rm V}=\frac{\widehat{\sigma}_1^2}{n_1}\buhat_1\trans  \bSigmahat_1\buhat_1+\frac{\widehat{\sigma}_2^2}{n_2}\buhat_2\trans  \bSigmahat_2\buhat_2.
$
With $\widehat{\Deltanew}$ and the variance estimator $\widehat{\rm V}$, we may construct 
a CI for the ITE $\Deltanew$ as  
\begin{equation}
\label{eq: ci}
{\rm CI}=\left(\widehat{\Deltanew}-{z_{\alpha/2}}\sqrt{\widehat{\rm V}},\quad \widehat{\Deltanew}+{z_{\alpha/2}}\sqrt{\widehat{\rm V}}\right) 
\end{equation}
and conduct HITS based on
\begin{equation}
\phi_{\alpha}={\mathbf 1}\left(\widehat{\Deltanew}-{z_{\alpha}}\sqrt{\widehat{\rm V}}>0\right), 
\label{eq: test}
\end{equation}
where $z_{\alpha}$ is the upper $\alpha$ quantile for the standard normal distribution.
That is, if $\widehat{\Deltanew} > {z_{\alpha}}\sqrt{\widehat{\rm V}}$, we would recommend subjects with $\bx_{\rm new}$ to receive treatment 1 over treatment 2, vice versa.

It is worth noting that the proposed HITS procedure utilizes the $\bxnew$ information in the construction of the projection direction $\buhat_{k}$, where both the constraints in \eqref{eq: constraint 1} and \eqref{eq: constraint 2} depend on the observation $\bx_{\rm new}$. For observations with different $\bx_{\rm new}$, the corresponding projection directions can be quite different.  Second, the method is computationally efficient as the bias correction step only requires the identification of two projection directions instead of performing debiased of $\widehat{\bbeta}_k$  coordinate by coordinate. 

%

\subsection{Application to Inference for Average Treatment Effect}
\label{sec: ATE}
In addition to the individualized treatment selection, the proposed method can also be applied to study the average treatment effect  (ATE). We follow the setting of \citet{athey2018approximate} where $k=1$ corresponds to the control group and $k=2$ corresponds to the treatment group. The average treatment over the treatment group is defined as $\bar{\bX}_{2}^{\intercal}(\bbeta_2-\bbeta_1)$ where $\bar{\bX}_2=\frac{1}{\nb}\sum_{i=1}^{\nb} \bX_{2,i}$. 
The statistical inference for the ATE $\bar{\bX}_{2}^{\intercal}(\bbeta_2-\bbeta_1)$ can be viewed as a special case of $x_{\rm new}^{\intercal}(\bbeta_2-\bbeta_1)$ with $x_{\rm new}=\bar{\bX}_{2}.$ Both the point estimator \eqref{eq: correction} and interval estimator \eqref{eq: ci} can be directly applied here to construct point and interval estimators for the ATE by taking $x_{\rm new}=\bar{\bX}_{2}.$

These estimators are different from those proposed in \citet{athey2018approximate}. The main distinction is the additional constraint \eqref{eq: constraint 2} proposed in the current paper. Without \eqref{eq: constraint 2}, \citet{athey2018approximate} requires either the {\em Bounded Loading} condition (Lemma 1 of \citet{athey2018approximate}) or the {\em Overlap} condition (Assumption 5 of \citet{athey2018approximate})  to guarantee the asymptotic limiting distribution of the corresponding ATE estimators. We state both conditions in the terminology of the current paper,  1){\em Bounded Loading.} $\bar{\bX}_{2}\Sigma_1^{-1}\bar{\bX}_{2}$ is assumed to be bounded;   2){\em Overlap.} There exists a constant $\eta>0$ such that $\eta\leq e(x)\leq 1-\eta$ for  all $x\in \R^{p}$ where $e(x)$ is the probability of receiving the treatment for an individual with covariates $x$.
Both conditions are actually limiting applications of the developed method to practical settings. As $\bar{\bX}_{2}\Sigma_1^{-1}\bar{\bX}_{2}$ is of the order $\sqrt{p/n}$, the bounded loading condition is not realistic in the high-dimensional setting $p\gg n$. In addition, if $e(x)$ is the commonly used logit or probit probability function, then the overlap condition only holds if the support of the propensity score is bounded.

\section{Theoretical Properties}
\label{sec: theory}
\subsection{Model Assumptions and Initial Estimators}
\label{sec: asyn}
We assume the following conditions on the random designs and the regression errors. 
\begin{enumerate}
	\item[(A1)] For $k=1,2$, the rows $\bX_{k,i} $ are i.i.d. $p$-dimensional {sub-gaussian} random vectors with mean $\bmu_k=\E \bX_{k,i}$ and the second order moment $\bSigma_k=\E (\bX_{k,i}\bX_{k,i}\trans $) where $c_0 \leq \lambda_{\min}\left(\bSigma_k\right) \leq \lambda_{\max}\left(\bSigma_k\right) \leq C_0$ for positive constants $C_0\geq c_0>0$. For $k=1,2$, the error vector $\bepsilon_k = (\epsilon_{k,1}, ..., \epsilon_{k,n_k})\trans$ is sub-gaussian  and satisfies the moment conditions $\E(\epsilon_{k,i}\mid \bX_{k,i})=0$ and $\E(\epsilon_{k,i}^2\mid\bX_{k,i})=\sigma_{k}^2$ for some unknown positive constant $0<\sigma_k^2<\infty$.
	\item[(A2)]  For $k=1,2$, the error vector $\bepsilon_k = (\epsilon_{k,1}, ..., \epsilon_{k,n_k})\trans$ is independent of $\Xbb_k$ and follow Gaussian distribution with mean zero and variance $\sigma_k^2$.
	\end{enumerate}
The assumption ${\ (A1)}$ is standard for the design and the regression error in the high-dimension literature. The condition ${\rm (A2)}$ on the error vectors is stronger but is only needed to establish the distributional results. This assumption is further relaxed in Section \ref{sec: extension}.

We consider the capped-$\ell_1$ sparse regression vectors with 
\begin{equation}
\sum_{j=1}^{p} \min\{ |{\bbeta}_{k,j}|/\sigma_{k}\lambda_0 ,1\} \leq s_k \quad \text{for}\quad k=1,2.
\label{eq: capped l1}
\end{equation}
where $\lambda_0=\sqrt{2\log p/n}$. As remarked in \citet{zhang2014confidence}, the capped-$\ell_1$ condition in \eqref{eq: capped l1} holds if $\bbeta_{k}$ is $\ell_0$ sparse with $\|\bbeta_{k}\|_0\leq s_k$ or $\ell_{q}$ sparse with ${\|\bbeta_{k}\|_{q}^q}/{(\sigma_{k}\lambda_0)^{q}}\leq s_k$ for $0<q\leq 1$. We introduce the following general conditions on the initial estimators.
\begin{enumerate}
	\item[(B1)] With probability at least $1-g(n_1,n_2)$ where $g(n_1,n_2)\rightarrow 0$, $\|\widehat{\bbeta}_k-\bbeta_k\|_1\lesssim s_k\sqrt{{ \log p}/{n_k}}$ for $k = 1, 2$,
	\item[(B2)] $\widehat{\sigma}_k^2$ is a consistent estimator of $\sigma_k^2$ for $k=1,2$, that is, 
$\max_{k=1,2}\left|{\widehat{\sigma}_k^2}/{\sigma_k^2}-1\right|\cip 0.$
	
\end{enumerate}
A variety of  estimators proposed in the high-dimensional regression literature for estimating the regression vector and the regression error variance are known to satisfy the above conditions under various conditions. See \citet{sun2012scaled,belloni2011square,bickel2009simultaneous,buhlmann2011statistics} and the reference therein for more details.

\subsection{Asymptotic Normality}
\label{sec: asyn}

Before stating the theorem, we present some intuitions for the estimation error of the proposed estimator, which relies on the following error decompositions of $\widehat{\bx_{\rm new}\trans {\bbeta_k}}$,
\begin{equation}
\widehat{\bx_{\rm new}\trans {\bbeta_k}}-\bx_{\rm new}\trans {\bbeta_k}=\buhat_k\trans  \frac{1}{n_k}\sum_{i=1}^{n_k} \bX_{k,i}\bepsilon_{k,i}+(\bSigmahat_k\buhat_k-\bx_{\rm new})\trans (\widehat{\bbeta}_k-\bbeta_k) .
\label{eq: decomposition a}
\end{equation}
This decomposition \eqref{eq: decomposition a} reflects the bias and variance decomposition of $\widehat{\bx_{\rm new}\trans {\bbeta_k}}$, where the first error term $\buhat_k\trans  \frac{1}{n_k}\sum_{i=1}^{n_k} \bX_{k,i}\bepsilon_{k,i}$ is the variance while the second error term $(\bSigmahat_k\buhat_k-\bx_{\rm new})\trans (\widehat{\bbeta}_k-\bbeta_k)
$ is the remaining stochastic bias. A similar bias and variance decomposition for the estimator $\widehat{\Deltanew}$ can be established by applying \eqref{eq: decomposition a} with $k=1,2$.
The following theorem establishes the rate of convergence for $\widehat{\Deltanew}$.
\begin{Theorem} Suppose that the conditions {\rm (A1)} and {\rm (B1)} hold and $s_k\leq c{{n_k}}/{\log p}$ for $k=1,2$, then with probability larger than $1-p^{-c}-g(n_1,n_2)-\frac{1}{t^2}$,
\begin{equation}
\begin{aligned}
\left|\widehat{\Deltanew}-\Deltanew\right|\lesssim t \|\bx_{\rm new}\|_2 \left(\frac{1}{\sqrt{n_1}}+\frac{1}{\sqrt{n_2}}\right)+
\|\bx_{\rm new}\|_2 \left(\frac{\|\bbeta_1\|_0 \log p}{n_1}+\frac{\|\bbeta_2\|_0 \log p}{n_2}\right).\end{aligned}
\label{eq: estimation error}
\end{equation}
\label{thm: estimation error}
\end{Theorem}
One of the terms on the right hand side of \eqref{eq: estimation error},  $\|\bx_{\rm new}\|_2 \left({\|\bbeta_1\|_0 \log p}/{n_1}+{\|\bbeta_2\|_0 \log p}/{n_2}\right)$, is an upper bound for the remaining bias of the proposed debiased estimator while $\|\bx_{\rm new}\|_2 \left({1}/{\sqrt{n_1}}+{1}/{\sqrt{n_2}}\right)$ is an upper bound for the corresponding variance. The following theorem shows that under the additional condition ${\rm (A3)}$ and the stronger sparsity condition, the proposed estimator has an asymptotic normal distribution.

\begin{Theorem}
Suppose that the conditions {\rm (A1), (A2)} and {\rm (B1)} hold and $s_k\leq c{\sqrt{n_k}}/{\log p}$ for $k=1,2$, then 
\begin{equation}
\frac{1}{\sqrt{\rm V}}\left(\widehat{\Deltanew}-\Deltanew\right)\overset{d}{\rightarrow} N(0,1)
\label{eq: limiting dist}
\end{equation}
where ${\rm V}$ is defined in \eqref{eq: variance}.
\label{thm: limiting distribution}
\end{Theorem}

A key component of establishing the limiting distribution for $\widehat{\Deltanew}$ is to show that the variance term in \eqref{eq: variance} dominates the upper bound for the bias term in \eqref{eq: estimation error}. We present this important component in the following Lemma, which characterizes the magnitude of the variance level ${\rm V}$ in \eqref{eq: variance}.
\begin{Lemma} Suppose that the condition {\rm (A1)} holds, then with probability larger than $1-p^{-c}$,
\begin{equation}
c_0\|\bx_{\rm new}\|_2\left({1}/{\sqrt{n_1}}+{1}/{\sqrt{n_2}}\right)
\leq \sqrt{\rm V}\leq C_0\|\bx_{\rm new}\|_2\left({1}/{\sqrt{n_1}}+{1}/{\sqrt{n_2}}\right),
\label{eq: dominating variance}
\end{equation}
for some positive constants $c_0, C_0>0$.
\label{lem: enhanced variance}
\end{Lemma}
We shall highlight that such a characterization of the variance, mainly the lower bound of \eqref{eq: dominating variance}, is only achieved through incorporating the novel additional constraint \eqref{eq: constraint 2}  to identify the projection direction. Without this additional constraint, as shown in Proposition \ref{prop: challenge}, the variance level can be exactly zero and hence the lower bound in \eqref{eq: dominating variance} does not hold.
\subsection{Hypothesis Testing and Confidence Interval}
\label{sec: testing}
We discuss two corollaries of Theorem \ref{thm: limiting distribution}, one for the hypothesis testing problem \eqref{eq: testing problem} related to the individualized treatment selection and the other for construction of CIs for $\Deltanew$.  Regarding the testing problem, we consider the following parameter space 
\begin{equation*}\textstyle
\Theta\left(s\right)=\left\{  \btheta=\mbox{\scriptsize $\left(\begin{aligned} & \bB_1,\bSigma_1\\ &\bB_2,\bSigma_2 \end{aligned}\right)$}: \sum_{j=1}^{p} \min\left(\frac{{|\bbeta}_{k,j}|}{\sigma_{k}\lambda_0} ,1\right)\leq s,\; 0<\sigma_k\leq M_0,\;\lambda_{\min}(\bSigma_k)\geq c_0, \; \text{for}\; k=1,2\right\},
\label{eq: signal para space}
\end{equation*}
for  some constants $M_0>0$ and $c_0>0$. Then we define the null hypothesis parameter space as 
\begin{equation}
\HH_0(s)=\left\{\btheta=\mbox{\scriptsize $\left(\begin{aligned} & \bB_1,\bSigma_1\\ &\bB_2,\bSigma_2 \end{aligned}\right)$}
\in \Theta\left(s\right):\bx_{\rm new}\trans \left(\bbeta_1-\bbeta_2\right)\leq 0\right\}
\label{eq: null space}
\end{equation}
and the local alternative parameter space as 
{\small
\begin{equation}
\HH_1(s, \delta_0)=\left\{\btheta=\mbox{\scriptsize $\left(\begin{aligned} & \bB_1,\bSigma_1\\ &\bB_2,\bSigma_2 \end{aligned}\right)$}\in \Theta\left(s\right):\bx_{\rm new}\trans \left(\bbeta_1-\bbeta_2\right)= {\delta_0}\|\bx_{\rm new}\|_2\left({1}/{\sqrt{n_1}}+{1}/{\sqrt{n_2}}\right)\right\},
\label{eq: local alternative}
\end{equation}
}
\noindent for $\delta_0 > 0$.
The following corollary provides the theoretical guarantee for the individualized treatment selection, where the type I error of the proposed HITS procedure in \eqref{eq: test} is controlled and the asymptotic power curve is also established.
\begin{Corollary} 
Suppose that the conditions {\rm (A1), (A2)} and {\rm (B1), (B2)} hold and $s_k\leq c{\sqrt{n_k}}/{\log p}$ for $k=1,2$, then 
for any $\bx\subnew\in \R^{p}$, the type I error of the proposed test $\phi_{\alpha}$ defined in \eqref{eq: test} is controlled as,
$
\limsup_{\min\{n_1,n_2\}\rightarrow \infty}\sup_{\btheta \in \HH_0}\PP_{\btheta}\left(\phi_{\alpha}=1\right)\leq \alpha. 
$
For any given $\btheta \in \HH_1(\delta_0)$ and any $\bx\subnew\in \R^{p}$, the asymptotic power of the test $\phi_{\alpha}$ is
\begin{equation}
\lim_{\min\{n_1,n_2\}\rightarrow \infty}\PP_{\btheta}\left(\phi_{\alpha}=1\right)=1-\Phi^{-1}\left(z_{\alpha}-\frac{\delta_0}{\sqrt{V}}\|\bx_{\rm new}\|_2\left({1}/{\sqrt{n_1}}+{1}/{\sqrt{n_2}}\right)\right). 
\label{eq: power}
\end{equation}
\label{Cor: hypothesis testing}
\end{Corollary}
Together with Lemma \ref{lem: enhanced variance}, we observe that the proposed test is powerful with $\delta_0\rightarrow\infty$, where $\delta_0$ controls the local alternative defined in \eqref{eq: local alternative}.
The main massage for real applications is that the individualized treatment selection would be easier if the collected data set has larger sample sizes $n_1$ and $n_2$ and also the future observation of interest has a smaller $\ell_2$ norm.  This phenomenon is especially interesting for the individualized treatment selection with high-dimensional covariates, where the corresponding norm $\|\bx_{\rm new}\|_2$ can be of different orders of magnitude; see Section \ref{sec: simulation} for numerical illustrations.

In addition to the hypothesis testing procedure, we also establish the coverage of the proposed CI in \eqref{eq: ci} for ITE $\Deltanew$, 
\begin{Corollary}
Suppose that {\rm (A1), (A2)} and {\rm (B1), (B2)} hold and $s_k\leq c{\sqrt{n_k}}/{\log p}$ for $k=1,2$. Then the coverage property of the proposed CI defined in \eqref{eq: ci} is
\[
\liminf_{\min\{n_1,n_2\}\rightarrow \infty}\PP_{\btheta}\left(\Deltanew\in {\rm CI}\right)\geq 1-\alpha \; \text{for any}\; \bx_{\rm new}\in \R^{p}.
\]
\label{Cor: coverage}
\end{Corollary}
Another important perspective of CI construction is the precision of the CIs, which can be measured by the length. It follows from Lemma \ref{lem: enhanced variance} that the length of the constructed CI in \eqref{eq: ci}  is controlled at the order of magnitude $\|\bx_{\rm new}\|_2 ({1}/{\sqrt{n_1}}+{1}/{\sqrt{n_2}})$, which means that the length depends on both the sample sizes $n_1$ and $n_2$ and also the $\ell_2$ norm of the future observation $\|\bx_{\rm new}\|_2$. For observations with different $\bxnew$, the lengths of the corresponding CIs for ITE $\Delta_{\rm new}$ can be quite different, where the length is determined by $\|x_{\rm new}\|_2$ and the numerical illustration is present in Section \ref{sec: simulation}. 

\begin{Remark} \rm 
It is helpful to compare some of the technical details with the related work \citet{zhu2017projection, javanmard2017flexible}. The type I error control in Theorem 1 of \citet{zhu2017projection} requires stronger model complexity assumptions than Corollary \ref{Cor: hypothesis testing}. Specifically, using our notation,  \citet{zhu2017projection}  requires $\log p=o(n^{1/8})$ and $s_k \ll n^{\frac{1}{4}}/\sqrt{\log p}$ while we only need $s_k \ll \sqrt{n}/\log p$. More fundamentally, \citet{zhu2017projection} does not establish the asymptotic limiting distribution as Theorem \ref{thm: limiting distribution} in the present paper. Instead of using the asymptotic limiting distribution, Theorems 2 and 4 in \citet{zhu2017projection} use inequalities to show that the asymptotic powers are close to 1 if the parameters in the alternative are well separated from the null under the $\ell_{\infty}$ norm by $n^{-\frac{1}{4}}$; as a consequence, the power function for the local neighborhood cannot be established as in \eqref{eq: power}. \citet{javanmard2017flexible} requires the loading to be sparse and particularly, in Lemma 2.4 , the loading $\bxnew$ is required to satisfy $\mu \|u_1\|_1<1$ where, using our notation, $\mu= \|\bxnew\|_2\sqrt{\log p/n}$ and $u_1=\bxnew.$ This imposes the condition $\|\bxnew\|_1\|\bxnew\|_2 < \sqrt{n/\log p}$ on the loading $\bxnew$; see more discussion after Proposition \ref{prop: challenge}.
\label{rem: tech}
\end{Remark}

\subsection{Further Extensions: Approximately Linear Models and Non-Gaussianity}
\label{sec: extension}
The inference results established under model \eqref{eq: linear models} can be further extended to approximate linear models \citep{belloni2011inference,belloni2012sparse},  
\begin{equation}
\bY_{k}=\Xbb_{k}\bbeta_{k}+{\bf r}_{k}+\bepsilon_{k}, \; k = 1, 2,
\label{eq: general}
\end{equation}
{where the high-dimensional vector  $\bbeta_k\in \R^{p}$ satisfies the capped $\ell_1$ sparsity \eqref{eq: capped l1} and the approximation error ${\bf r}_{k}=\left(r_{k,1},\cdots, r_{k,n_{k}}\right)^{\intercal} \in \R^{n_k}$ is defined with $r_{k,i}=\E\left(Y_{k,i}\mid \bX_{k,i}\right)-\bX_{k,i}^{\intercal}\bbeta_{k}$.} We also relax the Gaussian error assumption ${\rm (A2)}$ through modifying construction of the projection direction as follows,
\begin{equation}
\begin{aligned}
\buhat_k=\;\argmin_{\bu\in \R^{p}} \bu\trans \bSigmahat_k\bu \quad \text{subject to}\;
&\|\bSigmahat_k\bu-\bx_{\rm new}\|_{\infty}\leq  \|\bx_{\rm new}\|_2 \lambda_{k} \\
&\;|\bx_{\rm new}\trans \bSigmahat_k\bu-\|\bx_{\rm new}\|_2^2|\leq \|\bx_{\rm new}\|_2^2\lambda_{k},\\
&\;\|\bX_{k}\bu\|_{\infty}\leq \|\bx_{\rm new}\|_2\tau_{k} ,
\end{aligned}
\label{eq: relax}
\end{equation}
where $\lambda_{k}\asymp \sqrt{{\log p}/{n_k}}$ and $\sqrt{\log n_k}\lesssim \tau_{k}\ll \min\{\sqrt{n_1},\sqrt{n_2}\}.$ The additional constraint $\|\bX_{k}\bu\|_{\infty}\leq \|\bx_{\rm new}\|_2\tau_{k}$ has been proposed in \citet{javanmard2014confidence} to relax the Gaussian error assumption for establishing asymptotic normality for a single regression coefficient with $\bxnew=\be_j.$ The following result establishes the asymptotic normality of HITS under the general model \eqref{eq: general} with a small approximation error ${\bf r}_{k}$ and non-Gaussian error $\bepsilon_{k}.$
\begin{Proposition}
Suppose Condition {\rm (A1)} holds and for $k=1,2,$ $s_k\leq c{\sqrt{n_k}}/{\log p}$, $\|{\bf r}_{k}\|_2=o_{p}(1)$ and 
$\max_{k=1,2}\max_{1\leq i\leq n}\E(\epsilon_{k,i}^{2+\nu}\mid \bX_{k,i})\leq M_0$ for some constants $\nu>0$ and $M_0>0$. For the initial estimator $\widehat{\bbeta}_k$ given in \eqref{eq: Lasso estimator a} and the projection direction $\widehat{\bu}^{k}$ in \eqref{eq: relax}, 
the estimator $\widehat{\Deltanew}$ given in \eqref{eq: correction} satisfies the asymptotic limiting distribution \eqref{eq: limiting dist} under the model \eqref{eq: general}. 
\label{prop: limiting distribution relax}
\end{Proposition}
 
A few remarks are in order. Firstly, the asymptotic normality in Proposition \ref{prop: limiting distribution relax} holds for a broad class of estimators satisfying the condition ${\rm (B1)}$. For example, the Lasso estimator $\widehat{\bbeta}_k$ defined in \eqref{eq: Lasso estimator a} is an estimator satisfying this condition ${\rm (B1)}$. Other examples include the iterated Lasso estimator \citep{belloni2012sparse}, which is tuning free and shown to satisfy this condition by Theorem 1 of \citet{belloni2012sparse} and Proposition 9.7 of \citet{javanmard2017flexible} under the model \eqref{eq: general} with exact sparse $\bbeta_{k}.$ Secondly, to make the effect of the approximation errors ${\bf r}_{k}$ negligible for estimating $\bbeta_k$ under the model \eqref{eq: general}, the requirement is $\|{\bf r}_{k}\|_2=o_{p}(\|\bbeta_k\|_0)$ \citep{belloni2012sparse}; a stronger condition $\|{\bf r}_{k}\|_2=o_{p}(1)$ is imposed to guarantee the asymptotic normality, which is needed to show that the approximation error in estimating $\Deltanew$ is negligible in comparison to the main term that is the asymptotically normal. Thirdly, this limiting distribution result does not require independence between $\bepsilon_k$ and $\Xbb_k$ and the conditional moment conditions are sufficient for establishing the asymptotic normality. Lastly, this result can be used to construct CIs and conduct hypothesis testing as in \eqref{eq: ci} and \eqref{eq: test}, respectively, and the theoretical properties for  hypothesis testing  and confidence interval analogous to those in Corollaries \ref{Cor: hypothesis testing} and \ref{Cor: coverage} can be established.


%


\section{Optimality for Hypothesis Testing}
\label{sec: optimality}

We establish in this section the optimality of the proposed procedure in the hypothesis testing framework from two perspectives, minimaxity and adaptivity. 

\subsection{Optimality Framework for  Hypothesis Testing: Minimaxity and Adaptivity}
\label{sec: framework}
The performance of a testing procedure can be evaluated in terms of its size (type I error probability) and its power (or type II error probability). For a given test $\phi$ and a null parameter space $\HH_0(s)$,  its size is
\begin{equation}
{\alphab}(s,\phi)=\sup_{\theta\in \HH_0(s)} \E_{\theta} \phi. 
\label{eq: type I error}
\end{equation}
It has been shown in Corollary \ref{Cor: hypothesis testing} that the proposed test $\phi_{\alpha}$ satisfies ${\alphab}(s,\phi_{\alpha})\leq \alpha.$ 
To investigate the power, we consider the local alternative parameter space 
$
\HH_1(s,\tau)=\left\{\theta \in \Theta(s):\bx_{\rm new}^{\intercal}\left(\bbeta_1-\bbeta_2\right)=\tau\right\}, 
$
for a given $\tau>0$.
The power of a test $\phi$ over the parameter space $\HH_1(s,\tau)$ is 
 \begin{equation}
\omegab(s,\tau,\phi)=\inf_{\theta\in \HH_1(s,\tau)} \E_{\theta}\phi.
\label{eq: power definition}
\end{equation}
With a larger value of $\tau$, the alternative parameter space is further away from the null parameter space and hence it is easier to construct a test of higher power. The minimax optimality can be reduced to identifying the smallest $\tau$ such that the size is controlled over $\HH_0(s)$ and the corresponding power over $\HH_1(s,\tau)$ is large, that is,
\begin{equation}
\taumini=\argmin_{\tau}\left\{\tau: \sup_{\phi: \alphab(s,\phi)\leq \alpha}\omegab(s,\tau,\phi)\geq 1-\eta\right\},
\label{eq: detection boundary minimaxity}
\end{equation}
where $\eta\in[0,1)$ is a small positive constant controlling the type II error probability. The quantity $\taumini$ depends on $\bx_{\rm new}$, the sparsity level $\kt$ and the constants $\alpha,\eta\in(0,1)$. Throughout the discussion, we omit $\alpha$ and $\eta$ in the arguments of $\taumini$ for simplicity. 
This quantity $\taumini$ is referred to as the minimax detection boundary of the hypothesis testing problem \eqref{eq: testing problem}. In other words, $\taumini$ is the minimum distance such that there exists a test controlling size and achieving a good power. If a test $\phi$ satisfies the following property, 
\begin{equation}
\alphab(s,\phi)\leq \alpha \quad \text{and} \quad \omegab(s, \phi,\tau)\geq 1-\eta \quad \text{for}\;\; \tau \asymp \taumini
\label{eq: minimax optimal}
\end{equation}
then $\phi$ is defined to be minimax optimal. The minimax detection boundary in \eqref{eq: detection boundary minimaxity} is defined for a given sparsity level $s$, which is assumed to be known a priori. However, the exact sparsity level is typically unknown in practice. Hence, it is also of great importance to consider the optimality from the following two perspectives on adaptivity, 
\setlength{\leftmargini}{0.4in}

\begin{enumerate}
\item[Q1.] Whether it is possible to construct a test achieving the minimax detection boundary defined in \eqref{eq: detection boundary minimaxity} if the true sparsity level $\kt$ is unknown \item[Q2.] What is the optimal procedure in absence of accurate sparsity information?  
\end{enumerate}
To facilitate the definition of adaptivity, we consider the case of two sparsity levels, $\kt\leq \ku$. Here $\kt$ denotes the true sparsity level, which is typically not available in practice while $\ku$ denotes the prior knowledge of an upper bound for the sparsity level. If we do not have a good prior knowledge about the sparsity level $s$, then $\ku$ can be much larger than $\kt$. To answer the aforementioned questions on adaptivity, we assume that only the upper bound $\ku$ is known instead of the exact sparsity level $\kt$.
As a consequence, a testing procedure needs to be constructed with the size uniformly controlled over the parameter space $\HH_0(\ku),$
\begin{equation}
{\alphab}(\ku,\phi)=\sup_{\theta\in \HH_0(\ku)} \E_{\theta} \phi \leq \alpha. 
\label{eq: type I error adaptivity}
\end{equation}
In contrast to \eqref{eq: type I error}, the control of size is with respect to $\HH_0(\ku)$, a larger parameter space  than $\HH_0(\kt)$, due to the fact that the true sparsity level is unknown in conducting the hypothesis testing. 
Similar to \eqref{eq: detection boundary minimaxity}, we define the adaptive detection boundary $\tauadpt$ as
\begin{equation}
\tauadpt=\arg\min_{\tau}\left\{\tau: \sup_{\phi:\alphab(\ku,\phi)\leq \alpha}\omegab(\kt,\tau,\phi)\geq 1-\eta\right\}.
\label{eq: detection boundary adaptivity}
\end{equation}
Comparing \eqref{eq: detection boundary adaptivity} with \eqref{eq: detection boundary minimaxity}, the imprecise information about the sparsity level only affects the control of size, where the power functions in \eqref{eq: detection boundary adaptivity} and \eqref{eq: detection boundary minimaxity} are evaluated over the same parameter space, $\HH_1(\kt,\tau)$.
 If a test $\phi$ satisfies the following property,\begin{equation}
\alphab(\ku,\phi)\leq \alpha \quad \text{and} \quad \omegab(\kt,\tau,\phi)\geq 1-\eta \quad \text{for}\;\; \tau \asymp \tauadpt
\label{eq: adaptive optimal}
\end{equation}
then $\phi$ is defined to be adaptive optimal. 

The quantities $\taumini$ and $\tauadpt$ do not depend on the specific testing procedure but mainly reflect the difficulty of the testing problem \eqref{eq: testing problem}, which depends on the parameter space and also the loading vector $\bx_{\rm new}$. The question {Q1} can be addressed through comparing $\taumini$ and $\tauadpt$; if $\taumini \asymp \tauadpt,$ then the hypothesis testing problem \eqref{eq: testing problem} is defined to be adaptive, that is, even if one does not know the exact sparsity level, it is possible to construct a test as if the sparsity level is known; in contrast, if $\taumini \ll \tauadpt,$ the hypothesis testing problem \eqref{eq: testing problem} is defined to be not adaptive. The information on the sparsity level is crucial. In this case, the adaptive detection boundary itself is of great interest as it quantifies the effect of the knowledge of sparsity level. The question Q2 can be addressed using the adaptive detection boundary $\tauadpt$ and an adaptive optimal test satisfying \eqref{eq: adaptive optimal} would be the best that we can aim for if there is lack of accurate information on sparsity.

As a concluding remark, the minimax detection boundary characterizes the difficulty of the testing problem for the case of known sparsity level while the adaptive detection boundary characterizes a more challenging problem due to the unknown sparsity. The adaptive optimal test satisfying \eqref{eq: adaptive optimal} is more useful in practice than that of a minimax optimal test because the exact sparsity level is typically unknown in applications. 
\subsection{Detection Boundary for Testing Problem \eqref{eq: testing problem}}
\label{sec: results}
We now demonstrate the optimality of the proposed procedure by considering the following two types of loadings $\bx_{\rm new}$, Exact Loading and Decaying Loading.\\
%
\noindent (E) {\bf Exact Loading.} $\bx_{\rm new}$ is defined as an exact loading if it satisfies,
\begin{equation}
c_0\leq \max_{\{i: \bx_{\rm new, i}\neq 0\}}\left|\bx_{\rm new, i}\right|/\min_{\{i: \bx_{\rm new, i}\neq 0\}}\left|\bx_{\rm new, i}\right|\leq C_0,
\label{eq: exact loading}
\end{equation}
for some positive constants $C_0\geq c_0>0$. The condition \eqref{eq: exact loading} assumes that all non-zero coefficients of the loading vector $\bx_{\rm new}$ are of the same order of magnitude
and hence the complexity of an exact loading $\bx_{\rm new}$ is captured by its sparsity level.  We calibrate the sparsity levels of the regression vectors and the exact loading $\bx_{\rm new}$ as
\begin{equation}
\kt=p^{\gamma},\quad \ku=p^{\gamma_{u}}, \quad \|\bx_{\rm new}\|_0=p^{\gamma_{\rm new}} \; \text{for} \quad 0\leq \gamma< \gamma_{u} \leq 1,\; 0\leq  \gamma_{\rm new} \leq 1.
\label{eq: cal of beta}
\end{equation}
Based on the sparsity level of $\bx_{\rm new}$, we define the following types of loadings,
\begin{enumerate}
\item [(E1)] $\bx_{\rm new}$ is called an {\em exact sparse loading} if it satisfies \eqref{eq: exact loading} with $\gamma_{\rm new}\leq 2 \gamma$; 
\item [(E2)] $\bx_{\rm new}$ is called an {\em exact dense loading} if it satisfies \eqref{eq: exact loading} with $\gamma_{\rm new}>2 \gamma$; 
\end{enumerate} 

Exact loadings are commonly seen in the genetic studies, where the loading $\bx_{\rm new}$ represents a specific observation's SNP and only takes the value from $\{0,1,2\}$.

\medskip
\noindent (D) {\bf Decaying Loading.} Let 
$|x_{\rm new, (1)}|\geq \left|x_{\rm new, (2)}\right|\geq \cdots \geq \left|x_{\rm new, (p)}\right|$ be the sorted coordinates of $\left|\bx_{\rm new}\right|$.
We say that $\bx_{\rm new}$ is decaying at the rate $\delta$ if 
\begin{equation}
\left|x_{\rm new, (i)}\right| \asymp i^{-\delta} \quad \text{for some constant } \delta\geq 0.
\label{eq: decaying}
\end{equation}
Depending on the decaying rate $\delta$, we define the following two types of loadings,
\begin{enumerate}
\item [(D1)] $\bx_{\rm new}$ is called a {\em fast decaying loading} if it satisfies \eqref{eq: decaying} with $\delta\geq \frac{1}{2}$; 
\item [(D2)] $\bx_{\rm new}$ is called a {\em slow decaying loading} if it satisfies \eqref{eq: decaying} with  $0\leq \delta<\frac{1}{2}$.
\end{enumerate}



%
%
%

To simplify the presentation, we present the optimality results for the case $n_1\asymp n_2$, denoted by $n$ and the results can be extended to the case where $n_1$ and $n_2$ are of different orders. 
We focus on the exact loading and give a summary of the results for the decaying loading in Table \ref{tab: decaying theory}.
The detailed results about decaying loadings will be deferred to Section \ref{sec: decaying} in the supplement. The following theorem establishes the lower bounds for the adaptive detection boundary for exact loadings.
\begin{Theorem}
Suppose that  $s\leq \ku \lesssim {n}/{\log p}$. We calibrate $\kt, \ku$ and $\|\bx_{\rm new}\|_0$ by $\gamma$, $\gamma_{u}$ and $\gamma_{\rm new}$, respectively, as defined in \eqref{eq: cal of beta}.  
\begin{enumerate}
\item[\rm (E1)] If $\bx_{\rm new}$ is an exact sparse loading, then 
\begin{equation}
\tauadpt \gtrsim \frac{\sqrt{\|\bx_{\rm new}\|_0}\|\bx_{\rm new}\|_{\infty}}{\sqrt{n}} \asymp \frac{\|\bx_{\rm new}\|_2}{\sqrt{n}}; 
\label{eq: E1}
\end{equation}
\item[\rm (E2)] If $\bx_{\rm new}$ is an exact dense loading, then 
\begin{equation}
\tauadpt \gtrsim \begin{cases} \|x_{\rm new}\|_{\infty}\ku \sqrt{\frac{\log p}{n}}&\text{if}\quad \gamma_{\rm new}> 2\gamma_{u};\\
 \frac{\|\bx_{\rm new}\|_2}{\sqrt{n}} & \text{if}\quad \gamma_{\rm new} \le 2\gamma_{u}.
 \end{cases}
 \label{eq: E2}
\end{equation}
\end{enumerate}
\label{thm: exact sparse lower}
\end{Theorem}

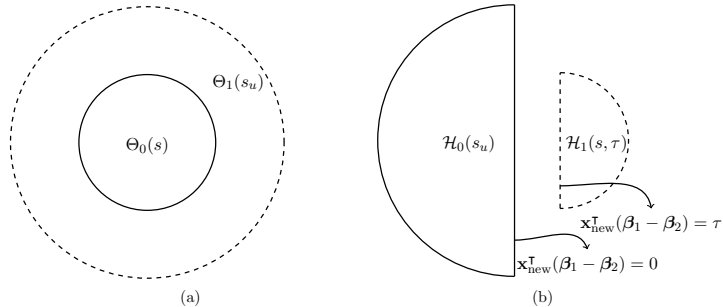
\begin{figure}[htb]
\scalebox{.6}{
  \begin{subfigure}[b]{0.48\textwidth}
   \begin{tikzpicture}
\draw[black,thick] (2,2) circle (1.5cm);
\draw[black,thick,dashed] (2,2) circle (3cm);
 \draw (2,1.6) node[above]{$\Theta_0(\kt)$};
 \draw (4,3) node[above]{$\Theta_1(\ku)$};
\end{tikzpicture}
    \caption{}
    \label{fig:1}
  \end{subfigure}
  \begin{subfigure}[b]{0.44\textwidth}
   \begin{tikzpicture}
   \draw[black,thick] (2,2) arc (270:90:3cm);
   \draw[black,thick,dashed] (3,6.5) arc (90:-90:1.5cm);
   \draw[black,thick] (2,2) -- (2,8); 
   \draw[black,thick,dashed] (3,3.5) -- (3,6.5); 
    \draw (1,4.6) node[above]{$\HH_0(\ku)$};
 \draw (3.8,4.6) node[above]{$\HH_1(\kt,\tau)$};
 \draw[->,thick] (2,2.8) to [out=0,in=-250] (3.6,2.6) node[anchor= north]
        {$\bx_{\rm new}^{\intercal}(\bbeta_1-\bbeta_2)=0$};
   \draw[->,thick] (3,4) to [out=0,in=100] (5,3.5) node[anchor= north]
        {$\bx_{\rm new}^{\intercal}(\bbeta_1-\bbeta_2)=\tau$};      
    \end{tikzpicture}
    \caption{}
    \label{fig:2}
  \end{subfigure}
  }
  \caption{(a) Null and alternative parameter spaces for the confidence set construction; (b) Null and alternative parameter spaces for the hypothesis testing problem.}
\label{fig: tool different}
\end{figure}

We shall point out here that establishing the adaptive detection boundaries in Theorem \ref{thm: exact sparse lower} requires technical novelty.
A closely related problem, adaptivity of confidence sets, has been carefully studied in the context of high-dimensional linear regression \citep{nickl2013confidence,cai2015regci,cai2018accuracy}. However, it requires new technical tools to establish the adaptive detection boundaries, due to the different geometries demonstrated in Figure \ref{fig: tool different}. 
The main idea of constructing the lower bounds in \citet{nickl2013confidence,cai2015regci,cai2018accuracy} is illustrated in Figure 2(a), where one interior point is first chosen in the smaller parameter space $\Theta(\kt)$ and the corresponding least favorable set is constructed in the larger parameter space $\Theta(\ku)$ such that they are not distinguishable. 

In comparison to Figure 2(a), the lower bound construction for the testing problem related to Figure 2(b) is more challenging due to the fact that the alternative parameter space $\HH_1(\ku,\tau)$ does not contain the indifference region $0<\bx_{\rm new}^{\intercal}(\bbeta_1-\bbeta_2)<\tau$.  A new technique, {\em transferring technique}, is developed for establishing the sharp lower bounds for the adaptive detection boundary. Define the index of $\bx_{\rm new}$ with the largest absolute value as $i_{\max}=\arg\max|\bx_{\rm new, i}|.$ In constructing the least favorable set in $\HH_0(\ku)$, we first perturb the regression coefficients at other locations except for $i_{\rm max}$ and then choose the regression coefficient at $i_{\rm max}$ such that $\bx_{\rm new,i_{\rm max}}(\bbeta_{1,i_{\rm max}}-\bbeta_{2,i_{\rm max}})>0$ and $\bx_{\rm new}^{\intercal}(\bbeta_1-\bbeta_2)\leq 0$; in construction of the corresponding least favorable set in $\HH_1(\kt,\tau)$, we simply set the regression coefficient with index $i_{\max}$ to be the same as the corresponding coefficient at $i_{\max}$ in $\HH_0(\ku)$  and set all other coefficients to be zero. The above construction is transferring the parameter space complexity from $\HH_0(\ku)$ to $\HH_1(\kt,\tau)$ by matching the regression coefficient at $i_{\max}$. Such a transferring technique can be of independent interest in establishing the adaptive detection boundaries  for other inference problems.

The following corollary presents the matched upper bounds for the detection boundaries established in Theorem \ref{thm: exact sparse lower} over certain regimes.
\begin{Corollary}
\label{cor: exact sparse adapt}
Suppose that  $s\leq \ku \lesssim {\sqrt{n}}/{\log p}$. 
\begin{enumerate}
\item[\rm (E1)] If the loading $\bx_{\rm new}$ is an exact sparse loading, then 
\begin{equation}
\tauadpt \asymp \taumini \asymp \frac{\|\bx_{\rm new}\|_2}{\sqrt{n}}
\label{eq: E1 match}
\end{equation}
\item[\rm (E2)]  If the loading $\bx_{\rm new}$ is an exact dense loading, then the results are divided into the following two cases,
\begin{enumerate}
\item[\rm (E2-a)] If $\gamma< \gamma_{u}< \frac{1}{2}\gamma_{\rm new},$ then 
\begin{equation}
\tauadpt \asymp\|x_{\rm new}\|_{\infty}\ku \sqrt{\frac{\log p}{n}} \gg \taumini \asymp\|x_{\rm new}\|_{\infty} \kt \sqrt{\frac{\log p}{n}}.
\label{eq: E2-a match}
\end{equation}
\item[\rm (E2-b)]  If $\gamma< \frac{1}{2}\gamma_{\rm new}\leq \gamma_{u} ,$ then
\begin{equation}
\tauadpt \asymp \frac{\|\bx_{\rm new}\|_2}{\sqrt{n}}\gg \taumini \asymp \|x_{\rm new}\|_{\infty}\kt \sqrt{\frac{\log p}{n}}.
\label{eq: E2-b match}
\end{equation}
\end{enumerate}
\end{enumerate}
\end{Corollary}
The question Q1 about the possibility of adaptivity of the testing problem \eqref{eq: testing problem} can be addressed by the above corollary, where the testing problem is adaptive for the exact sparse loading case (E1) but not adaptive for the exact dense loading case (E2). The specific cut-off for the ``dense" and ``sparse" cases occurs at $\gamma_{\rm new}=2\gamma$.
For the case (E2), depending on the value of $\gamma_{u}$, the adaptive detection boundaries can be quite different. The case (E2-a) corresponds to the case that the exact sparsity level is unknown but the upper bound $\ku$ is relatively precise (both $\gamma$ and $\gamma_u$ are below $1/2 \cdot \gamma_{\rm new}$), then we can utilize the proposed procedure $\phi_{\alpha}$ with the sparsity information $\ku$ to construct a testing procedure matching the adaptive detection boundary. See the detailed construction in Section \ref{sec: sparsity procedure} of the supplement. In contrast, the case (E2-b) corresponds to the setting where the prior knowledge of the upper bound $\ku$ is quite rough. For such a case, the proposed testing procedure $\phi_{\alpha}$ defined in \eqref{eq: test} achieves the adaptive detection boundary $\tauadpt$, but not the minimax detection boundary $\taumini$. 

Beyond answering Q1, we can also address the question Q2 with the following corollary, which considers the practical setting that there is limited information on sparsity and presents a unified optimality result for the case of exact loadings. 
\begin{Corollary}
\label{cor: optimality}
Suppose that  $s\leq \ku \lesssim {\sqrt{n}}/{\log p}$ and $\gamma_{u}\geq \gamma_{\rm new}/2$. Then the testing procedure $\phi_{\alpha}$ in \eqref{eq: test} achieves the adaptive detection boundary $\tauadpt \asymp {\|\bx_{\rm new}\|_2}/{\sqrt{n}}$ for any $\bx_{\rm new}$ satisfying \eqref{eq: exact loading}. 
\end{Corollary}
The above corollary states that, in absence of accurate sparsity information, the proposed procedure $\phi_{\alpha}$ is an adaptive optimal test for all exact loadings $\bx_{\rm new}$. 
\subsection{Comparison with Existing Optimality Results on CI}
\label{sec: comparison}


It is helpful to compare the established optimality results to the related work \citet{cai2015regci} on the minimaxity and adaptivity of confidence intervals for the linear contrast in the one-sample high-dimensional regression. 
Beyond the technical difference highlighted in Figure \ref{fig: tool different}, we also observed the following three distinct features between the present paper and  \citet{cai2015regci}.

\begin{enumerate}
\item The current paper closes the gap between the sparse loading regime and the dense loading regime in \citet{cai2015regci}, where the lower bounds for the exact sparse loading only covered the case $\gamma_{\rm new}\leq \gamma$ instead of the complete regime $\gamma_{\rm new}\leq 2\gamma$ defined in this paper. 
\item In comparison to \citet{cai2015regci}, the current paper considers the additional setting (E2-b), which corresponds to the case where the knowledge on the sparsity level is rough. This additional result is not only of technical interest, but has broad implications to practical applications. It addresses the important question, ``what is the optimal testing procedure in a practical setting where no accurate sparsity information is available? " As shown in Corollary \ref{cor: optimality}, the proposed procedure $\phi_{\alpha}$ is an adaptive optimal test for all exact loadings $\bx_{\rm new}.$
\item In addition, Theorem \ref{thm: general lower} develops the technical tools  for a general loading $\bxnew$,  which includes the loadings not considered in \citet{cai2015regci}. Specifically, we summarize in Table \ref{tab: decaying theory} the optimality results for the decaying loading defined in \eqref{eq: decaying}. As shown in Table \ref{tab: decaying theory}, the fast decaying loading (D1) is similar to the exact sparse loading (E1) while the slow decaying loading (D2) is similar to the exact dense loading (E2). In contrast, the decaying loading has the distinct setting (D2-c) from the exact loading case where the hypotheses \eqref{eq: testing problem} can be tested adaptively if both $\gamma$ and $\gamma_{u}$ are above $1/2$. See the detailed discussion in Section \ref{sec: decaying} of the supplement.
\end{enumerate}
\begin{table}[htb]
\renewcommand{\arraystretch}{1.3}
\begin{tabular}{|c|c|c|c|c|c|c|}
\hline
$\delta$&Setting& $\gamma,\gamma_u$&$\taumini$&Rel&$\tauadpt$ &Adpt\\
\hline
$[{1}/{2},\infty)$& (D1)& $\gamma<\gamma_u$& ${\|\bx_{\rm new}\|_2}/{\sqrt{n}}$&$\asymp$&${\|\bx_{\rm new}\|_2}/{\sqrt{n}}$&Yes\\ 
\hline 
\multirow{3}{*}{$[0,{1}/{2})$}&(D2-a)&$\gamma< \gamma_{u}\leq \frac{1}{2}$&${\kt^{1-2\delta}}\left(\log p\right)^{\frac{1}{2}-\delta}/{\sqrt{{{n}}}}$&$\ll$&${\ku^{1-2\delta}}\left(\log p\right)^{\frac{1}{2}-\delta}/{\sqrt{{{n}}}}$&No\\
&(D2-b)&$\gamma< \frac{1}{2}\leq \gamma_{u}$&${\kt^{1-2\delta}}\left(\log p\right)^{\frac{1}{2}-\delta}/{\sqrt{{{n}}}}$&$\ll$&${\|\bx_{\rm new}\|_2}/{\sqrt{n}}$&No\\
&(D2-c)&$\frac{1}{2}\leq \gamma<\gamma_{u}$& ${\|\bx_{\rm new}\|_2}/{\sqrt{n}}$&$\asymp$&${\|\bx_{\rm new}\|_2}/{\sqrt{n}}$&Yes\\
    \hline
\end{tabular}
\caption{Optimality for the decaying loading $\bx_{\rm new}$ defined in \eqref{eq: decaying} over the regime $s\lesssim \ku\lesssim {\sqrt{n}}/{\log p}$. The column indexed with ``Rel" compares $\taumini$ and $\tauadpt$ and the column indexed with ``Adpt" reports whether the testing problem is adaptive in the corresponding setting.}
\label{tab: decaying theory}
\end{table}

\section{Uncertainty Quantification related to High-dimensional Prediction}
\label{sec: prediction}

As mentioned in the introduction, the hypothesis testing method developed in the current paper can also be used for the prediction problem in a single high-dimensional regression. Consider the regression model with i.i.d observations $\{(X_{i\cdot},y_i)\}_{1\leq i\leq n}$ satisfying
\begin{equation}
y_i=X_{i\cdot}^{\intercal}\bbeta+\epsilon_i \quad \text{where} \; \epsilon_i\stackrel{iid}{\sim} N(0,\sigma^2)\;\; \text{for}\;\; 1\leq i\leq n,
\label{eq: single regression}
\end{equation} 
and $\{\epsilon_i\}_{1\leq i\leq n}$ is independent of the design matrix $X$.
The problem of interest is inference for the conditional expectation $\E(y_i\mid X_{i\cdot}=\bx\subnew)=\bx\subnew^{\intercal}\bbeta.$ Uncertainty quantification for $\bx\subnew^{\intercal}\bbeta$ is a one-sample version of the testing problem \eqref{eq: test}. Due to its importance and for clarity, we  present a separate result on this prediction problem. We use $\widehat{\bbeta}$ to denote the scaled Lasso estimator of $\bbeta$ based on the observations $\{(X_{i\cdot},y_i)\}_{1\leq i\leq n}$ and construct the following bias-corrected point estimator,
$
\widehat{\bx\subnew\trans\bbeta}=\bx\subnew\trans\bbetahat+\buhat^{\intercal} \frac{1}{n}\sum_{i=1}^{n} \bX_{i\cdot}(y_{i}-\bX_{i\cdot}\trans \widehat{\bbeta}) 
$
with the projection direction defined as 
\begin{align*}
\buhat=\;\argmin_{\bu\in \R^{p}} \bu\trans \bSigmahat \bu \quad \text{subject to}\;
& \left \|\bSigmahat \bu-\bx_{\rm new}\right\|_{\infty}\leq  \|\bx_{\rm new}\|_2 \lambda\\
&\;\left |\bx_{\rm new}\trans \bSigmahat \bu-\|\bx_{\rm new}\|_2^2 \right|\leq \|\bx_{\rm new}\|_2^2\lambda, 
\end{align*}
where $\widehat{\Sigmab}=\frac{1}{n}\sum_{i=1}^{n}X_{i\cdot} X_{i\cdot}^{\intercal}$ and $\lambda\asymp \sqrt{{\log p}/{n}}$. The key difference between this construction and the projection construction for the single regression coefficient in \citet{zhang2014confidence,van2014asymptotically,javanmard2014confidence,athey2018approximate} is the additional constraint $\left |\bx_{\rm new}\trans \bSigmahat \bu-\|\bx_{\rm new}\|_2^2 \right|\leq \|\bx_{\rm new}\|_2^2\lambda,$ which guarantees the asymptotic limiting distribution for any $\bx_{\rm new}\in \R^{p}$. We consider the capped-$\ell_1$ sparsity as in \eqref{eq: capped l1}, $
\sum_{j=1}^{p} \min\{ |{\bbeta}_{j}|/\sigma\lambda_0 ,1\} \leq s, $  and introduce the following general condition for the initial estimator $\widehat{\bbeta}$ and then establish the limiting distribution for the point estimator $\widehat{\bx\subnew\trans\bbeta}$ in Corollary \ref{Cor: prediction}.
\begin{enumerate}
\item[(P)] With probability at least $1-g(n)$ where $g(n)\rightarrow 0$, $\|\widehat{\bbeta}-\bbeta\|_1\lesssim s\sqrt{{ \log p}/{n}}.$
\end{enumerate}
\begin{Corollary}
Suppose that the regression model \eqref{eq: single regression} holds where $s\leq c{\sqrt{n}}/{\log p}$ and the rows $X_{i\cdot} $ are i.i.d. $p$-dimensional sub-gaussian random vectors with mean $\bmu=\E X_{i\cdot}$ and the second order moment $\bSigma=\E (X_{i\cdot}X_{i\cdot}\trans $) satisfying $c_0 \leq \lambda_{\min}\left(\bSigma\right) \leq \lambda_{\max}\left(\bSigma\right) \leq C_0$ for positive constants $C_0,c_0>0$. For any initial estimator $\widehat{\bbeta}$ satisfying the condition $\rm (P)$, then 
$
\frac{1}{\sqrt{{{\sigma}^2}\buhat\trans  \bSigmahat\buhat/n}}\left(\widehat{\bx\subnew\trans\bbeta}-\bx\subnew\trans\bbeta\right)\overset{d}{\rightarrow} N(0,1).
$
\label{Cor: prediction}
\end{Corollary}
Based on this corollary, we use $\widehat{\rm V}={\widehat{\sigma}^2}\buhat\trans  \bSigmahat\buhat/n$ to estimate the variance of $\widehat{\bx\subnew\trans\bbeta}$ and construct the CI, 
$
{\rm CI}_{\rm Pred}=\left(\widehat{\bx\subnew\trans\bbeta}-{z_{\alpha/2}}\sqrt{{{\widehat{\sigma}^2}\buhat\trans  \bSigmahat\buhat}/n},\quad \widehat{\bx\subnew\trans\bbeta}+{z_{\alpha/2}}\sqrt{{{\widehat{\sigma}^2}\buhat\trans  \bSigmahat\buhat}/n}\right).
$
If $\widehat{\sigma}^2$ is a consistent estimator of $\sigma^2$, then this constructed CI is guaranteed to have coverage for $\bx_{\rm new}^{\intercal}\beta$ for any $\bx_{\rm new}\in \R^{p}.$
%
%
The optimality theory established in Section \ref{sec: optimality} can be easily extended to the one-sample case. 

\section{Simulation Studies}
\label{sec: simulation}

In this section, we present numerical studies comparing the performance of HITS estimation and inference to existing methods. Before presenting the numerical results, we first introduce the equivalent dual form to find the projection direction defined in \eqref{eq: constraint 1} and \eqref{eq: constraint 2}. 
\begin{Proposition}
The constrained optimizer $\buhat_k\in \R^{p}$ for $k=1,2$ defined in \eqref{eq: constraint 1} and \eqref{eq: constraint 2} can be computed in the form of $\buhat_k=-\frac{1}{2}[\bvhat^{k}_{-1}+\frac{\bx_{\rm new}}{\|\bx_{\rm new}\|_2}\bvhat^{k}_{1}],$ where $\bvhat^{k}\in \R^{p+1}$ is defined as 
\begin{equation}
\bvhat^{k}=\argmin_{\bv \in \R^{p+1}} \left\{\frac{1}{4}\bv\trans \Hbb^{\intercal}\bSigmahat_k\Hbb \bv+\bx_{\rm new}^{\intercal} \Hbb \bv+\lambda_{k}\|\bx_{\rm new}\|_2 \cdot\|\bv\|_1 \right\}
\label{eq: dual problem}
\end{equation}
with 
$\Hbb=\left[\begin{matrix} \frac{\bx_{\rm new}}{\|\bx_{\rm new}\|_2}, \mathbb{I}_{p\times p} \end{matrix} \right] \in \R^{p \times (p+1)}$.
\label{prop: dual form}
\end{Proposition}
Proposition \ref{prop: dual form} allows us to transform the constrained minimization problem to the unconstrained minimization problem in \eqref{eq: dual problem}, which can be solved via standard penalized least squares algorithms. 

We first consider the setting with exact sparse regression vectors and loadings $\bxnew$ generated from Gaussian distributions in Section \ref{sec: general loading}. Results for the setting with decaying loadings are given  in Section \ref{sec: decaying loading} in the supplementary materials. In Section \ref{sec: approx sparse}, we consider settings of approximately sparse regression. Throughout, we let $p = 501$ including intercept and $n_1 = n_2 = n$ with various choices of $n$. For simplicity, we generate the covariates $(\bX_{k,i})_{-1}$  from the same multivariate normal distribution with zero mean and covariance $\bSigma = [0.5^{1+|j-l|}]_{(p-1)\times(p-1)}$. 

\subsection{Exact Sparse Regression with General Loading}
\label{sec: general loading}
We consider the exact sparse regression in the following. To simulate $\bY_1$ and $\bY_2$, we generate $\epsilon_{k,i}$ from the standard normal and set $\beta_{1,1}=-0.1, \beta_{1,j}=-{\mathbf 1}(2 \le j \le 11)0.4(j-1), \beta_{2,1}=-0.5$, and $\beta_{2,j}=0.2(j-1) {\mathbf 1}(2 \le j \le 6)$. 
We consider the case with the loading $\bx\subnew$ being a dense vector, generated via two steps. In the first step, we generate $\bx_{\rm basis} = [1, \bx_{{\rm basis},-1}\trans]\trans \in \R^{p}$ with $\bx_{{\rm basis},-1}\sim N(0,\bSigma) .$
In the second step, we generate $\bx_{\rm new}$ based on $\bx_{\rm basis}$ in two specific settings, 
\begin{enumerate}[label=(\arabic*)]
\item generate $\bx_{\rm new}$ as a shrunk version of $x_{\rm basis}$ with
\begin{equation}
x_{{\rm new},j}={\Ssc} \cdot {{\mathbf 1}(j \ge 12)}\cdot x_{{\rm basis},j}  , \mbox{\quad for $j = 1, ..., p$}
\label{eq: loading gen}
\end{equation} 
and ${\Ssc} \in\{1,0.5,0.2,0.1\}$. 
\item let 
$x_{{\rm new},j}={\mathbf 1}(j = 1) -\frac{2}{3}{\mathbf 1}(j = 2) + {\Ssc}\cdot {\mathbf 1}(j \ge 12)\cdot x_{{\rm basis},j}$  for $j = 1, ..., p$
and ${\Ssc} \in\{1,0.5,0.2,0.1\}$. 
\end{enumerate}
Under the above configurations, the scale parameter $\Ssc$ controls the magnitude of the noise variables in $\bx\subnew$. As $\Ssc$ increases, $\|\bx\subnew\|_2$ increases but $\Delta\subnew$ remains the same for all choices of $\Ssc$. 
Setting 1 corresponds to an alternative setting with $\Deltanew = 1.082$ and setting 2 corresponds to the null setting with $\Delta\subnew = 0$.

We report the simulation results based on $1,000$ replications for each setting in Tables \ref{tab: Gaussian setting 1} and \ref{tab: Gaussian setting 2}. 
Under Setting 1, as the sample size $n$ increases and as the magnitude of the noise variables decreases, the statistical inference problem becomes ``easier" in the sense that the CI length and root mean square error (RMSE) get smaller, the empirical rejection rate (ERR) gets closer to $100\%$, where ERR denotes the proportion of null hypotheses being rejected out of the total $1,000$ replications and is an empirical measure of power under the alternative. 
This is consistent with the established theoretical results. The most challenging setting for HITS is the case with ${\Ssc}=1$, where the noise variables are of high magnitude. As a result, the HITS procedure has a lower power in detecting the treatment effect even when $n = 1,000$. When ${\Ssc}$ drops to $0.2$, the power of the HITS is about $72\%$ when $n=200$ and $95\%$ when $n = 400$. Across all sample sizes considered including when $n = 100$, the empirical coverage of the CIs are close to the nominal level. 
 

In Table \ref{tab: Gaussian setting 1},  HITS is compared with the plug-in Lasso estimator (shorthanded as Lasso) and plug-in debiased estimator (shorthanded as Deb) in terms of RMSE. For Lasso, the regression vectors are estimated by the scaled Lasso in the R package {\rm FLARE} \citep{li2015flare}; For Deb, the regression vectors are estimated by the debiased estimators \citet{javanmard2014confidence} using the code at \url{https://web.stanford.edu/~montanar/sslasso/code.html}. Across all settings,  HITS always outperforms Deb; in comparison to Lasso, HITS has substantially smaller bias but at the expense of larger variance, reflecting the bias-variance trade-off.  Specifically, when $\Ssc$ is small (taking values in $\{0.2,0,1\}$), HITS generally has a smaller RMSE than Lasso. When $\Ssc = 1,0.5$ in which case $\bx\subnew$ is dense, HITS has a much larger variability compared to  Lasso. This suggests that under the challenging dense scenario, a high price is paid to ensure the validity of the interval estimation.

We further compare HITS with the two plug-in estimators in the context of hypothesis testing. The Lasso estimator is not useful in hypothesis testing or CI construction due to the fact that the bias component is as large as the variance. In contrast, the variance component of both HITS and Deb dominates the corresponding bias component. Due to this reason, we only report the empirical comparison between the HITS method and the Deb method. As illustrated in the coverage property, the empirical coverage of CIs based on the Deb estimator is about 10\% below the nominal level while our proposed CI achieves the nominal level. 

In Table \ref{tab: Gaussian setting 2}, we report the results in Setting 2 where the null hypothesis holds. 
We observe a similar pattern as in Setting 1 for estimation accuracy and relative performance compared to the Lasso and Deb estimators. All the ERRs in this case, which correspond to the empirical size, are close to the nominal level $5\%$ for the HITS method while the corresponding ERRs cannot be controlled for the Deb estimators.


\begin{table}[!htbp]
\centering
\scalebox{0.8}{
\begin{tabular}{|r|r|rr|rr|r|rrr|rrr|rrr|}
  \hline
  &&\multicolumn{2}{c}{ERR}\vline& \multicolumn{2}{c}{Coverage}\vline&Len&\multicolumn{3}{c}{HITS}\vline&\multicolumn{3}{c}{Lasso}\vline&\multicolumn{3}{c}{Deb}\vline\\
  \hline
  ${\Ssc}$&n& HITS & Deb & HITS & Deb & HITS & RMSE& Bias & SE & RMSE& Bias & SE  & RMSE& Bias & SE  \\ 
  \hline
\multirow{5}{*}{1}&100& 0.10 & 0.25 & 0.97 & 0.81 & 9.33 & 2.10 & 0.05 & 2.10 & 0.90 & 0.61 & 0.66 & 2.56 & 0.07 & 2.56 \\ 
  &200 & 0.11 & 0.26 & 0.97 & 0.86 & 7.65 & 1.79 & 0.01 & 1.79 & 0.61 & 0.41 & 0.45 & 1.97 & 0.02 & 1.97 \\ 
  &400 & 0.20 & 0.30 & 0.97 & 0.85 & 5.38 & 1.30 & 0.03 & 1.30 & 0.43 & 0.31 & 0.30 & 1.62 & 0.02 & 1.62 \\ 
  &600 & 0.23 & 0.36 & 0.97 & 0.87 & 4.47 & 1.02 & 0.07 & 1.02 & 0.34 & 0.24 & 0.24 & 1.34 & 0.05 & 1.34 \\ 
  &1000 & 0.34 & 0.32 & 0.97 & 0.85 & 3.49 & 0.83 & 0.02 & 0.83 & 0.26 & 0.19 & 0.19 & 1.49 & 0.05 & 1.49 \\ 
  \hline
 \multirow{5}{*}{0.5}&100& 0.16 & 0.36 & 0.96 & 0.83 & 4.80 & 1.10 & 0.13 & 1.09 & 0.81 & 0.64 & 0.49 & 1.27 & 0.03 & 1.27 \\ 
   &200& 0.30 & 0.46 & 0.95 & 0.83 & 3.90 & 0.96 & 0.08 & 0.96 & 0.51 & 0.40 & 0.33 & 1.09 & 0.08 & 1.09 \\ 
   &400& 0.49 & 0.57 & 0.94 & 0.86 & 2.74 & 0.67 & 0.09 & 0.67 & 0.34 & 0.25 & 0.22 & 0.82 & 0.05 & 0.81 \\ 
   &600 & 0.59 & 0.62 & 0.96 & 0.84 & 2.30 & 0.57 & 0.02 & 0.57 & 0.29 & 0.23 & 0.17 & 0.73 & 0.02 & 0.73 \\ 
  &1000 & 0.76 & 0.52 & 0.96 & 0.87 & 1.80 & 0.45 & 0.02 & 0.45 & 0.23 & 0.18 & 0.14 & 0.76 & 0.12 & 0.75 \\ 
  \hline
\multirow{5}{*}{0.2}&100& 0.59 & 0.77 & 0.94 & 0.80 & 2.27 & 0.60 & 0.04 & 0.60 & 0.72 & 0.58 & 0.42 & 0.65 & 0.07 & 0.65 \\ 
  &200 & 0.72 & 0.82 & 0.95 & 0.83 & 1.81 & 0.45 & 0.03 & 0.45 & 0.50 & 0.41 & 0.28 & 0.51 & 0.00 & 0.51 \\ 
   &400 & 0.95 & 0.95 & 0.95 & 0.85 & 1.28 & 0.32 & 0.05 & 0.32 & 0.32 & 0.26 & 0.20 & 0.39 & 0.04 & 0.38 \\ 
   &600  & 0.98 & 0.98 & 0.94 & 0.84 & 1.10 & 0.28 & 0.02 & 0.28 & 0.27 & 0.22 & 0.16 & 0.32 & 0.00 & 0.32 \\ 
  &1000& 1.00 & 0.98 & 0.95 & 0.88 & 0.85 & 0.21 & 0.02 & 0.21 & 0.21 & 0.17 & 0.12 & 0.33 & 0.01 & 0.33 \\ 
  \hline
\multirow{5}{*}{0.1}&100 & 0.75 & 0.91 & 0.91 & 0.80 & 1.67 & 0.48 & 0.06 & 0.48 & 0.70 & 0.56 & 0.42 & 0.51 & 0.07 & 0.50 \\ 
  &200 & 0.94 & 0.97 & 0.93 & 0.80 & 1.29 & 0.35 & 0.01 & 0.35 & 0.49 & 0.40 & 0.29 & 0.38 & 0.05 & 0.38 \\ 
  &400& 1.00 & 1.00 & 0.94 & 0.83 & 0.91 & 0.24 & 0.01 & 0.24 & 0.33 & 0.28 & 0.19 & 0.28 & 0.02 & 0.28 \\ 
   &600 & 1.00 & 1.00 & 0.96 & 0.87 & 0.80 & 0.19 & 0.02 & 0.19 & 0.28 & 0.24 & 0.16 & 0.22 & 0.02 & 0.22 \\ 
  &1000& 1.00 & 1.00 & 0.94 & 0.84 & 0.62 & 0.16 & 0.02 & 0.16 & 0.20 & 0.16 & 0.12 & 0.24 & 0.01 & 0.24 \\ 
  \hline
\end{tabular}
}
\caption{\small  Performance of HITS, in comparison with the Deb Estimator, with respect to ERR as well as the empirical coverage (Coverage) and length (Len) of the CIs under dense setting 1 where $\Deltanew=1.082$. Reported also are the RMSE, bias and the standard error (SE) of the HITS estimator compared to the  Lasso and Deb estimators.}
\label{tab: Gaussian setting 1}
\end{table}

\begin{table}[!htbp]
\centering
\scalebox{0.8}{
\begin{tabular}{|r|r|rr|rr|r|rrr|rrr|rrr|}
  \hline
  &&\multicolumn{2}{c}{ERR}\vline& \multicolumn{2}{c}{Coverage}\vline&Len&\multicolumn{3}{c}{HITS}\vline&\multicolumn{3}{c}{Lasso}\vline&\multicolumn{3}{c}{Deb}\vline\\
  \hline
  ${\Ssc}$&n& HITS & Deb & HITS & Deb & HITS & RMSE& Bias & SE & RMSE& Bias & SE  & RMSE& Bias & SE  \\ 
  \hline
\multirow{5}{*}{1}&100& 0.02 & 0.10 & 0.98 & 0.83 & 9.18 & 1.97 & 0.20 & 1.96 & 0.56 & 0.16 & 0.53 & 2.37 & 0.21 & 2.36 \\ 
  &200 & 0.03 & 0.10 & 0.97 & 0.84 & 7.61 & 1.75 & 0.07 & 1.75 & 0.38 & 0.15 & 0.35 & 1.98 & 0.10 & 1.97 \\ 
  &400 & 0.03 & 0.09 & 0.96 & 0.87 & 5.35 & 1.31 & 0.10 & 1.31 & 0.26 & 0.10 & 0.24 & 1.58 & 0.08 & 1.58 \\ 
  &600& 0.03 & 0.11 & 0.97 & 0.87 & 4.45 & 1.03 & 0.04 & 1.03 & 0.21 & 0.07 & 0.20 & 1.32 & 0.02 & 1.32 \\ 
  &1000& 0.03 & 0.10 & 0.97 & 0.83 & 3.49 & 0.82 & 0.05 & 0.81 & 0.16 & 0.06 & 0.15 & 1.53 & 0.08 & 1.53 \\ 
  \hline
\multirow{5}{*}{0.5}&100& 0.02 & 0.13 & 0.97 & 0.82 & 4.68 & 1.00 & 0.04 & 1.00 & 0.38 & 0.21 & 0.31 & 1.24 & 0.02 & 1.24 \\ 
   &200 & 0.04 & 0.13 & 0.97 & 0.84 & 3.82 & 0.91 & 0.04 & 0.91 & 0.26 & 0.15 & 0.21 & 1.02 & 0.02 & 1.02 \\ 
 &400& 0.03 & 0.07 & 0.96 & 0.87 & 2.70 & 0.62 & 0.07 & 0.62 & 0.17 & 0.09 & 0.14 & 0.76 & 0.08 & 0.75 \\ 
&600& 0.03 & 0.09 & 0.97 & 0.86 & 2.24 & 0.52 & 0.02 & 0.52 & 0.15 & 0.09 & 0.12 & 0.68 & 0.04 & 0.68 \\ 
 &1000& 0.04 & 0.13 & 0.95 & 0.83 & 1.75 & 0.45 & 0.00 & 0.45 & 0.11 & 0.06 & 0.09 & 0.78 & 0.02 & 0.78 \\ 
  \hline
\multirow{5}{*}{0.2}&100& 0.06 & 0.18 & 0.97 & 0.80 & 1.96 & 0.46 & 0.11 & 0.44 & 0.33 & 0.22 & 0.24 & 0.53 & 0.09 & 0.52 \\ 
  &200 & 0.05 & 0.13 & 0.96 & 0.85 & 1.62 & 0.38 & 0.02 & 0.38 & 0.23 & 0.17 & 0.16 & 0.44 & 0.03 & 0.44 \\ 
  &400& 0.03 & 0.12 & 0.96 & 0.85 & 1.13 & 0.27 & 0.00 & 0.27 & 0.16 & 0.11 & 0.11 & 0.34 & 0.01 & 0.34 \\ 
  &600& 0.03 & 0.08 & 0.96 & 0.88 & 0.94 & 0.22 & 0.01 & 0.22 & 0.13 & 0.09 & 0.09 & 0.27 & 0.01 & 0.27 \\ 
   &1000 & 0.03 & 0.09 & 0.97 & 0.88 & 0.74 & 0.18 & 0.01 & 0.18 & 0.10 & 0.07 & 0.07 & 0.31 & 0.01 & 0.31 \\ 
  \hline
\multirow{5}{*}{0.1}&100 & 0.07 & 0.20 & 0.93 & 0.77 & 1.12 & 0.29 & 0.05 & 0.29 & 0.31 & 0.21 & 0.22 & 0.33 & 0.04 & 0.32 \\ 
   &200 & 0.04 & 0.12 & 0.96 & 0.84 & 0.94 & 0.23 & 0.01 & 0.23 & 0.21 & 0.15 & 0.15 & 0.25 & 0.01 & 0.25 \\ 
   &400 & 0.04 & 0.11 & 0.96 & 0.87 & 0.66 & 0.16 & 0.00 & 0.16 & 0.16 & 0.12 & 0.10 & 0.19 & 0.01 & 0.19 \\ 
  &600& 0.03 & 0.10 & 0.95 & 0.87 & 0.55 & 0.13 & 0.01 & 0.13 & 0.13 & 0.09 & 0.09 & 0.16 & 0.00 & 0.16 \\ 
  &1000 & 0.03 & 0.08 & 0.96 & 0.87 & 0.43 & 0.10 & 0.00 & 0.10 & 0.10 & 0.07 & 0.06 & 0.17 & 0.00 & 0.17 \\ 
  \hline
\end{tabular}
}
\caption{\small  Performance of HITS, in comparison with the Deb Estimator, with respect to ERR as well as the empirical coverage (Coverage) and length (Len) of the CIs under  dense setting 2 where $\Delta\subnew = 0$. Reported also are the RMSE, bias and the standard error (SE) of the HITS estimator compared to the  Lasso and Deb stimators.}
\label{tab: Gaussian setting 2}
\end{table}

\vspace{-3mm}

\subsection{Approximate Sparse Regression}
\label{sec: approx sparse}
For approximately sparse regression vectors, we generate the first few coefficients of $\beta_{1}$ and $\beta_{2}$ as in Section \ref{sec: general loading},  $\beta_{1,1}=-0.1, \beta_{1,j}=-0.4(j-1)$ for $2 \le j \le 11$ and $\beta_{2,1}=-0.5$, $\beta_{2,j}=0.2(j-1)$ for $2 \le j \le 6$ and then generate the remaining coefficients under he following two settings.
\begin{enumerate}[label=(\arabic*)]
\item {\bf Approximate sparse with decaying coefficients:} $\beta_{1,j}=(j-1)^{-\delta_1}$ for $12 \le j \le 501$ and $\beta_{2,j}=0.5\cdot (j-1)^{-\delta_1}$ for $7 \le j \le 501$ and we vary $\delta_1$ across $\{0.5,1,2,3\}.$ 
\item {\bf Capped-$\ell_1$ sparse:} $\beta_{1,j}=\delta_2\cdot \lambda_0$ and $\beta_{2,j}=\beta_{1,j}/2$ for $11\leq j\leq 50$
with $\lambda_0=\sqrt{2 {\log p}/{n}}$ and $\beta_{1,j}=\beta_{2,j}=0$ for $51\leq j\leq 501.$ We vary $\delta_2$ across $\{0.5,0.2,0.1,0.05\}$
\end{enumerate}

%


For the decaying coefficients setting,  the decay rate $\delta_1$ controls the sparsity.
For the capped-$\ell_1$ sparse setting,  the sparsity is measured by capped-$\ell_1$ sparsity defined in \eqref{eq: capped l1}, where $s_{k}$ denotes the capped-$\ell_1$ sparsity of $\beta_k$, for $k=1,2$.
We vary $\delta_2$ over $\{0.5,0.2,0.1,0.05\}$ and the upper bound for $s_1$ ranges over $11+\{20,8,4,2\}$ and the upper bound for $s_2$ ranges over $6+\{10,4,2,1\}.$ In both cases, we consider the dense loading $\bxnew$ generated in \eqref{eq: loading gen} with $\mathcal{S}=0.2.$

The simulation results are reported in  in Table \ref{tab: approximate sparse 1}. The results from the approximate sparse settings are largely consistent with those from the exact sparse setting.
 We observe that the proposed CIs achieve the $95\%$ coverage while CIs constructed based on the Deb estimators do not have desired coverage and the Lasso estimators have a dominant bias component.  Additionally, we note that, 1) for the case that the coefficients decay slowly (upper part of Table \ref{tab: approximate sparse 1} with $\delta_1=0.5$), the HITS CI over-covers since the variance of the point estimator is over-estimated due to the relatively dense regression vectors; once $\delta_1$ becomes larger, say  $\delta_1\ge 1$, the coverage is achieved at the desired level; 
 2) the HITS CIs are longer than those based on the Deb estimator;
 however, the Deb CIs do not have the correct coverage.

In addition, HITS is computationally more efficient. The average of the ratio of the computational time of Deb over that for HITS  is reported under the column indexed with ``TRatio ". 
The numerical results demonstrate that, for both the exact sparse and approximate sparse settings, HITS not only has the desired coverage property for arbitrary loading $\bxnew$, but also is computationally efficient.

\begin{table}[!htbp]
\centering
\scalebox{0.725}{
\begin{tabular}{|r|r|rr|rr|rr|rrr|rrr|rrr|r|}
\multicolumn{18}{c}{Approximate sparse with decaying coefficients}\\
\hline
  &&\multicolumn{2}{c}{ERR}\vline& \multicolumn{2}{c}{Coverage}\vline&\multicolumn{2}{c}{Len} \vline &\multicolumn{3}{c}{HITS}\vline&\multicolumn{3}{c}{Lasso}\vline&\multicolumn{3}{c}{Deb}\vline& \\
  \hline
  $\delta_1$ & n& HITS & Deb & HITS & Deb & HITS &Deb& RMSE& Bias & SE & RMSE& Bias & SE  & RMSE& Bias & SE& TRatio  \\ 
  \hline
\multirow{4}{*}{0.5}  &100 & 0.01 & 0.50 & 1.00 & 0.76 & 9.99 & 2.77 & 1.06 & 0.19 & 1.04 & 1.28 & 0.95 & 0.87 & 1.16 & 0.06 & 1.16 & 13.58 \\ 
 &200 & 0.05 & 0.67 & 1.00 & 0.81 & 5.84 & 2.15 & 0.67 & 0.03 & 0.67 & 0.82 & 0.60 & 0.57 & 0.79 & 0.04 & 0.78 & 10.08 \\ 
 & 400 & 0.47 & 0.90 & 1.00 & 0.90 & 3.21 & 1.56 & 0.44 & 0.05 & 0.43 & 0.53 & 0.35 & 0.40 & 0.50 & 0.02 & 0.50 & 4.93 \\ 
 & 600 & 0.82 & 0.96 & 0.99 & 0.89 & 2.35 & 1.31 & 0.36 & 0.05 & 0.36 & 0.40 & 0.25 & 0.32 & 0.40 & 0.01 & 0.40 & 4.24 \\ 
 \hline
\multirow{4}{*}{1}  &100 &0.53 & 0.77 & 0.94 & 0.79 & 2.44 & 1.72 & 0.62 & 0.07 & 0.62 & 0.74 & 0.58 & 0.46 & 0.69 & 0.05 & 0.69 & 14.97 \\ 
  &200 & 0.77 & 0.88 & 0.97 & 0.85 & 1.91 & 1.44 & 0.46 & 0.04 & 0.46 & 0.47 & 0.38 & 0.28 & 0.50 & 0.07 & 0.50 & 11.98 \\ 
  &400 & 0.96 & 0.96 & 0.96 & 0.85 & 1.33 & 1.13 & 0.34 & 0.02 & 0.33 & 0.34 & 0.27 & 0.21 & 0.40 & 0.01 & 0.40 & 5.70 \\ 
  & 600 & 0.99 & 0.98 & 0.96 & 0.84 & 1.14 & 0.97 & 0.28 & 0.03 & 0.28 & 0.26 & 0.20 & 0.17 & 0.33 & 0.02 & 0.33 & 4.76 \\ 
  \hline 
\multirow{4}{*}{2}  &100& 0.49 & 0.71 & 0.93 & 0.80 & 2.31 & 1.70 & 0.61 & 0.12 & 0.60 & 0.77 & 0.63 & 0.45 & 0.67 & 0.00 & 0.67 & 14.97 \\ 
 & 200 & 0.74 & 0.85 & 0.95 & 0.86 & 1.83 & 1.42 & 0.47 & 0.01 & 0.47 & 0.49 & 0.40 & 0.28 & 0.51 & 0.04 & 0.51 & 12.18 \\ 
 &400 & 0.96 & 0.96 & 0.94 & 0.87 & 1.29 & 1.12 & 0.32 & 0.03 & 0.32 & 0.33 & 0.27 & 0.19 & 0.39 & 0.04 & 0.39 & 5.78 \\ 
 &600 & 0.99 & 0.98 & 0.95 & 0.87 & 1.10 & 0.96 & 0.27 & 0.03 & 0.27 & 0.27 & 0.22 & 0.16 & 0.32 & 0.02 & 0.32 & 4.80 \\ 
  \hline 
\multirow{4}{*}{3}  &100& 0.50 & 0.72 & 0.91 & 0.78 & 2.29 & 1.68 & 0.62 & 0.11 & 0.62 & 0.77 & 0.63 & 0.44 & 0.67 & 0.01 & 0.67 & 14.97 \\ 
 &200 & 0.77 & 0.87 & 0.95 & 0.82 & 1.82 & 1.42 & 0.48 & 0.01 & 0.48 & 0.49 & 0.40 & 0.29 & 0.53 & 0.06 & 0.53 & 12.11 \\ 
 &400 & 0.96 & 0.94 & 0.97 & 0.85 & 1.30 & 1.11 & 0.32 & 0.01 & 0.32 & 0.33 & 0.27 & 0.19 & 0.38 & 0.01 & 0.38 & 5.73 \\ 
 &600 & 0.99 & 0.99 & 0.95 & 0.86 & 1.10 & 0.96 & 0.27 & 0.03 & 0.27 & 0.27 & 0.22 & 0.16 & 0.33 & 0.02 & 0.33 & 4.83 \\ 
   \hline
\multicolumn{18}{c}{   Capped-$\ell_1$ sparse}\\
     \hline
  &&\multicolumn{2}{c}{ERR}\vline& \multicolumn{2}{c}{Coverage}\vline&\multicolumn{2}{c}{Len} \vline &\multicolumn{3}{c}{HITS}\vline&\multicolumn{3}{c}{Lasso}\vline&\multicolumn{3}{c}{Deb}\vline&\\
    \hline
  $\delta_2$ & n& HITS & Deb & HITS & Deb & HITS &Deb& RMSE& Bias & SE & RMSE& Bias & SE  & RMSE& Bias & SE& TRatio  \\ 
  \hline
\multirow{4}{*}{0.5}  &100 & 0.28 & 0.66 & 0.98 & 0.76 & 3.84 & 1.97 & 0.76 & 0.14 & 0.75 & 0.92 & 0.70 & 0.60 & 0.79 & 0.02 & 0.79 & 13.78 \\ 
  &200 & 0.70 & 0.87 & 0.98 & 0.87 & 2.18 & 1.51 & 0.47 & 0.01 & 0.47 & 0.55 & 0.43 & 0.34 & 0.53 & 0.04 & 0.53 & 10.80 \\ 
  &400 & 0.97 & 0.96 & 0.97 & 0.87 & 1.39 & 1.16 & 0.33 & 0.05 & 0.32 & 0.33 & 0.26 & 0.21 & 0.40 & 0.04 & 0.40 & 5.27 \\ 
  &600 & 0.99 & 0.99 & 0.97 & 0.85 & 1.15 & 0.98 & 0.27 & 0.02 & 0.27 & 0.27 & 0.21 & 0.17 & 0.32 & 0.01 & 0.32 & 4.43 \\ 
  \hline
\multirow{4}{*}{0.2}& 100 & 0.46 & 0.73 & 0.96 & 0.79 & 2.60 & 1.75 & 0.62 & 0.11 & 0.61 & 0.78 & 0.62 & 0.47 & 0.69 & 0.00 & 0.69 & 14.40 \\ 
  &200 & 0.76 & 0.87 & 0.96 & 0.85 & 1.91 & 1.44 & 0.47 & 0.03 & 0.47 & 0.51 & 0.41 & 0.31 & 0.52 & 0.06 & 0.52 & 11.49 \\ 
  &400 & 0.95 & 0.94 & 0.94 & 0.83 & 1.32 & 1.12 & 0.34 & 0.05 & 0.34 & 0.34 & 0.27 & 0.21 & 0.41 & 0.05 & 0.41 & 5.43 \\ 
  &600 & 0.99 & 0.99 & 0.95 & 0.87 & 1.11 & 0.96 & 0.28 & 0.02 & 0.28 & 0.27 & 0.21 & 0.17 & 0.33 & 0.01 & 0.33 & 4.54 \\ 
  \hline 
\multirow{4}{*}{0.1}& 100  & 0.50 & 0.71 & 0.95 & 0.79 & 2.41 & 1.70 & 0.61 & 0.12 & 0.59 & 0.78 & 0.64 & 0.44 & 0.65 & 0.02 & 0.65 & 14.50 \\ 
  &200 & 0.81 & 0.88 & 0.95 & 0.83 & 1.85 & 1.42 & 0.47 & 0.04 & 0.47 & 0.49 & 0.39 & 0.30 & 0.53 & 0.09 & 0.52 & 11.62 \\ 
 &400 & 0.97 & 0.96 & 0.95 & 0.83 & 1.29 & 1.12 & 0.33 & 0.04 & 0.32 & 0.33 & 0.27 & 0.20 & 0.38 & 0.03 & 0.38 & 5.50 \\ 
 &600 & 0.98 & 0.98 & 0.97 & 0.88 & 1.10 & 0.96 & 0.27 & 0.00 & 0.27 & 0.28 & 0.23 & 0.16 & 0.32 & 0.00 & 0.32 & 4.59 \\ 
 \hline
\multirow{4}{*}{0.05}& 100&0.53 & 0.70 & 0.94 & 0.79 & 2.33 & 1.70 & 0.60 & 0.10 & 0.60 & 0.75 & 0.61 & 0.44 & 0.66 & 0.03 & 0.66 & 14.45 \\ 
 &200 & 0.78 & 0.87 & 0.96 & 0.86 & 1.82 & 1.42 & 0.45 & 0.03 & 0.45 & 0.49 & 0.39 & 0.29 & 0.50 & 0.06 & 0.50 & 11.59 \\ 
 &400 & 0.94 & 0.94 & 0.93 & 0.84 & 1.29 & 1.12 & 0.33 & 0.00 & 0.33 & 0.34 & 0.28 & 0.20 & 0.39 & 0.00 & 0.39 & 5.47 \\ 
 &600 & 0.99 & 0.98 & 0.96 & 0.87 & 1.10 & 0.96 & 0.27 & 0.01 & 0.27 & 0.28 & 0.22 & 0.16 & 0.33 & 0.01 & 0.33 & 4.63 \\ 
   \hline
\end{tabular}
}
\caption{\small   Performance of HITS, in comparison with the Deb Estimator, with respect to ERR as well as the empirical coverage (Coverage) and length (Len) of the CIs under approximate sparse regression settings. Reported also are the RMSE, bias and the standard error (SE) of the HITS estimator compared to the  Lasso and Deb estimators; the ratio of computational time of Deb to HITS (``TRatio").
}
\label{tab: approximate sparse 1}
\end{table}



\def\betahat{\widehat{\beta}}

\section{Real Data Analysis}
\label{sec: real data}
Tumor Necrosis Factor (TNF) is an inflammatory cytokine important for immunity and inflammation. TNF blockade therapy has found its success in treating RA \citep{taylor2009anti}. However,  the effect of anti-TNF varies greatly among patients and multiple genetic markers have been identified as predictors of anti-TNF response \citep{padyukov2003genetic,liu2008genome,chatzikyriakidou2007combined}. We seek to estimate ITE of anti-TNF in reducing inflammation for treating RA using EHR data from PHS as described in Section \ref{sec-intro}. Here, the inflammation is measured by the inflammation marker, C-reactive Protein (CRP). Since a higher value of CRP is more indicative of a worse treatment response, we define $Y=-\log\mbox{CRP}$. 

The analyses include $n = 183$ RA patients who are free of coronary artery disease, out of which $n_1= 92$ were on the combination therapy of anti-TNF and methotraxate and $n_2 = 91$ on methotraxate alone. To sufficiently control for potential confounders, we extracted a wide range of predictors from the EHR and included both potential confounders and predictors of CRP in  $\bX$, resulting a total of $p=171$ predictors. Examples of predictors include diagnostic codes of RA and comorbidities such as systemic lupus erythematosus (SLE) and diabetes, past history of lab results including CRP,  rheumatoid factor (RF), and anticyclic citrullinated peptide (CCP), prescriptions of other RA medications including Gold and Plaquenil, as well as counts of NLP mentions for a range of clinical terms including disease conditions and medications. Since counts of diagnosis or medication codes, referred to as codified (COD) mentions, are highly correlated with the corresponding NLP mentions in the narrative notes, we combine the counts of COD and NLP mentions of the same clinical concept to represent its frequency. The predictors also include a number of single-nucleotide polymorphism (SNP) markers and genetic risk scores identified as associated with RA risk or progression. All count variables were transformed via $x\mapsto \log(1+x)$ and lab results were transformed by $x \mapsto \log(x)$ since their distributions are highly skewed. Missing indicator variables were created for past history of lab measurements since the availability of lab results can be indicative of disease severity. We assume that conditional on $\bX$, the counterfactual outcomes are independent of the treatment actually received. 

We applied the proposed HITS procedures to infer about the benefit of anti-TNF for individual patients. Out of the $p=171$ predictors, 8 of which were assigned with non-zero coefficients in either treatment groups. The leading predictors, as measured by the magnitude of {difference between two Lasso estimators} $\{\bbetahat_{1,j} - \bbetahat_{2,j},  j = 1, ..., p\}$, include counts of SLE COD or NLP mentions, indicator of no past history of CRP measurements, and SNPs including rs12506688, rs8043085 and rs2843401. {Confidence intervals for $\bbeta_1,\bbeta_2$ and $\bbeta_1-\bbeta_2$ based on debiased estimators are also reported.}  These predictors are generally consistent with results previously reported in clinical studies. The anti-TNF has been shown as effective among patients with presentations of both RA and SLE \citep{danion2017long}. The rs8043085 SNP located in the RASGRP1 gene is associated with an increased risk of sero-positive RA \citep{eyre2012high} and  the combination therapy has been previously reported as being more beneficial for sero-postive RA than for sero-negative RA \citep{seegobin2014acpa}. The rs2843401 SNP in the MMLE1 gene has been reported as protective of RA risk \citep{eyre2012high}, which appears to be associated with lower benefit of anti-TNF. The 
rs12506688 is in the RB-J gene which is a key upstream negative regulator of TNF-induced osteoclastogenesis. 

\begin{table}[htb]
\centering
\begin{tabular}{lllllll}
  \hline
 & $\widehat{\bbeta}_1$ & CI$_{\bbeta_1}$ & $\widehat{\bbeta}_2$ & CI$_{\bbeta_2}$ & $\widehat{\bbeta}_1-\widehat{\bbeta}_2$ & CI$_{\bbeta_1-\bbeta_2}$ \\ 
  \hline
Echo & 0.02 & [0.02, 0.18] & -0.03 & [-0.19, -0.03] & 0.04 & [0.08, 0.33] \\ 
  rs2843401 & -0.03 & [-0.07, -0.02] & 0 & [-0.02, 0.03] & -0.03 & [-0.10, -0.01] \\ 
  rs12506688 & -0.08 & [-0.18, -0.07] & 0 & [-0.03, 0.08] & -0.08 & [-0.23, -0.07] \\ 
  rs8043085 & 0 & [-0.03, 0.15] & -0.05 & [-0.21, -0.03] & 0.05 & [0.06, 0.29] \\ 
  race black & 0 & [-0.30, 0.62] & -0.02 & [-0.70, 0.22] & 0.02 & [-0.10, 0.90] \\ 
  prior CRP missing & -0.17 & [-0.7, -0.16] & 0 & [-0.23, 0.31] & -0.17 & [-1.24, 0.30] \\ 
  Gold & -0.01 & [-0.13, -0.01] & 0 & [-0.11, 0.02] & -0.01 & [-0.12, 0.06] \\ 
  SLE & 0 & [-0.07, 0.08] & -0.16 & [-0.34, -0.18] & 0.16 & [0.14, 0.40] \\ 
   \hline
\end{tabular}
\caption{\small  {Lasso} Estimates of $\bbeta_1$, $\bbeta_2$ and $\bbeta_1-\bbeta_2$ for the predictors of CRP along with their 95\% CIs. All predictors not included the table received zero Lasso estimates for both $\bbeta_1$ and $\bbeta_2$. } 
\end{table}

\begin{figure}[ht]
  \begin{subfigure}[b]{0.48\textwidth}
    \includegraphics[width=2.5in,height=2in]{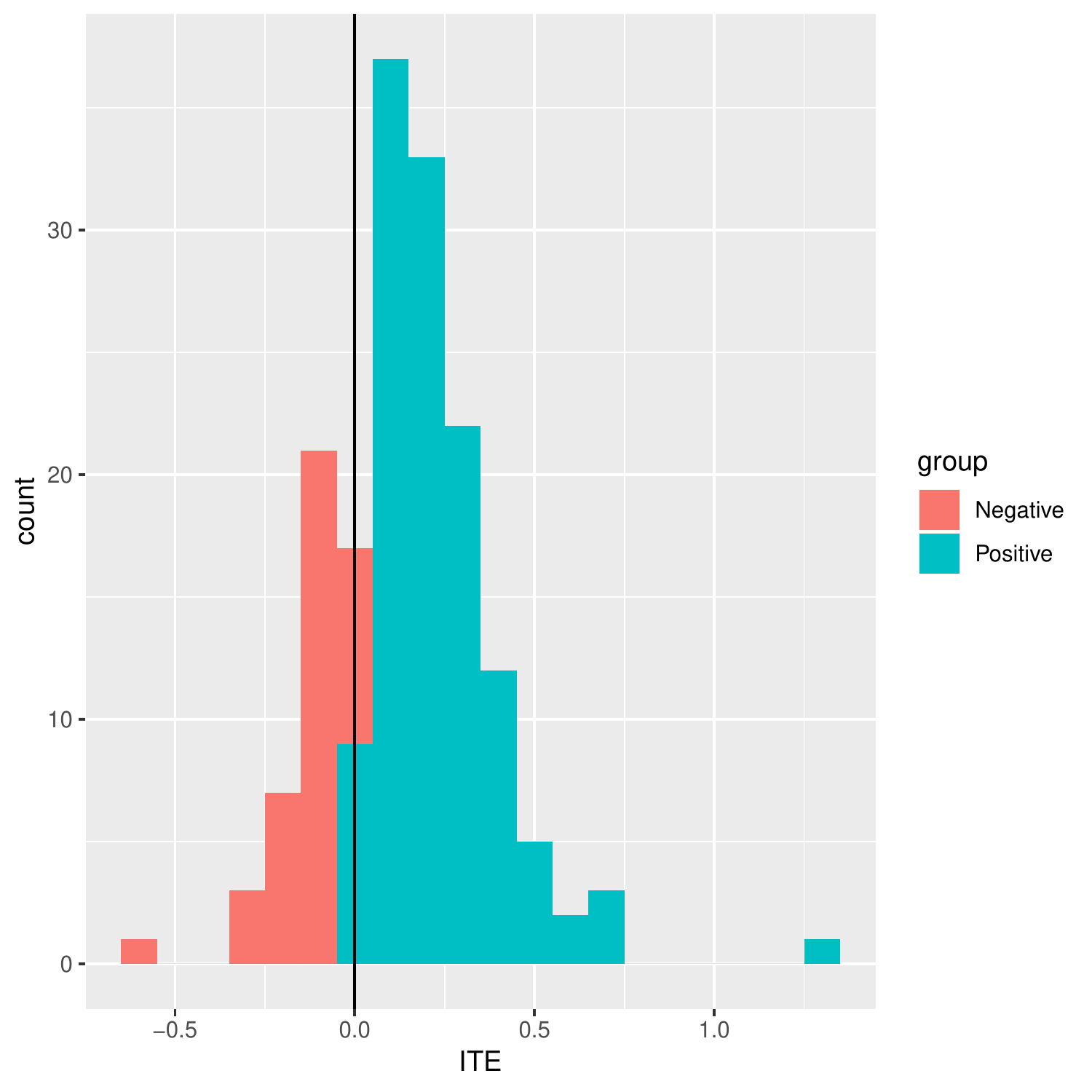}\;\;
    \caption{}
    \label{fig:1}
  \end{subfigure}
  \begin{subfigure}[b]{0.48\textwidth}
    \includegraphics[width=2.5in,height=2in]{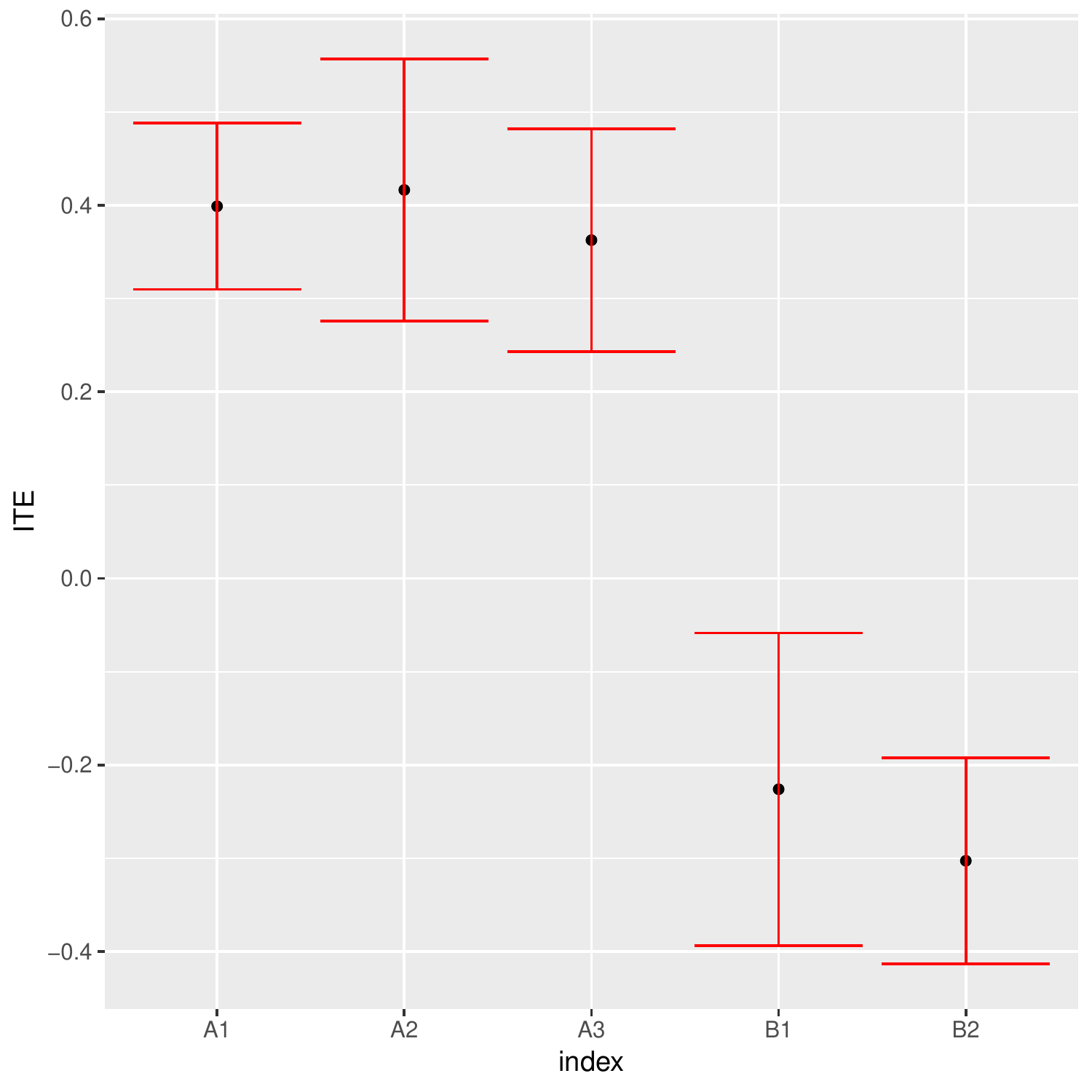}
    \caption{}
    \label{fig:2}
  \end{subfigure}
  \caption{(a) Histogram of the estimated ITE for the observed set of $\bx\subnew$ where the vertical line represents the median value; (b) point estimate and 95\% CIs for 5 choices of $\bx\subnew$ where the x-axis indexes $\bxnew\trans(\bbetahat_{1}-\bbetahat_2)$.}
\label{fig:example}
\end{figure}

We obtained  estimates of $\Delta\subnew$ for the observed set of $\bx\subnew$. As shown in Figure \ref{fig:example}(a), the predicted ITE ranges from -1.3 to 0.6 with median -0.14. About 72\% of the patients in this population appear to benefit from combination therapy. We also obtained CIs for a few examples of $\bx\subnew$, including (A) those with rs12506688 $= 0$, rs2843401 $= 0$, prior CRP not missing,  rs8043085 $> 0$, and $\ge 1$ SLE mention; and (B) those with rs12506688 $> 0$, rs2843401 $> 0$, prior CRP missing, rs8043085 $=0$, and no SLE mention. There are  three such patients in (A) (indexed by A1, A2, A3) and two in (B), (indexed by B1, B2). The point estimates and their corresponding 95\% CIs are shown in Figure \ref{fig:example}(b). The estimated ITEs were around -0.4 for A1-A3 with 95\% CIs all below $-0.2$, suggesting that adding anti-TNF is beneficial for these patients. On the contrary, anti-TNF may even be detrimental for B1 and B2 whose estimated ITEs are 0.23 (95\% CI: [0.058, 0.39]) and 0.30 (95\% CI: [0.19, 0.41]), respectively. These results support prior findings that the benefit of combination therapy is heterogeneous across patients.


\section{Discussions}
\label{sec: discussion}

%
We introduced the HITS procedures for inference on ITE with high-dimensional covariates. Both the theoretical and numerical properties of  HITS  are established. Unlike the debiasing methods proposed in the literature, HITS has the major advantage of not requiring the covariate vector $\bx\subnew$ to be sparse or of other specific structures. A key innovation lies in the novel construction of the projection direction with an additional constraint  \eqref{eq: constraint 2}. 
 We elaborate the importance of this step to further illustrate the challenges of statistical inference for dense loading.  The following result shows that the algorithm without \eqref{eq: constraint 2} fails to correct the bias of $\bx_{\rm new}^{\intercal}\bbeta_1$ for a certain class of $\bx_{\rm new}$.
\begin{Proposition} 
\label{prop: challenge}
The minimizer $\widetilde{\bu}_1$ in \eqref{eq: initial correction} is zero if either of the following conditions on $\bx_{\rm new}$ is satisfied: {\rm (F1)} ${\|\bx_{\rm new}\|_2}/{\|\bx_{\rm new}\|_{\infty}}\geq {1}/{\lambda_1}$; {\rm (F2)} The non-zero coordinates of $\bx_{\rm new}$ are of the same order of magnitude and $\|\bx_{\rm new}\|_0\geq C\sqrt{{n_1}/{\log p}}$ for some positive constant $C>0$.
\end{Proposition}
Since ${\|\bx_{\rm new}\|_2}/{\|\bx_{\rm new}\|_{\infty}}$ can be viewed as a measure of sparsity of $\bx_{\rm new}$, both Conditions {\rm (F1)} and {\rm (F2)} state that the optimization algorithm \eqref{eq: initial correction} fails to produce a non-zero projection direction if the loading $\bx_{\rm new}$ is dense to certain degree. That is, without the additional constraint \eqref{eq: constraint 2}, the projection direction  $\widetilde{\bu}_1$  does not correct the bias of estimating $\bx_{\rm new}^{\intercal}\bbeta_1$.

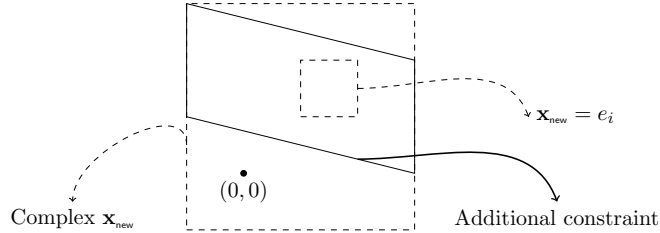
\begin{figure}[htb]
\scalebox{.75}{
   \begin{tikzpicture}
   \draw (0,0) [dashed] -- (1,0) -- (1,1) -- (0,1) -- (0,0);
   \draw (-2, -2) [dashed]-- (2,-2) -- (2,2) -- (-2,2) -- (-2,-2);
   \draw (-2, 0)-- (-2,2) -- (2,1)--(2,-1) -- (-2,0); 
   \draw[fill=black] (-1,-1) circle (0.05) node[below] {$(0,0)$};
    \draw[->,thick] (1,-0.75) to [out=0,in=-250] (4.5,-1.5) node[anchor= north]
        {Additional constraint};
     \draw[->,dashed] (1,0.5) to [out=0,in=-250] (4,0) node[anchor= west]
        {$\bxnew=e_i$};
       \draw[->,dashed] (-2,-0.5) to [out=90,in=90] (-4,-1.5) node[anchor= north]
        {Complex $\bxnew$};   
    \end{tikzpicture}
  }
  \caption{Geometric illustration of Proposition \ref{prop: challenge}: the solid parallelogram corresponds to the feasible set of \eqref{eq: constraint 1} and \eqref{eq: constraint 2} for a dense $\bxnew$ while the large dashed triangle corresponds to that of \eqref{eq: initial correction}; the small dashed square corresponds to the feasible set of \eqref{eq: initial correction} for $\bxnew=e_i$.}
\label{fig: additional constraint}
\end{figure}

We shall provide some geometric insights about Proposition \ref{prop: challenge}. 
The feasible set for constructing $\widetilde{\bu}$  in \eqref{eq: initial correction} depends on both $\bxnew$ and $\|\bxnew\|_2$. If $p$ is large and $\bxnew\in \R^{p}$ is dense, this feasible set is significantly enlarged in comparison to the feasible set corresponding to the simpler case $\bxnew={\bf e}_i$. As illustrated in Figure \ref{fig: additional constraint}, the larger and smaller dashed squares represent the feasible sets for a dense $\bxnew$ and $\bxnew={\bf e}_i$, respectively.  Since zero vector is contained in  the feasible set for a dense $\bxnew$, the optimizer in \eqref{eq: initial correction} is zero and the bias correction is not effective.  With the additional constraint \eqref{eq: constraint 2}, even in the presence of dense $\bxnew$, the feasible set is largely shrunk to be the solid parallelogram, as the intersection of the larger dashed square and the parallel lines introduced by the constraint \eqref{eq: constraint 2}.  Interestingly, the additional constraint \eqref{eq: constraint 2}  simply restricts the feasible set from one additional direction determined by $\bxnew$ and automatically enables a unified inference procedure for an arbitrary $\bx_{\rm new}$. 

The high dimensional outcome modeling adopted in HITS allows us to extensively adjust for confounders to overcome treatment by indication bias frequently encountered in observational studies. The present paper focused on the supervised setting. For EHR applications, in addition to the labeled data with outcome variables observed, there are often also a large amount of unlabeled data available where only the covariates are observed. It is known that for certain inference problems, the unlabeled data can be used to significantly improve the inference accuracy \citep{cai2018semi}. Inference for ITE in the semi-supervised setting warrants future research.

\section{Proofs}
\label{sec:proof}
We  present  in this section the proof for the optimality results, Theorem \ref{thm: exact sparse lower} and Corollary \ref{cor: exact sparse adapt} and also the proof of  Lemma \ref{lem: enhanced variance} for non-vanishing variance. For reasons of space, the proofs of all other results are deferred to Section \ref{sec: sup proof} in the supplementary material.
\subsection{Proof of Theorem \ref{thm: exact sparse lower} and Corollary \ref{cor: exact sparse adapt}}

In the following theorem, we first introduce a general machinery for establishing the detection boundary $\tauadpt$ for the hypothesis testing problem \eqref{eq: testing problem}.
\begin{Theorem}[Detection Boundary Lower Bound ]
Suppose $\kt \leq \ku\lesssim \min\{p,\frac{n}{\log p}\}$. Re-order $\bxnew$ such that $\left|x_{\rm new,1}\right|\geq \left|x_{\rm new,2}\right|\geq \cdots \geq \left|x_{\rm new,p}\right|$. For any $(q,L)$ satisfies $q\leq \ku$ and $L\leq \|x_{\rm new}\|_0$, the adaptation detection boundary $\tauadpt$ satisfies
$
\tauadpt\geq \tau^{*}
$
where 
{\small \begin{equation}
\tau^{*}= C\frac{1}{\sqrt{n}}\cdot\max\left\{\sqrt{\sum_{j=1}^{s} x^2_{\rm new,j}},\sum_{j=\max\{L-q+2,1\}}^{L} \left|x_{\rm new,j}\right|\sqrt{\left({\log \left({c L}/{q^2}\right)}\right)_{+}}\right\}.
\label{eq: detection boundary}
\end{equation}}
\label{thm: general lower}
\end{Theorem}

To establish the lower bounds in Theorem \ref{thm: exact sparse lower} and Corollary \ref{cor: exact sparse adapt}, we simply apply the lower bound for the general detection boundary in \eqref{eq: detection boundary} to the case of exact loadings. Specifically, $\tau^{*}$ is reduced to the following expression by taking $L=\|\bx_{\rm new}\|_0$ and $q=\min\{s_u, \sqrt{\|\bx_{\rm new}\|_0}\},$
{\small
\begin{equation}
\tau^{*}= \frac{\|x_{\rm new}\|_{\infty}}{\sqrt{n}}\cdot\max\left\{\min\{\sqrt{s},\sqrt{\|x_{\rm new}\|_0}\},\min\{s_u, \sqrt{\|\bx_{\rm new}\|_0}\}\sqrt{\left({\log \left(\frac{c \|\bx_{\rm new}\|_0}{\min\{s_u, \sqrt{\|\bx_{\rm new}\|_0}\}^2}\right)}\right)_{+}}\right\}.
\label{eq: detection boundary exact}
\end{equation}
}
For the case (E1), we have $\|\bx_{\rm new}\|_0\leq s^2\leq s_{u}^2$ and $\tau^{*}\asymp {\|\bx_{\rm new}\|_{\infty}}\sqrt{\|\bx_{\rm new}\|_0/n}$; hence the lower bound \eqref{eq: E1} follows.  For the case (E2), if $\gamma_{\rm new}> 2\gamma_{u}$, we have $\tau^{*}\asymp \frac{\|x_{\rm new}\|_{\infty}}{\sqrt{n}}s_{u}\sqrt{\log p}$; if $\gamma_{\rm new}\leq  2\gamma_{u}$, we have $\tau^{*}\asymp \frac{\|\bx_{\rm new}\|_{\infty}}{\sqrt{n}}\sqrt{\|\bx_{\rm new}\|_0}.$
Hence, the lower bounds in \eqref{eq: E2} follow.

By applying Corollary \ref{Cor: hypothesis testing}, we establish that the detection boundaries $\tauadpt$ in \eqref{eq: E1 match} and \eqref{eq: E2-b match} are achieved by the hypothesis testing procedure $\phi_{\alpha}$ defined in \eqref{eq: test}. All the other detection boundaries will be achieved by the procedure $\phi(q,\ku)$ defined in \eqref{eq: test with sparsity} in the supplement. 
\subsection{Proof of Lemma \ref{lem: enhanced variance}}
Part of \eqref{eq: dominating variance}, $\sqrt{\rm V}\leq C_0\|\bx_{\rm new}\|_2\left({1}/{\sqrt{n_1}}+1/{\sqrt{n_2}}\right)$, is a consequence of the high probability concentration,
$
\min_{k\in\{1,2\}}\mathbf{P}\left(\|\bSigmahat_k\buhat_k-\bx_{\rm new}\|_{\infty}\leq C\|\bx_{\rm new}\|_2\sqrt{{\log p}/{n_k}}\right)\geq 1-p^{-c} 
$
which is the second high probability inequality of Lemma 4 established in \citet{cai2015regci}. Hence, $\bSigma_1^{-1}\bx_{\rm new}$ satisfies the constraints \eqref{eq: constraint 1} and \eqref{eq: constraint 2} and 
$
{\rm V}\leq \frac{\sigma_1^2}{n_1} \bx_{\rm new}^{\intercal}\bSigma_1^{-1}\bSigmahat_1\bSigma_1^{-1}\bx_{\rm new}+\frac{\sigma_2^2}{n_2} \bx_{\rm new}^{\intercal}\bSigma_2^{-1}\bSigmahat_2\bSigma_2^{-1}\bx_{\rm new}.
$
By Lemma 10 (specifically, the last high probability inequality) of \citet{cai2018semi}, with probability larger than $1-p^{-c}$, we have 
\begin{equation}
\left|\frac{\bx_{\rm new}^{\intercal}\bSigma_1^{-1}\bSigmahat_1\bSigma_1^{-1}\bx_{\rm new}}{\bx_{\rm new}^{\intercal}\bSigma_1^{-1}\bx_{\rm new}}-1\right|\lesssim \sqrt{\frac{\log p}{n_1}}\;\;\text{and}\;\;\left|\frac{\bx_{\rm new}^{\intercal}\bSigma_2^{-1}\bSigmahat_2\bSigma_2^{-1}\bx_{\rm new}}{\bx_{\rm new}^{\intercal}\bSigma_2^{-1}\bx_{\rm new}}-1\right|\lesssim \sqrt{\frac{\log p}{n_2}}
\end{equation}
Then we establish 
$
\mathbb{P}\left(\sqrt{\rm V}\leq C_0\|\bx_{\rm new}\|_2\left({1}/{\sqrt{n_1}}+{1}/{\sqrt{n_2}}\right)\right)\geq 1-p^{-c}.$

The proof of the lower bound $\sqrt{\rm V}\geq c_0\|\bx_{\rm new}\|_2$ is similar to that of Lemma 3.1 of \citet{javanmard2014confidence} through constructing another optimization algorithm. The main difference is that the proof in \citet{javanmard2014confidence} is for an individual regression coefficient and the following proof for a general linear contrast mainly relies on the additional constraint \eqref{eq: constraint 2}, instead of \eqref{eq: constraint 1}. To be specific,  we define a proof-facilitating optimization problem,
\begin{equation}
\begin{aligned}
\bar{\bf u}_1=&\;\min_{u\in \R^{p}} \bu^{\intercal}\bSigmahat_1\bu \quad
\text{subject to}\; \left |\bx_{\rm new}^{\intercal}\bSigmahat_1\bu-\|\bx_{\rm new}\|_2^2 \right|\leq \|\bx_{\rm new}\|_2^2\lambda_1 \end{aligned}
\label{eq: proof-facilitating}
\end{equation}
Note that $\buhat_1$ satisfies the feasible set of \eqref{eq: proof-facilitating} and hence
\begin{equation}
\begin{aligned}
&\buhat_1^{\intercal}\bSigmahat_1\buhat_1\geq \bar{\bf u}_1^{\intercal}\bSigmahat_1\bar{\bf u}_1
\geq \bar{\bf u}_1^{\intercal}\bSigmahat_1\bar{\bf u}_1+t((1-\lambda_1)\|\bx_{\rm new}\|_2^2-\bx_{\rm new}^{\intercal}\bSigmahat_1\bar{\bf u}_1) \; \text{for any}\; t\geq 0,
\end{aligned}
\label{eq: proof-facilitating inequality 1}
\end{equation}
where the last inequality follows from the constraint of \eqref{eq: proof-facilitating}. For any given $t\geq 0$, 
\begin{equation}
\begin{aligned}
\bar{\bf u}_1^{\intercal}\bSigmahat_1\bar{\bf u}_1+t((1-\lambda_1)\|\bx_{\rm new}\|_2^2-\bx_{\rm new}^{\intercal}\bSigmahat_1\bar{\bf u}_1) 
\geq \min_{\bu\in \R^{p}}{\bu}^{\intercal}\bSigmahat_1{\bu}+t((1-\lambda_1)\|\bx_{\rm new}\|_2^2-\bx_{\rm new}^{\intercal}\bSigmahat_1{\bu}).
\end{aligned}
\label{eq: proof-facilitating inequality 2}
\end{equation}
By solving the minimization problem of the right hand side of \eqref{eq: proof-facilitating inequality 2}, we have the minimizer $u^{*}$ satisfies $\bSigmahat_1u^{*}=\frac{t}{2} \bSigmahat_1\bx_{\rm new}$ and hence the minimum of the right hand side of \eqref{eq: proof-facilitating inequality 2} is 
$-\frac{t^2}{4}\bx_{\rm new}^{\intercal}\bSigmahat_1\bx_{\rm new}+t(1-\lambda_1)\|\bx_{\rm new}\|_2^2.$
Combined with \eqref{eq: proof-facilitating inequality 1} and \eqref{eq: proof-facilitating inequality 2}, we have 
$\buhat_1^{\intercal}\bSigmahat_1\buhat_1\geq \max_{t\geq 0}\left[-\frac{t^2}{4}\bx_{\rm new}^{\intercal}\bSigmahat_1\bx_{\rm new}+t(1-\lambda_1)\|\bx_{\rm new}\|_2^2 \right]$ and the minimum is achieved at $t^{*}=2\frac{(1-\lambda_1)\|\bx_{\rm new}\|_2^2}{\bx_{\rm new}^{\intercal}\bSigmahat_1\bx_{\rm new}}>0$ and hence
$
\buhat_1^{\intercal}\bSigmahat_1\buhat_1\geq \frac{(1-\lambda_1)^2\|\bx_{\rm new}\|_2^4}{\bx_{\rm new}^{\intercal}\bSigmahat_1\bx_{\rm new}}.
$
By Lemma 10 of \citet{cai2018semi}, with probability larger than $1-p^{-c}$, we have 
$
\left|\frac{\bx_{\rm new}^{\intercal}\bSigmahat_1\bx_{\rm new}}{\bx_{\rm new}^{\intercal}{\bSigma}_1\bx_{\rm new}}-1\right|\lesssim \sqrt{{\log p}/{n_1}}
$
and hence 
$
\buhat_1^{\intercal}\bSigmahat_1\buhat_1\geq c\|\bx_{\rm new}\|_2^2. 
$
Similarly, we establish 
$
\buhat_2^{\intercal}\bSigmahat_2\buhat_2\geq c\|\bx_{\rm new}\|_2^2
$
and hence
$
\mathbb{P}(\sqrt{\rm V}\geq c_0\|\bx_{\rm new}\|_2\left({1}/{\sqrt{n_1}}+{1}/{\sqrt{n_2}}\right))\geq 1-p^{-c}.
$

\section*{Acknowledgement}
{{The research of Tianxi Cai was supported in part by NIH grants R21 CA242940 and R01 HL089778.}} The research of Tony Cai was supported in part by NSF Grant DMS-1712735 and NIH grants R01-GM129781 and R01-GM123056. The research of Zijian Guo was supported in part by NSF Grants DMS-1811857 and DMS-2015373. 
\setlength{\bibsep}{0.1pt plus 0.5ex}
\bibliographystyle{chicago.bst}
\bibliography{HDRef}

\newpage
\appendix
\renewcommand{\thefigure}{A\arabic{figure}}
\setcounter{figure}{0}
\section{Detection Boundary for Decaying Loading}
\label{sec: decaying}
In the following, we consider the optimality result about decaying loading. Specifically, we calibrate the $i$-th largest element $\bx_{\rm new, (i)}$ by the decaying rate parameter $\delta$ as defined in \eqref{eq: decaying}. A larger value of $\delta$ means that the loading decays faster; for the case $\delta=0$, the loading is not decaying at all. 
\begin{Theorem}
\label{thm: decaying lower}
Suppose that  $\kt\leq \ku \lesssim \frac{n}{\log p}$. We calibrate $\kt, \ku$ and the decaying of $\bxnew$ by $\gamma$, $\gamma_{u}$ and $\delta$, respectively, as defined in \eqref{eq: cal of beta} and \eqref{eq: decaying}.
\begin{enumerate}
\item[(D1)] If $\bxnew$ is {\it a fast decaying loading} with $\delta\geq \frac{1}{2},$ then 
\begin{equation}
\tauadpt \gtrsim \frac{1}{\sqrt{n}} \cdot (1+\sqrt{\log \kt}\cdot \mathbf{1}(\delta=\frac{1}{2}))
\label{eq: lower D1}
\end{equation}
\item[(D2)] If $\bxnew$ is {\it a slow decaying loading} with $0\leq \delta< \frac{1}{2},$ then \begin{equation}
\tauadpt \gtrsim \begin{cases} c_{p}\frac{\ku^{1-2\delta}}{\sqrt{{{n}}}}\left(\log p\right)^{\frac{1}{2}-\delta} &\text{if}\quad \gamma_{u}< \frac{1}{2}\\
 \sqrt{\frac{p^{1-2\delta}}{{n}}}& \text{if}\quad \gamma_{u}\geq \frac{1}{2}
 \end{cases}
 \label{eq: lower D2}
\end{equation}
where $c_p=\sqrt{\frac{\log(\log p)}{\log p}}.$
\end{enumerate}
\end{Theorem}
Similar to the exact sparse loading in Theorem \ref{thm: exact sparse lower}, the detection lower bounds in the above Theorem can be attained under regularity conditions. The following corollary presents the matched upper bound for the detection boundaries established in Theorem \ref{thm: decaying lower} over certain regimes.
\begin{Corollary}
\label{cor: decaying adapt}
Suppose that  $s\leq \ku \lesssim \frac{\sqrt{n}}{\log p}$. 
\begin{enumerate}
\item[\rm (D1)] If $\bxnew$ is {\it a fast decaying loading} with $\delta\geq \frac{1}{2},$ then 
\begin{equation}
\tauadpt \asymp \taumini \asymp \frac{\|\bx_{\rm new}\|_2}{\sqrt{n}}
\label{eq: D1 match}
\end{equation}
In particular, for $\delta=\frac{1}{2}$, the detection boundary holds if $\gamma>0$; otherwise the detection boundary holds up to a $\sqrt{\frac{\log p}{\log s}}$ factor.
\item[\rm (D2)]  If $\bxnew$ is {\it a slow decaying loading} with $0 \leq \delta<\frac{1}{2},$ then the minimaxity detection boundary and adaptive detection boundary hold up to a $\sqrt{\log p}$ order
\begin{enumerate}
\item[\rm (D2-a)]  If the true sparsity  $\kt$ and the knowledge of $\ku$ satisfies  satisfies $\gamma< \gamma_{u}\leq \frac{1}{2},$ then
\begin{equation}
\tauadpt \asymp \frac{\ku^{1-2\delta}}{\sqrt{{{n}}}}\left(\log p\right)^{\frac{1}{2}-\delta} \gg \taumini \asymp \frac{\kt^{1-2\delta}}{\sqrt{{{n}}}}\left(\log p\right)^{\frac{1}{2}-\delta} .
\label{eq: D2-a match}
\end{equation}
\item[\rm (D2-b)] If the true sparsity $\kt$ and the knowledge of $\ku$ satisfies $\gamma< \frac{1}{2}\leq \gamma_{u},$ then 
\begin{equation}
\tauadpt \asymp \sqrt{\frac{p^{1-2\delta}}{{n}}} \gg \taumini \asymp \frac{\kt^{1-2\delta}}{\sqrt{{{n}}}}\left(\log p\right)^{\frac{1}{2}-\delta}.
\label{eq: D2-b match}
\end{equation}
\item[\rm (D2-c)] If the true sparsity $\kt$ and the knowledge of $\ku$ satisfies $\frac{1}{2}\leq \gamma< \gamma_{u},$ then
\begin{equation}
\tauadpt \asymp \taumini  \asymp \sqrt{\frac{p^{1-2\delta}}{{n}}}.
\label{eq: D2-c match}
\end{equation}
\end{enumerate}
\end{enumerate}
\end{Corollary}

We will provide some remarks for the above corollary. As an analogy to the exact sparse loading, (D1), (D2-a) and (D2-b) correspond to (E1), (E2-a) and (E2-b), respectively. 
\begin{enumerate}
\item[\rm (D1)] This corresponds to a large class of fast decaying loadings. in this case, even without the exact information about the sparsity level, we can conduct the hypothesis testing procedure as well as we know the exact sparsity level. 
\item [\rm (D2)] For the case of slowly decaying loading $\bxnew$, we first discuss the following two cases,
\begin{enumerate}
\item[\rm (D2-a)] This is similar to (E2-a), where the prior knowledge of sparsity $\ku$ is relatively precise. We can use the sparsity level $\ku$ to construct a testing procedure matching the adaptive detection boundary. See the proof of Corollary \ref{cor: decaying
 adapt} for details.
\item[\rm (D2-b)] This is similar to (E2-b), where the prior knowledge of sparsity $\ku$ is rough. For such a case, the proposed testing procedure $\phi_{\alpha}$ defined in \eqref{eq: test} achieves the adaptive detection boundary. 
\end{enumerate}
\end{enumerate}
Although the decaying loading shares some similarity with the exact sparse loading, there still exist significant distinctions in terms of the exact detection boundary. Interestingly, there exists an additional case (D2-c), which correspond to the case that the true sparsity itself is relatively dense. In this case, although the true sparsity level is high and the knowledge of sparsity is not available, the hypothesis testing problem itself is adaptive, which means, without any knowledge on the true sparsity level, we can conduct the hypothesis testing problem as well as the case of known sparsity. 


We conclude this section by establishing a uniform optimality result of the proposed testing procedure $\phi_{\alpha}$ in \eqref{eq: test} over the decaying loading $\bx_{\rm new}$, which parallels the corollary \ref{cor: optimality} in the main paper for the case of exact loading,
\begin{Corollary}
\label{cor: optimality decaying}
Suppose that  $s\leq \ku \lesssim \frac{\sqrt{n}}{\log p}$ and $\gamma_{u}\geq \frac{1}{2}$. Then the testing procedure $\phi_{\alpha}$ in \eqref{eq: test} achieves the adaptive detection boundary $\tauadpt \asymp \frac{\|\bx_{\rm new}\|_2}{\sqrt{n}}$ for any $\bx_{\rm new}$ satisfying \eqref{eq: decaying}. 
\end{Corollary}

The above Corollary states that, in absence of accurate sparsity information, the proposed procedure $\phi_{\alpha}$ is an adaptive optimal test for all decaying loadings $\bx_{\rm new}$. 
\section{Sparsity-assisted Hypothesis Testing Procedure}
\label{sec: sparsity procedure}
In this section, we consider the setting that there is additional information on the sparsity and present the method of constructing confidence interval for $\Deltanew$ and conducting hypothesis testing for \eqref{eq: testing problem} with incorporating the given sparsity information. 
Without loss of generality, we can assume the loading $\bxnew$ is ordered as follows,
\begin{equation}
\left|\bx_{\rm new,1}\right|\geq \left|\bx_{\rm new, 2}\right|\geq \cdots \geq |\bx_{\rm new, p}|.
\label{eq: aligned xi}
\end{equation}
For $k=1,2$, we define $\widetilde{\bbeta}_{k,j}$ to be the de-biased estimator introduced by \citet{javanmard2014confidence,zhang2014confidence,van2014asymptotically} with the corresponding covariance matrix of $\widetilde{\bbeta}_{k,\cdot}\in \R^{p}$ as $\widehat{\rm Var}^{k}$. Define the vector ${\xi}$ as a sparsified version of $\bxnew$,
\begin{equation}
{\xi}_j=x_{\rm new, j} \quad \text{for} \;\;1\leq j\leq q, \quad {\xi}_j=0 \quad \text{for} \;\; q+1\leq  j\leq p,
\end{equation}
where $q$ is an integer to be specified later. 
Define $\dd=\bbeta_1-\bbeta_2$ and the index sets $G_1$ and $G_2$ as
\begin{equation*}
G_1=\left\{j: q+1\leq j\leq p,\; \max\left\{\left|\widetilde{\bbeta}_{1,j}/\sqrt{\widehat{\rm Var}^{1}_{j,j}}\right|, \left|\widetilde{\bbeta}_{2,j}/\sqrt{\widehat{\rm Var}^{2}_{j,j}}\right|\right\}>\sqrt{{2.01\log (2p)}} \right\}
\end{equation*}
and 
\begin{equation*}
G_2=\{j: q+1\leq j\leq p, j \not\in G_1\}
\end{equation*}

%
%
We define the estimator $\widehat{\xi^{\intercal}\dd}$ as in \eqref{eq: correction} with $\bxnew$ replaced with the sparsified loading $\xi$. In particular, the projection directions
$\check{\bf\uu}_k$ for $k=1,2$ are constructed as
\begin{align*}
\check{\bf\uu}_k=\;\argmin_{\bu\in \R^{p}} \bu\trans \bSigmahat_k\bu \quad \text{subject to}\;
& \left \|\bSigmahat_k\bu-\xi\right\|_{\infty}\leq  \|\xi\|_2 \lambda_{k}\\
&\;\left |\xi\trans \bSigmahat_k\bu-\|\xi\|_2^2 \right|\leq \|\xi\|_2^2\lambda_{k} ,
\end{align*}
where $\lambda_{k}\asymp \sqrt{{\log p}/{n_k}}$. 
 We propose the following estimator 
\begin{equation}
\check{\Delta}_{\rm new}=\widehat{\xi^{\intercal}\dd} +\sum_{j \in G_1} x_{\rm new, j}\left(\widetilde{\bbeta}_{1,j}-\widetilde{\bbeta}_{2,j}\right) .
\label{eq: est with sparsity}
\end{equation}
Note that, with probability larger than $1-p^{-c}$,  
\begin{equation}
\max_{j\in G_1}\left|\left(\widetilde{\bbeta}_{1,j}-\widetilde{\bbeta}_{2,j}\right)-\dd_{j}\right| \leq \sqrt{\widehat{\rm Var}^{1}_{j,j}+\widehat{\rm Var}^{2}_{j,j}} \sqrt{2.01 \log p}.
\label{eq: bias bound 1}
\end{equation}
and 
\begin{equation}
\max_{j \in G_2} |\bbeta_{1,j}| \leq \sqrt{\widehat{\rm Var}^{1}_{j,j}}2\sqrt{2.01 \log (2p)} \quad \text{and}\quad \max_{j \in G_2} |\bbeta_{2,j}| \leq \sqrt{\widehat{\rm Var}^{2}_{j,j}}2\sqrt{2.01 \log (2p)}.
\label{eq: event 2}
\end{equation}
As a consequence, we apply \eqref{eq: event 2} and obtain 
\begin{equation}
\begin{aligned}
&\sum_{j\in G_2} \left|{\bbeta_{1,j}}\right| \leq \sqrt{2\frac{\log p}{n_1}}\sigma_1 \sum_{j\in G_2}   \min\left\{\left|\frac{\bbeta_{1,j}}{\sqrt{2\frac{\log p}{n_1}}\sigma_1}\right| ,\frac{\sqrt{\widehat{\rm Var}^{1}_{j,j}}2\sqrt{2.01 \log (2p)}}{\sqrt{2\frac{\log p}{n_1}}\sigma_1}\right\} \\
&\leq \sqrt{2\frac{\log p}{n_1}}\sigma_1 \sum_{j\in G_2}   \min\left\{\left|\frac{\bbeta_{1,j}}{\sqrt{2\frac{\log p}{n_1}}\sigma_1}\right|,1\right\}\cdot \max\left\{1,\max_{j\in G_2}\frac{\sqrt{\widehat{\rm Var}^{1}_{j,j}}2\sqrt{2.01 \log (2p)}}{\sqrt{2\frac{\log p}{n_1}}\sigma_1}\right\}\\
&\leq \max\left\{\sqrt{2\frac{\log p}{n_1}}\sigma_1, \max_{j\in G_2}{\sqrt{\widehat{\rm Var}^{1}_{j,j}}2\sqrt{2.01 \log (2p)}} \right\} s_{u}
\end{aligned}
\end{equation}
where the last inequality follows from the assumption of capped $\ell_1$ sparse of $\bbeta_{1}$. Similarly, we have 
\begin{equation}
\begin{aligned}
\sum_{j\in G_2} \left|{\bbeta_{2,j}}\right| \leq \max\left\{\sqrt{2\frac{\log p}{n_2}}\sigma_2, \max_{j\in G_2}{\sqrt{\widehat{\rm Var}^{2}_{j,j}}2\sqrt{2.01 \log (2p)}}\right\} s_{u}
\end{aligned}
\end{equation}
Hence, we have 
\begin{equation}
\begin{aligned}
&\left|\sum_{j\in G_2} x_{\rm new, j}(\bbeta_{1,j}-\bbeta_{2,j})\right|\leq \max_{j\in G_2}|x_{\rm new,j}|\cdot \left(\sum_{j\in G_2} \left|{\bbeta_{1,j}}\right|+\sum_{j\in G_2} \left|{\bbeta_{2,j}}\right|\right)\\
&\leq \max_{j\in G_2}|x_{\rm new,j}|\cdot  s_{u} \sum_{k=1}^{2}\max\left\{\sqrt{2\frac{\log p}{n_k}}\sigma_2, \max_{j\in G_2}{\sqrt{\widehat{\rm Var}^{k}_{j,j}}2\sqrt{2.01 \log (2p)}}\right\} 
\end{aligned}
\label{eq: bias bound 2}
\end{equation}
Combining the bounds \eqref{eq: bias bound 1} and \eqref{eq: bias bound 2}, we have 
\begin{equation}
\begin{aligned}
&\left|\sum_{j \in G_1} x_{\rm new, j}\left(\widetilde{\bbeta}_{1,j}-\widetilde{\bbeta}_{2,j}\right)-\sum_{j=q+1}^{p}x_{\rm new, j} \dd_j\right| \leq S(\ku)
\end{aligned}
\label{eq: comp 2}
\end{equation}
where 
\begin{equation}
\begin{aligned}
&{S}(\ku)=\sum_{j \in G_1} \sqrt{\widehat{\rm Var}^{1}_{j,j}+\widehat{\rm Var}^{2}_{j,j}} \sqrt{2.01 \log p} \\
&+\max_{j\in G_2}|x_{\rm new,j}|\cdot  s_{u} \sum_{k=1}^{2}\max\left\{\sqrt{2\frac{\log p}{n_k}}\sigma_2, \max_{j\in G_2}{\sqrt{\widehat{\rm Var}^{k}_{j,j}}2\sqrt{2.01 \log (2p)}}\right\} 
\end{aligned}
\end{equation}
It follows from Theorem \ref{thm: limiting distribution} that 
\begin{equation}
\frac{\widehat{\xi^{\intercal}\dd}-\xi^{\intercal}\dd}{\sqrt{\widetilde{V}}}\rightarrow N(0,1)
\label{eq: comp 1}
\end{equation}

We construct the CI as
\begin{equation}
{\rm CI}(q,\ku)=\left(\check{\Delta}_{\rm new}-z_{\alpha/2}\sqrt{\widetilde{V}}-{S}(\ku),\check{\Delta}_{\rm new}+z_{\alpha/2}\sqrt{\widetilde{V}}+{S}(\ku)\right)
\label{eq: comb thresholded CI}
\end{equation}
with $\widetilde{\rm V}$ defined as \begin{equation}
\widetilde{\rm V}=\frac{\widehat{\sigma}_1^2}{n_1}\check{\bf\uu}_1\trans  \bSigmahat_1\check{\bf\uu}_1+\frac{\widehat{\sigma}_2^2}{n_2}\check{\bf\uu}_2\trans  \bSigmahat_2\check{\bf\uu}_2.
\label{eq: variance est sparsity}
\end{equation}

We propose the following decision rule,
\begin{equation}
\phi(q,\ku)={\mathbf 1}\left(\check{\Delta}_{\rm new}-z_{\alpha/2}\sqrt{\widetilde{V}}-{S}(\ku)>0\right).
\label{eq: test with sparsity}
\end{equation}

Combining \eqref{eq: comp 1} and \eqref{eq: comp 2}, we establish the coverage property of the confidence interval ${\rm CI}(q,\ku)$ proposed in \eqref{eq: comb thresholded CI} and also control the type I error of the testing procedure $\phi(q,\ku)$ defined in \eqref{eq: test with sparsity}.  It remains to control the length of $z_{\alpha/2}\sqrt{\widetilde{V}}+{S}(\ku)$. We focus on the decaying loading $|x_{\rm new,j}|\asymp j^{-\delta}$ for $0\leq \delta<\frac{1}{2}$. Following from \eqref{eq: sum bound}, we have 
\begin{equation}
\sqrt{\widetilde{V}} \asymp \|\xi\|_2\lesssim \sqrt{\sum_{j=1}^{q} |x_{\rm new, j}|^2} \sqrt{\frac{1}{n_1}+\frac{1}{n_2}}\lesssim q^{\frac{1}{2}-\delta}\sqrt{\frac{1}{n_1}+\frac{1}{n_2}}
\end{equation}
and 
\begin{equation}
{S}(\ku)\lesssim q^{-\delta} s_{u} \sqrt{\log p} \sqrt{\frac{1}{n_1}+\frac{1}{n_2}}
\end{equation}
We take $q=\lfloor \ku^2\log p \rfloor$ for both decaying loading and the exact loading and have 
\begin{equation}
\left|z_{\alpha/2}\sqrt{\widetilde{V}}+{S}(\ku)\right|\lesssim (\ku^2 \log p)^{\frac{1}{2}-\delta} \sqrt{\frac{1}{n_1}+\frac{1}{n_2}}
\end{equation}
 For the exact loading, we can also take $q=0$.

\section{Additional Proofs}
\label{sec: sup proof}
We present the proofs of Theorems \ref{thm: estimation error} and \ref{thm: limiting distribution} in Section \ref{sec: estimation and limiting} and the proof of Proposition \ref{prop: limiting distribution relax} in Section \ref{sec: relax};  we present the proof of Proposition \ref{prop: dual form} in Section \ref{sec: dual proof};
we present the proof of Corollaries \ref{Cor: hypothesis testing} and \ref{Cor: coverage}  in Section \ref{sec: coverage}; we present the proof of Proposition \ref{prop: challenge} in Section \ref{sec: challenge}; we present the proof of Theorem \ref{thm: general lower} in Section \ref{sec: general lower}; we present the proof of Theorem \ref{thm: decaying lower} and Corollary \ref{cor: decaying adapt} in Section \ref{sec: decaying lower}.
\subsection{Proof of Theorems \ref{thm: estimation error} and \ref{thm: limiting distribution}}
\label{sec: estimation and limiting}
By combining the error decompositions for $k=1$ and $k=2$ in \eqref{eq: decomposition a}, we have 
\begin{equation}
\begin{aligned}
\esta-\estb&-\bx_{\rm new}^{\intercal}\left({\bbeta_1}-\bbeta_2\right)=\buhat_1^{\intercal} \frac{1}{n}\sum_{i=1}^{n} \bXi\epsilon_{1,i}-\buhat_2^{\intercal} \frac{1}{n}\sum_{i=1}^{n} \bXib\epsilon_{2,i}\\
&+\left(\bSigmahat_1\buhat_1-\bx_{\rm new}\right)^{\intercal}(\widehat{\bbeta}_1-\bbeta_1)-\left(\bSigmahat_2\buhat_2-\bx_{\rm new}\right)^{\intercal}\left(\widehat{\bbeta}_2-\bbeta_2\right)
\end{aligned}
\label{eq: key expression}
\end{equation}
By H{\"o}lder's inequality, for $k=1,2$, we have
$$\left|(\bSigmahat_k\buhat_k-\bx_{\rm new})^{\intercal}(\widehat{\bbeta}_k-\bbeta_k)\right|\leq \|\bSigmahat_k\buhat_k-\bx_{\rm new}\|_{\infty}\cdot \|\widehat{\bbeta}_k-\bbeta_k\|_1\lesssim \|\bx_{\rm new}\|_2 \sqrt{\frac{\log p}{n_{k}}}\cdot s_k\sqrt{\frac{\log p}{n_k}},$$
where the second inequality follows from the optimization constraint \eqref{eq: constraint 1} and the condition ${\rm (B1)}$. Hence, we establish that, with probability larger than $1-g(n_1,n_2)$,
\begin{equation}
\begin{aligned}
& \left|\left(\bSigmahat_1\buhat_1-\bx_{\rm new}\right)^{\intercal}(\widehat{\bbeta}_1-\bbeta_1)-\left(\bSigmahat_2\buhat_2-\bx_{\rm new}\right)^{\intercal}\left(\widehat{\bbeta}_2-\bbeta_2\right)\right|\\
\lesssim\ &  \|\bx_{\rm new}\|_2 \left(\frac{\|\bbeta_1\|_0 \log p}{n_1}+\frac{\|\bbeta_2\|_0 \log p}{n_2}\right).
\end{aligned}
\label{eq: bias bound}
\end{equation}
\noindent \underline{Proof of Theorem \ref{thm: estimation error}} Under the assumption ${\rm (A1)}$
\begin{equation}
\E_{\cdot\mid \Xbb}\left(\buhat_1^{\intercal} \frac{1}{n_1}\sum_{i=1}^{n_1} \bXi\epsilon_{1,i}-\buhat_2^{\intercal} \frac{1}{n_2}\sum_{i=1}^{n_2} \bXib\epsilon_{2,i}\right)^2\lesssim{\rm V},
\end{equation}
where ${\rm V}$ is defined in \eqref{eq: variance}. By \eqref{eq: dominating variance}, with probability larger than $1-p^{-c}-\frac{1}{t^2}$, then
\begin{equation}
\left|\buhat_1^{\intercal} \frac{1}{n_1}\sum_{i=1}^{n_1} \bXi\epsilon_{1,i}-\buhat_2^{\intercal} \frac{1}{n_2}\sum_{i=1}^{n_2} \bXib\epsilon_{2,i}\right|\lesssim t \|\bx_{\rm new}\|_2 \left({1}/{\sqrt{n_1}}+{1}/{\sqrt{n_2}}\right)
\label{eq: variance bound}
\end{equation}
Combing \eqref{eq: bias bound} and \eqref{eq: variance bound}, we establish Theorem \ref{thm: estimation error}.\\

\noindent \underline{Proof of Theorem \ref{thm: limiting distribution}} We establish Theorem \ref{thm: limiting distribution} by assuming the Gaussian error assumption ${\rm (A2)}$. Conditioning on $\Xbb$, we establish the following by applying Condition ${\rm (A2)}$,
$
\buhat_1^{\intercal} \frac{1}{n_1}\sum_{i=1}^{n_1} \bXi\epsilon_{1,i}-\buhat_2^{\intercal} \frac{1}{n_2}\sum_{i=1}^{n_2} \bXib\epsilon_{2,i}\sim N(0,{\rm V})
$
where ${\rm V}$ is defined in \eqref{eq: variance}. After normalization, we have
\begin{equation}
\frac{1}{\sqrt{\rm V}} \left(\buhat_1^{\intercal} \frac{1}{n_1}\sum_{i=1}^{n_1} \bXi\epsilon_{1,i}-\buhat_2^{\intercal} \frac{1}{n_2}\sum_{i=1}^{n_2} \bXib\epsilon_{2,i}\right)\mid \Xbb \sim N(0,1)
\label{eq: conditional distribution}
\end{equation}
and then after integrating with respect to $\Xbb$, we have
\begin{equation} \frac{1}{\sqrt{\rm V}} \left(\buhat_1^{\intercal} \frac{1}{n_1}\sum_{i=1}^{n_1} \bXi\epsilon_{1,i}-\buhat_2^{\intercal} \frac{1}{n_2}\sum_{i=1}^{n_2} \bXib\epsilon_{2,i}\right)\sim N(0,1)
\label{eq: unconditional distribution}
\end{equation}
Combing \eqref{eq: bias bound} with \eqref{eq: dominating variance}, we show that with probability larger than $1-p^{-c}-g(n_1,n_2)$, 
$$
\frac{1}{\sqrt{\rm V}}\left|\left(\bSigmahat_1\buhat_1-\bx_{\rm new}\right)^{\intercal}(\widehat{\bbeta}_1-\bbeta_1)-\left(\bSigmahat_2\buhat_2-\bx_{\rm new}\right)^{\intercal}\left(\widehat{\bbeta}_2-\bbeta_2\right)\right|\leq \frac{\|\bbeta_1\|_0 \log p}{\sqrt{n_1}}+\frac{\|\bbeta_2\|_0 \log p}{\sqrt{n_2}}.
$$
Together with \eqref{eq: unconditional distribution}, we establish Theorem \ref{thm: limiting distribution}.
\subsection{Proof of Proposition \ref{prop: limiting distribution relax}}
\label{sec: relax}
This proposition is a modification of proofs of Theorem \ref{prop: limiting distribution relax} and Theorem \ref{thm: limiting distribution} presented in \ref{sec: estimation and limiting}. We shall focus on three parts that are different from the proof in \ref{sec: estimation and limiting}.  Firstly, we modify the error decomposition of \eqref{eq: key expression} as follows, by including the additional approximation error,
\begin{equation}
\begin{aligned}
\esta-\estb&-\bx_{\rm new}^{\intercal}\left({\bbeta_1}-\bbeta_2\right)=\buhat_1\trans \frac{1}{n_1}\sum_{i=1}^{n_1} \bXi \left(\epsilon_{1,i}+{\bf r}_{1,i}\right)-\buhat_2\trans \frac{1}{n_2}\sum_{i=1}^{n_2} \bXi \left(\epsilon_{2,i}+{\bf r}_{2,i}\right)\\
&+\left(\bSigmahat_1\buhat_1-\bx_{\rm new}\right)^{\intercal}(\widehat{\bbeta}_1-\bbeta_1)-\left(\bSigmahat_2\buhat_2-\bx_{\rm new}\right)^{\intercal}\left(\widehat{\bbeta}_2-\bbeta_2\right)
\end{aligned}
\label{eq: key expression general}
\end{equation}
%
We can control the additional error terms involving ${\bf r}_1$ as
$$| \buhat_1\trans \frac{1}{n_1}\sum_{i=1}^{n_1} \bXi {\bf r}_{1,i}| \leq \sqrt{\frac{1}{n_1^2}\sum_{i=1}^{n_1} ( \buhat_1\trans \bXi)^2} \|{\bf r}_{1}\|_2=o_{p}(\sqrt{{\rm V}}),$$ which follows from the assumption on ${\bf r}_1.$
Similar argument is applied to the term involved with ${\bf r}_2.$

Secondly, we check the Lindeberg's condition and establish the asymptotic normality as a modification of \eqref{eq: unconditional distribution} by allowing for non-Gaussian errors. 
We write $$\frac{1}{\sqrt{\rm V}} \left(\buhat_1^{\intercal} \frac{1}{n_1}\sum_{i=1}^{n_1} \bXi\epsilon_{1,i}-\buhat_2^{\intercal} \frac{1}{n_2}\sum_{i=1}^{n_2} \bXib\epsilon_{2,i}\right)=\sum_{k=1}^{2}\sum_{i=1}^{n_k} W_{k,i}$$ with 
\begin{equation}
W_{1,i}=\frac{1}{n_1\sqrt{\rm V}}\buhat_1^{\intercal}\bXi\epsilon_{1,i} \; \text{for}\; 1\leq i\leq n_1 \quad \text{and}\quad W_{2,i}=-\frac{1}{n_2\sqrt{\rm V}}\buhat_2^{\intercal}\bXib\epsilon_{2,i} \; \text{for}\; 1\leq i\leq n_1
\end{equation}


Conditioning on $\bX$, then $\{W_{1,i}\}_{1\leq i\leq n_1}$ and $\{W_{2,i}\}_{1\leq i\leq n_2}$ are independent random variables with $\E(W_{1,i}\mid \bX)=0$ for $1\leq l\leq n_1$, $\E(W_{2,i}\mid \bX)=0$ for $1\leq k\leq n_2$ and $\sum_{k=1}^{2}\sum_{i=1}^{n_k}{\rm Var}(W_{k,i}\mid \bX)=1$.
To establish the asymptotic normality, it is sufficient to check the Lindeberg's condition, that is, for any constant $c>0$,
\begin{equation*}
\begin{aligned}
\sum_{k=1}^{2}\sum_{i=1}^{n}\E\left(W_{ki}^2 \mathbf{1}{\left\{\left|W_{ki}\right|\geq c\right\}}\mid \bX\right)&\leq \sum_{k=1}^{2}\sum_{i=1}^{n}\frac{\sigma_k^2}{n_k^2{\rm V}}(\buhat_k^{\intercal}\bX_{k,i})^2\E\left(\frac{\epsilon_{k,i}^2}{\sigma_k^2} \mathbf{1}{\left\{\left|\epsilon_{ki}\right|\geq \frac{c n_{k}\sqrt{\rm V}}{\|\bxnew\|_2\tau_k}\right\}}\mid \bX \right)\\
&\leq \max_{1\leq k \leq 2}\max_{1\leq i\leq n_k}\E\left(\frac{\epsilon_{k,i}^2}{\sigma_k^2} \mathbf{1}{\left\{\left|\epsilon_{ki}\right|\geq \frac{c n_{k}\sqrt{\rm V}}{\|\bxnew\|_2\tau_k}\right\}}\mid \bX \right)\\
&\lesssim \left(\frac{c n_{k}\sqrt{\rm V}}{\|\bxnew\|_2\tau_k}\right)^{-\nu}
\end{aligned}
\label{eq: lin condition}
\end{equation*}
where the first inequality follows from the optimization constraint $\|\bX_{k}\bu\|_{\infty}\leq \|\bx_{\rm new}\|_2\tau_{k}$
in \eqref{eq: relax}, the second inequality follows from the definition of ${\rm V}$ in \eqref{eq: variance} and the last inequality follows from the condition
$\max_{k=1,2}\max_{1\leq i\leq n}\E(\epsilon_{k,i}^{2+\nu}\mid \bX_{k,i})\leq M_0.$
Hence, conditioning on $\bX$, we establish the asymptotic normality of $\frac{1}{\sqrt{\rm V}} \left(\buhat_1^{\intercal} \frac{1}{n_1}\sum_{i=1}^{n_1} \bXi\epsilon_{1,i}-\buhat_2^{\intercal} \frac{1}{n_2}\sum_{i=1}^{n_2} \bXib\epsilon_{2,i}\right).$ By calculating its characteristic function, we can apply bounded convergence theorem to establish 
$$\frac{1}{\sqrt{\rm V}} \left(\buhat_1^{\intercal} \frac{1}{n_1}\sum_{i=1}^{n_1} \bXi\epsilon_{1,i}-\buhat_2^{\intercal} \frac{1}{n_2}\sum_{i=1}^{n_2} \bXib\epsilon_{2,i}\right)\overset{d}{\rightarrow} N(0,1).$$

Thirdly, we need to verify that $\widehat{\beta}$ satisfies condition ${\rm (B1)}.$ In particular, we state the following lemma, which is essentially rewriting Theorem 6.3. of \citet{buhlmann2011statistics} in the terminology of the current paper. 

\begin{Lemma} For $k=1,2$, the Lasso estimator $\widehat{\bbeta}_k$ defined in \eqref{eq: Lasso estimator a} with 
$
A\geq 8e \|\bepsilon_{k,i}\|_{\psi_2} \max_{1\leq j\leq p}\frac{\|\bX_{k,i}\|_{\psi_2}}{{\bSigma_{k,j,j}}},
$
satisfies 
\begin{equation*}
\| \widehat{\bbeta}_k - \bbeta_k \|_1 \lesssim \frac{1}{\sqrt{n_k \log p}} \| {\bf r}_k\|_2^2 + \sqrt{\frac{\log p}{n_k}} \sum_{j=1}^{p} \min\{ |{\bbeta}_{k,j}|/\sigma_{k}\lambda_0 ,1\} \lesssim s_k  \sqrt{\frac{\log p}{n_k}}. 
\end{equation*}
where $\|\bepsilon_{k,i}\|_{\psi_2}$ and  $\|\bX_{k,i}\|_{\psi_2}$ denote the sub-gaussian norms of $\bepsilon_{k,i} \in \R$ and $\bX_{k,i}\in \R^{p}$, respectively, and $\bSigma_{k,j,j}$ denotes the $j$-th diagonal entry of $\bSigma_{k}.$
\label{prop: approximate sparse}
\end{Lemma}
To apply Theorem 6.3. of \citet{buhlmann2011statistics}, it is sufficient to check 
$$\max_{1\leq j\leq p} \frac{1}{W_{k,j}} \left|\frac{1}{n_k }\sum_{i=1}^{n_k}\bepsilon_{k,i} \bX_{k,i,j}\right|\leq A\sqrt{\frac{\log p}{n_k}},$$
where $\bX_{k,i,j}$ denote the $j$-th variable of $i$-th observation in the $k$-th treatment group.
Note that $\E \bepsilon_{k,i} \bX_{k,i,j}=0$ and $\bepsilon_{k,i} \bX_{k,i,j}$ is random variable with sub-exponential norm smaller than $2\|\bepsilon_{k,i}\|_{\psi_2} \|\bX_{k,i}\|_{\psi_2}.$  By equation (73) of \citet{javanmard2014confidence} or Corollary 5.17 of \citet{vershynin2010introduction}, we establish that, with probability larger than $1-p^{-c},$ for some positive constant $c>0,$ 
$$
\max_{1\leq j\leq p} \left|\frac{1}{n_k }\sum_{i=1}^{n_k}\bepsilon_{k,i} \bX_{k,i,j}\right|\leq 8 e\sqrt{\frac{\log p}{n_k}}\|\bepsilon_{k,i}\|_{\psi_2} \|\bX_{k,i}\|_{\psi_2}
$$
By definition of $G_3$ and Lemma 4 in \citet{cai2015regci}, we have $P\left(\left|\frac{W_{k,j}}{\bSigma_{k,j,j}}-1\right|\gtrsim \sqrt{\log p/n_k}\right)\lesssim p^{-c}$. Hence, with probability larger than $1-p^{-c}$ for some positive $c>0,$ we have  
\begin{equation*}
\max_{1\leq j\leq p} \frac{1}{W_{k,j}} \left|\frac{1}{n_k }\sum_{i=1}^{n_k}\bepsilon_{k,i} \bX_{k,i,j}\right|\leq 8 e\sqrt{\frac{\log p}{n_k}}\|\bepsilon_{k,i}\|_{\psi_2} \max_{1\leq j\leq p}\frac{\|\bX_{k,i}\|_{\psi_2}}{{\bSigma_{k,j,j}}}.
\end{equation*}


\subsection{Proof of Proposition \ref{prop: dual form}}
\label{sec: dual proof}
In the following proof, we omit the index $k$ to simply the notation, that is, $\bu=\bu_{k}$, $\bSigmahat=\bSigmahat_{k}$ and $\lambda=\lambda_{k}$.
We introduce the corresponding Lagrange function, 
\begin{equation}
\begin{aligned}
&L(\bu,\tau,\eta,\tau_0,\eta_0)=\bu^{\intercal}\bSigmahat \bu+\tau^{\intercal}\left(\bSigmahat \bu-\bx_{\rm new}- \|\bx_{\rm new}\|_2 \lambda{\bf 1}\right)
+\eta^{\intercal}\left(\bx_{\rm new}-\bSigmahat \bu -\|\bx_{\rm new}\|_2 \lambda{\bf 1}\right)\\
&+\tau_0\left(\frac{\bx_{\rm new}^{\intercal}}{\|\bx_{\rm new}\|_2} \bSigmahat \bu-(1+\lambda)\|\bx_{\rm new}\|_2\right)+\eta_0\left((1-\lambda)\|\bx_{\rm new}\|_2-\frac{\bx_{\rm new}^{\intercal}}{\|\bx_{\rm new}\|_2} \bSigmahat u\right)
\end{aligned}
\end{equation}
where $\tau\in \R^{p}$, $\eta\in \R^{p}$ and $\left\{\tau_i,\eta_i\right\}_{0\leq i\leq p}$ are positive constants. 
Then we derive the dual function $g(\tau,\eta,\tau_0,\eta_0)=\arg\min_{\bu}L(u,\tau,\eta,\tau_0,\eta_0)$. By taking the first order-derivative of $L(u,\tau,\eta,\tau_0,\eta_0)$, we establish that the minimizer $u^{*}$ of $L(u,\tau,\eta,\tau_0,\eta_0)$ satisfies 
\begin{equation}
2\bSigmahat \bu^{*}+\bSigmahat\left[\left(\tau-\eta\right)+\left(\tau_0-\eta_0\right)\frac{\bx_{\rm new}}{\|\bx_{\rm new}\|_2}\right]=0.
\label{eq: relationship}
\end{equation}
By introducing $\gamma=\tau-\eta$ and $\gamma_0=\tau_0-\eta_0$, we have the expression of $L(u^*,\tau,\eta,\tau_0,\eta_0)$ as 
\begin{equation*}
\begin{aligned}
g(\gamma,\eta,\gamma_0,\eta_0)&=-\frac{1}{4}\left[\gamma+\gamma_0\frac{\bx_{\rm new}}{\|\bx_{\rm new}\|_2}\right]^{\intercal}\bSigmahat\left[\gamma+\gamma_0\frac{\bx_{\rm new}}{\|\bx_{\rm new}\|_2}\right]-\bx_{\rm new}^{\intercal}\gamma-\|\bx_{\rm new}\|_2 \lambda{\bf 1}^{\intercal}(\gamma+2\eta)\\  
&-\|\bx_{\rm new}\|_2\gamma_0-\lambda_n\|\bx_{\rm new}\|_2(\gamma_0+2\eta_0), \quad \text{where}\; \eta_i\geq -\gamma_i \;\text{and}\; \eta_i\geq 0 \; \text{for}\; 0\leq i\leq p
\end{aligned}
\end{equation*}
The computation of the maximum over $\eta_0$ and $\{\eta_i\}_{1\leq i\leq p}$ is based on the following observation, if $\gamma_i\geq 0$, then $\max_{\eta_i\geq \max\{0,-\gamma_i\}}{\gamma_i+2\eta_i}=\gamma_i$; if $\gamma_i< 0$, then $\max_{\eta_i\geq \max\{0,-\gamma_i\}}{\gamma_i+2\eta_i}=-\gamma_i$; Hence,
\begin{equation}
\max_{\eta_i\geq \max\{0,-\gamma_i\}}{\gamma_i+2\eta_i}=\left|\gamma_i\right|.
\label{eq: key maximum}
\end{equation}
By applying \eqref{eq: key maximum}, we establish 
\begin{equation}
\begin{aligned}
\max_{\eta,\eta_0}g(\gamma,\eta,\gamma_0,\eta_0)&=-\frac{1}{4}\left[\gamma+\gamma_0\frac{\bx_{\rm new}}{\|\bx_{\rm new}\|_2}\right]^{\intercal}\bSigmahat\left[\gamma+\gamma_0\frac{\bx_{\rm new}}{\|\bx_{\rm new}\|_2}\right]\\
&-\bx_{\rm new}^{\intercal}\left(\gamma+\gamma_0\frac{\bx_{\rm new}}{\|\bx_{\rm new}\|_2}\right)-\lambda \|\bx_{\rm new}\|_2 \left(|\gamma_0|+\|\gamma\|_1\right)\\  
\end{aligned}
\end{equation}
Then it is equivalent to solve the dual problem defined in \eqref{eq: dual problem}.
By \eqref{eq: relationship}, we establish
\begin{equation*}
\buhat_k=\bvhat^{k}_{-1}+\frac{\bx_{\rm new}}{\|\bx_{\rm new}\|_2}\bvhat^{k}_{1}.
\label{eq: computation expression}
\end{equation*}
\subsection{Proof of Corollaries \ref{Cor: hypothesis testing} and \ref{Cor: coverage}}
\label{sec: coverage}

Note that 
$
\left|\frac{\widehat{\rm V}}{{\rm V}}-1\right|\leq \sum_{j=1}^{2}\left|\frac{\widehat{\sigma}_j^2}{\sigma_j^2}-1\right|
$
and hence $\frac{\widehat{\rm V}}{{\rm V}}\cip1$ follows from the condition {\rm (B2)}. Together with Theorem \ref{thm: limiting distribution}, we establish these two corollaries.

\subsection{Proof of Proposition \ref{prop: challenge}} 
\label{sec: challenge}
Under the condition {\rm (F1)}, the projection ${\bf u}={\bf 0}$ belongs to the feasible set in \eqref{eq: initial correction} and hence the minimizer $\butilde_1$ of \eqref{eq: initial correction} is zero since $\bSigmahat_1$ is semi-positive-definite matrix.

If the non-zero coordinates of the loading $\bx_{\rm new}$ are of the same order of magnitude, we have  $\|\bx_{\rm new}\|_2\asymp \|\bx_{\rm new}\|_0 \|\bx_{\rm new}\|_{\infty}$. Then the condition $\|\bx_{\rm new}\|_0\geq C\sqrt{{n_1}/{\log p}}$ will imply the condition {\rm (F1)}.

\subsection{Proof of Theorem \ref{thm: general lower}}
\label{sec: general lower}
Suppose that we observe a random variable $Z$ which has a distribution $\mathbf{P}_{\theta}$ where the parameter $\theta$ belongs to the parameter space $\HH$. Let $\pi_{i}$ denote the prior distribution supported on the parameter space $\HH_{i}$ for $i=0,1$.  Let $f_{\pi_{i}}\left(z\right)$ denote the density function of the marginal distribution of $Z$  with the prior $\pi_{i}$ on $\HH_{i}$ for $i=0,1$. More specifically, $f_{\pi_{i}}\left(z\right)=\int f_{\theta}\left(z\right) \pi_{i}\left(\theta\right)d\theta,$ for $i=0,1.$ 
Denote by $\mathbb{P}_{\pi_{i}}$ the marginal distribution of $Z$ with the prior $\pi_{i}$ on $\HH_{i}$ for $i=0,1$.  For any function $g$, we write $\E_{{\pi_{\HH_0}}}\left(g\left(Z\right)\right)$ for the expectation of $g\left(Z\right)$ with respect to the marginal distribution of $Z$ with the prior $\pi_{\HH_{0}}$ on $\HH_{0}$. We define the $\chi^2$ distance between two density functions $f_{1}$ and $f_{0}$  by
\begin{equation}
\chi^2(f_{1},f_{0})=\int \frac{\left(f_1(z)-f_0(z)\right)^2}{f_0(z)} dz=\int \frac{f^2_{1}(z)}{f_{0}(z)} dz-1
\label{eq: chisq distance}
\end{equation}
and the total variation distance by
$\TV(f_{1},f_{0})=\int \left|f_1(z)-f_0(z)\right| dz.$
It is well known that 
\begin{equation}
\TV(f_{1},f_{0})\leq \sqrt{\chi^2(f_{1},f_{0})}.
\label{eq: relation between chisq and TV}
\end{equation}
\begin{Lemma}
Suppose that $\pi_i$ is a prior on the parameter space $\FF_i$ for $i=0,1$, then we have 
\begin{equation}
\inf_{\theta\in \FF_1} \E_{\theta}\phi\leq L_1\left(f_{\pi_1},f_{\pi_0}\right)+ \sup_{\theta\in \FF_0} \E_{\theta}\phi
\label{eq: basic bound}
\end{equation}
In addition, suppose that $L_1\left(f_{\pi_1},f_{\pi_0}\right)<1-\alpha-\eta$ for $0< \alpha<\frac{1}{2}$, $\FF_0\subset \HH_0(\ku)$ and $\FF_1\subset \HH_1(\kt,\tau),$ then
\begin{equation}
\tauadpt \geq \min_{\theta_0\in \FF_0, \theta_1\in \FF_1}\left|\T(\theta_0)-\T(\theta_1)\right|.
\label{eq: lower bound detection}
\end{equation}
\label{lem: general lower}
\end{Lemma}
\noindent{\bf Proof of Lemma \ref{lem: general lower}} It follows from the definition of $L_1(f_1,f_0)$ that 
\begin{equation}
\E_{\pi_1}\phi-\E_{\pi_0}\phi\leq  L_1\left(f_{\pi_1},f_{\pi_0}\right).
\label{eq: basic 1}
\end{equation}
Then \eqref{eq: basic bound} follows from 
$$\inf_{\theta\in \FF_1} \E_{\theta}\phi\leq \E_{\pi_1}\phi\leq L_1\left(f_{\pi_1},f_{\pi_0}\right)+ \E_{\pi_0}\phi\leq L_1\left(f_{\pi_1},f_{\pi_0}\right)+ \sup_{\theta\in \FF_0} \E_{\theta}\phi,
$$
where the first and last inequalities follows from the definition of $\inf$ and $\sup$ and the second inequality follows from \eqref{eq: basic 1}.  The lower bound in \eqref{eq: lower bound detection} follows from the definition of $\tauadpt$ and the fact that 
$$\omegab(\kt,\tau,\phi)\leq \inf_{\theta\in \FF_1} \E_{\theta}\phi\leq L_1\left(f_{\pi_1},f_{\pi_0}\right)+ \sup_{\theta\in \FF_0} \E_{\theta}\phi\leq 1-\eta.$$

To establish the lower bound results, we divide the whole proof into two parts, where the first proof depends on the location permutation and the second proof does not depend on this.\\
\noindent {\bf Permuted Location Lower Bound} We first establish the following lower bound through permuting the locations of non-zero coefficients, 
\begin{equation}
\tauadpt\geq \frac{1}{\sqrt{n}}\sum_{j=\max\{L-q+2,1\}}^{L} \left|x_{\rm new,j}\right|\sqrt{\left({\log \left({c L}/{q^2}\right)}\right)_{+}}.
\label{eq: permuted lower}
\end{equation}
For this case, we assume that $q\leq \sqrt{c L}$; otherwise the lower bound in \eqref{eq: permuted lower} is trivial.
To simplify the notation of the proof, we fix $\bbeta_2=0$ and denote $\bbeta_1=\bgamma$. In addition, we set $\bSigma_1=\bSigma_2={\rm I}$ and $\sigma_2=1$. Without loss of generality, we set $x_{\rm new,i}\geq 0$ and $x_{\rm new,i}\geq x_{\rm new,i+1}$. 
By applying Lemma \ref{lem: general lower}, we need to construct two parameters spaces $\FF_0$ and $\FF_1$ with considering the following three perspectives,
\begin{enumerate}
\item $\FF_0\subset \HH_0(\ku)$ and $\FF_1\subset \HH_1(\kt,\tau)$.
\item to constrain the distribution distance $L_1\left(f_{\pi_1},f_{\pi_0}\right)$
\item to maximize the functional distance $\min_{\theta_0\in \FF_0, \theta_1\in \FF_1}\left|\T(\theta_0)-\T(\theta_1)\right|$
\end{enumerate}
To establish the lower bound \eqref{eq: permuted lower}, we construct the following parameter spaces,
\begin{equation}
\begin{aligned}
\FF_0&=\left\{\btheta=\left(\begin{aligned} & \bgamma, 1,{\rm I}\\ &{\bf 0},1,{\rm I} \end{aligned}\right): \bgamma_1=\rho\cdot \frac{\sum_{j=L-q+2}^{L}x_{\rm new,j}}{x_{\rm new,1}}, \|\bgamma_{-1}\|_0=q-1, \bgamma_{j} \in \left\{0,-\rho\right\} \; \text{for} \; 2\leq j\leq L\right\}\\
\FF_1&=\left\{\btheta=\left(\begin{aligned} & \bgamma, 1,{\rm I}\\ &{\bf 0},1,{\rm I} \end{aligned}\right): \bgamma_1=\rho\cdot \frac{\sum_{j=L-q+2}^{L}x_{\rm new,j}}{x_{\rm new,1}}, \bgamma_{-1}={\bf 0}\right\}
\end{aligned}
\end{equation}
For $\theta\in \FF_0,$ we have $\Delta_{\rm new}=\rho\cdot \left(\sum_{j=L-q+2}^{L}x_{\rm new,j}-\sum_{j \in \supp(\bgamma_{-1})}x_{\rm new,j}\right)\leq 0$, which is due to the definition of $\supp(\bgamma_{-1})$; For $\theta\in \FF_1,$ we have  $\Delta_{\rm new}=\rho\cdot \sum_{j=L-q+2}^{L}x_{\rm new,j}\geq 0$.
Hence, we have shown that 
\begin{equation}
\FF_0 \subset \HH_0(\ku) \quad \text{and} \quad \FF_1 \subset \HH_1(\kt,\tau) \quad \text{for}\quad \tau=\rho\cdot \sum_{j=L-q+2}^{L}x_{\rm new,j}
\end{equation}
To establish the distributional difference, we introduce $\pi_0$ to be the uniform prior on the parameter space $\FF_0$ and $\pi_1$ to denote the mass point prior on the parameter space $\FF_1$. Since $L_1$ distance is symmetric, we have 
\begin{equation}
\TV(f_{\pi_1},f_{\pi_0})\leq \sqrt{\chi^2(f_{\pi_0},f_{\pi_1})}.
\label{eq: symmetric l1 and chisq}
\end{equation}
As a remarkable difference from the typical lower bound construction, the null parameter space $\FF_0$ is composite but the alternate parameter space $\FF_1$ is simple. We use the symmetric property of the $L_1$ distance to control the distributional difference between this composite null and simple alternative in \eqref{eq: symmetric l1 and chisq}. We take $\rho= \frac{1}{2}\sqrt{\frac{2\log[(L-1)/(q-1)^2]}{n}}.$ By Lemma 3 and Lemma 4 in \citet{cai2018accuracysup}, we rephrase (3.33) in \citet{cai2018accuracysup} as
\begin{equation*}
\chi^2(f_{\pi_0},f_{\pi_1})+1\leq \exp\left(\frac{(q-1)^2}{L-q}\right)\left(1+\frac{1}{\sqrt{L-1}}\right)^{q-1} \leq \exp\left(\frac{(q-1)^2}{L-q}+\frac{q-1}{\sqrt{L-1}}\right)
\end{equation*}
The above inequality is further upper bounded by $\exp(\frac{1}{2}w^2+w)$ for $w=\frac{q}{\sqrt{L}}$. Under the condition $\frac{q}{\sqrt{L}}\leq c$, we have 
$L_1(f_{\pi_1},f_{\pi_0})\leq \sqrt{\exp(\frac{1}{2}c^2+c)-1}.$ By taking $c=\sqrt{1+2 \log [c_0^2+1]}-1$, we have $L_1(f_{\pi_1},f_{\pi_0})<c_0.$
Then it suffices to control the functional difference $\min_{\theta_0\in \FF_0, \theta_1\in \FF_1}\left|\T(\theta_0)-\T(\theta_1)\right|$, where $\T=\Delta_{\rm new}$ and hence we have
\begin{equation}
\min_{\theta_0\in \FF_0, \theta_1\in \FF_1}\left|\T(\theta_0)-\T(\theta_1)\right| \gtrsim \sqrt{\frac{2\log[L/q^2]}{n}} \cdot \sum_{j=L-q+2}^{L}x_{\rm new,j}
\end{equation}

\noindent {\bf Fixed Location Lower Bound} We will establish the following lower bound, \begin{equation}
\tauadpt\geq \frac{1}{\sqrt{n}}\cdot \sqrt{\sum_{j=1}^{s} x^2_{\rm new,j}}.
\label{eq: fixed lower}
\end{equation}
In this case, we do not perturb the location of non-zeros in constructing the null and alternative space but only perturb the coefficients corresponding to $s$-largest coefficients. 
To simplify the notation of the proof, we fix $\bbeta_2=0$ and denote $\bbeta_1=\bgamma$. In addition, we set $\bSigma_1=\bSigma_2={\rm I}$ and $\sigma_2=1$. Without loss of generality, we set $x_{\rm new,i}\geq 0$ and $x_{\rm new,i}\geq x_{\rm new,i+1}$. To establish the lower bound \eqref{eq: permuted lower}, we construct the following parameter space,
\begin{equation}
\begin{aligned}
\FF_0&=\left\{\btheta=\left(\begin{aligned} & 0, 1,{\rm I}\\ &{\bf 0},1,{\rm I} \end{aligned}\right)\right\}\\
\FF_1&=\left\{\btheta=\left(\begin{aligned} & \bgamma, 1,{\rm I}\\ &{\bf 0},1,{\rm I} \end{aligned}\right): \bgamma_j=\rho \cdot \frac{x_{\rm new,j}}{\sqrt{\sum_{j=1}^{s} x_{\rm new,j}^2}} \quad \text{for}\;  1\leq j\leq s\right\}
\end{aligned}
\end{equation}
For $\theta\in \FF_0,$ we have $\Delta_{\rm new}=0$; For $\theta\in \FF_1,$ we have  $\Delta_{\rm new}=\rho\cdot \sqrt{\sum_{j=1}^{s} x_{\rm new,j}^2}\geq 0$. 
Hence, we have shown that 
\begin{equation}
\FF_0 \subset \HH_0(\ku) \quad \text{and} \quad \FF_1 \subset \HH_1(\kt,\tau) \quad \text{for}\quad \tau=\rho\cdot \sqrt{\sum_{j=1}^{s} x_{\rm new,j}^2}
\end{equation}
Let $\pi_0$ and $\pi_1$ denote the point mass prior over the parameter space $\FF_0$ and $\FF_1$, respectively. It follows from (7.25) in \citet{cai2015regci} that 
\begin{equation}
\chi^2(f_{\pi_1},f_{\pi_0})\leq \exp\left(2 n \rho^2 \right)-1
\end{equation}
By taking $\rho=\sqrt{\frac{\log (1+c_0^2)}{2n}},$ we have $L_1(f_{\pi_1},f_{\pi_0})\leq c_0.$ Then it suffices to control the functional difference $\min_{\theta_0\in \FF_0, \theta_1\in \FF_1}\left|\T(\theta_0)-\T(\theta_1)\right|$, where $\T=\Delta_{\rm new}$ and hence we have
\begin{equation}
\min_{\theta_0\in \FF_0, \theta_1\in \FF_1}\left|\T(\theta_0)-\T(\theta_1)\right| \gtrsim \frac{1}{\sqrt{n}}\cdot \sqrt{\sum_{j=1}^{s} x_{\rm new,j}^2}.
\end{equation}

\subsection{Proof of Theorem \ref{thm: decaying lower} and Corollary \ref{cor: decaying adapt}}
\label{sec: decaying lower}
The lower bound is an application of the general detection boundary \eqref{eq: detection boundary}, which is translated to the following lower bound,
\begin{equation}
\tau\leq\tau^{*}= \frac{1}{\sqrt{n}}\cdot\max\left\{\sqrt{\sum_{j=1}^{s} j^{-2\delta}},\sum_{j=\max\{L-q+2,1\}}^{L} j^{-\delta}\sqrt{\left({\log \left({c L}/{q^2}\right)}\right)_{+}}\right\}.
\label{eq: detection boundary decaying}
\end{equation}
We also need the following fact, for integers $l_1>2$ and $l_2>l_1$
\begin{equation*}
\int_{l_1}^{l_2+1}x^{-\delta} dx \leq \sum_{j=l_1}^{l_2}j^{-\delta}\leq \int_{l_1-1}^{l_2}x^{-\delta} dx
\end{equation*}
Hence, we further have 
\begin{equation}
\sum_{j=l_1}^{l_2}j^{-\delta} \in \begin{cases} 
\frac{1}{\delta-1}[(l_1-1)^{1-\delta}-l_2^{1-\delta},l_1^{1-\delta}-(l_2+1)^{1-\delta}] & \; \text{for}\; \delta>1\\
[\log \frac{l_2+1}{l_1}, \log \frac{l_2}{l_1-1}]& \; \text{for}\; \delta=1\\
\frac{1}{1-\delta}[(l_2+1)^{1-\delta}-l_1^{1-\delta},l_2^{1-\delta}-(l_1-1)^{1-\delta}] & \; \text{for}\; \delta<1\\
\end{cases}
\label{eq: sum bound}
\end{equation}
For the case (D1), we first consider the case $2\delta>1$ and hence $\sum_{j=1}^{s} j^{-2\delta}=1+\sum_{j=2}^{s} j^{-2\delta}\asymp 1$; for the case $2\delta=1$, we have $\sum_{j=1}^{s} j^{-2\delta}=1+\sum_{j=2}^{s} j^{-2\delta}\asymp \log s$; Hence, the lower bound \eqref{eq: lower D1} follows.

For the case (D2), we first consider $\gamma_{u}\geq \frac{1}{2}$, we take $q=\sqrt{p}$ in \eqref{eq: detection boundary decaying} and have 
\begin{equation}
\sum_{j=\max\{p-q+2,1\}}^{p} j^{-\delta}\geq \frac{1}{1-\delta}\left((p+1)^{1-\delta}-(p-\sqrt{p}-2)^{1-\delta}\right)\asymp (p-c\sqrt{p})^{-\delta}\sqrt{p}\asymp p^{\frac{1}{2}-\delta}.
\label{eq: first inter}
\end{equation}
For the case $\gamma_{u}< \frac{1}{2}$, we take $L=s_u^2 \log p$, then we have 
$$\sum_{j=\max\{L-s_u+2,1\}}^{L} j^{-\delta}\geq \frac{1}{1-\delta}\left((L+1)^{1-\delta}-(L-s_u+2)^{1-\delta}\right)\asymp (L-c s_u)^{-\delta}s_u\asymp \ku^{1-2\delta} (\log p)^{-\delta}.$$
Hence we have
$$\sum_{j=\max\{L-\ku+2,1\}}^{L} j^{-\delta}\sqrt{\left({\log \left({c L}/{\ku^2}\right)}\right)_{+}}\geq  \ku^{1-2\delta} (\log p)^{\frac{1}{2}-\delta} \sqrt{\frac{\log(\log p)}{\log p}}.$$
Combined with \eqref{eq: first inter}, we establish the lower bound \eqref{eq: lower D2}. 
Since
\begin{equation*}
\|\bx_{\rm new}\|_2=\sqrt{\sum_{j=1}^{p}j^{-2\delta}}\asymp\begin{cases} 
 1& \; \text{for}\; \delta>1/2\\
\sqrt{\log p}& \; \text{for}\; \delta=1/2\\
p^{\frac{1}{2}-\delta}& \; \text{for}\; \delta<1/2\\
\end{cases}
\end{equation*}
we apply Corollary \ref{Cor: hypothesis testing} to establish the upper bounds and show that the detection boundaries $\tauadpt$ in \eqref{eq: D1 match}, \eqref{eq: D2-b match} and \eqref{eq: D2-c match} are achieved by the hypothesis testing procedure $\phi_{\alpha}$ defined in \eqref{eq: test}. In addition, the detection boundaries $\taumini$ and $\tauadpt$ in \eqref{eq: D2-a match} and $\taumini$ in \eqref{eq: D2-b match} are achieved by the hypothesis testing procedure introduced in \eqref{eq: test with sparsity}.


\section{Additional Simulation Results}

\subsection{Additional Simulation Results for Decaying Loading}
\label{sec: decaying loading}
%

In this session, we  present additional simulation results for the decaying loading case, where $\bx_{\rm new}$ is generated as,
$
x_{{\rm new},j}={\rm Ratio} \cdot j^{-\delta}, $
where $\delta \in \{0, 0.1, 0.25, 0.5\}$ and ${\rm Ratio} \in \{0.25,0.375,0.5\}.$ We report the performance of the proposed HITS method for the decaying loading in Table \ref{tab: decaying setting complete}. Specifically, with an increasing sample size, the empirical power reaches $100\%$, the empirical coverage rate reaches $95\%$ and the length of CIs gets shorter. A similar bias-and-variance tradeoff is observed, where across all settings, in comparison to the plug-in Lasso estimator, both the proposed HITS estimator and the plug-in debiased estimator attained substantially smaller bias but at the expense of larger variability. 
For the slow or no decay settings with $\delta = 0$ or $0.1$, the proposed HITS estimator has uniformly higher power and shorter length of CIs than the debiased plug-in estimator $\Deltanewtilde$ while the coverage of the CIs constructed based on both estimators are close to $95\%$.  In the relatively faster decay setting with $\delta = 0.5$,  $\Deltanewtilde$ and our proposed $\widehat{\Delta\subnew}$ perform more similarly. This is not surprising since the case of fast decaying loading is similar to the sparse loading case and the plugging-in of the debiased estimators can be shown to work if the loading is sufficiently sparse (or decaying sufficiently fast). However, we shall emphasize that $\Deltanewtilde$ is substantially more computationally intensive than $\widehat{\Delta\subnew}$. The calculation of $\widehat{\Delta\subnew}$ requires four fittings of Lasso-type algorithms twice whereas $\Deltanewtilde$ requires $2p+2$ fittings. 

\begin{table}[htp]
\centering
\scalebox{0.7}{
\begin{tabular}{|r|rr|rr|rr|rr|rrr|rrr|rrr|}
  \hline
   & &&\multicolumn{2}{c}{ERR}\vline& \multicolumn{2}{c}{Coverage}\vline&\multicolumn{2}{c}{Len}\vline&\multicolumn{3}{c}{HITS}\vline&\multicolumn{3}{c}{Lasso}\vline&\multicolumn{3}{c}{Deb}\vline\\
  \hline
$\delta$ & Ratio&n& HITS & Deb & HITS & Deb & HITS &Deb& RMSE& Bias & SE & RMSE& Bias & SE  & RMSE& Bias & SE  \\ 
  \hline
\multirow{9}{*}{0} & \multirow{3}{*}{0.25}&200&0.86 & 0.79 & 0.96 & 0.95 & 1.54 & 1.75 & 0.39 & 0.03 & 0.39 & 0.20 & 0.13 & 0.15 & 0.45 & 0.01 & 0.45 \\ 
   & &300&0.99 & 0.95 & 0.96 & 0.97 & 1.11 & 1.39 & 0.26 & 0.03 & 0.26 & 0.16 & 0.11 & 0.12 & 0.33 & 0.01 & 0.33 \\ 
   & &400&1.00 & 0.98 & 0.96 & 0.96 & 1.00 & 1.19 & 0.25 & 0.02 & 0.25 & 0.13 & 0.09 & 0.10 & 0.30 & 0.01 & 0.30 \\ 
  \hline
   &  \multirow{3}{*}{0.375}& 200&  0.93 & 0.90 & 0.96 & 0.96 & 2.29 & 2.60 & 0.55 & 0.04 & 0.55 & 0.27 & 0.21 & 0.18 & 0.62 & 0.02 & 0.62 \\ 
   & &300&0.99 & 0.96 & 0.95 & 0.94 & 1.63 & 2.06 & 0.42 & 0.04 & 0.42 & 0.22 & 0.17 & 0.15 & 0.53 & 0.01 & 0.53 \\ 
   & &400&1.00 & 0.99 & 0.95 & 0.95 & 1.46 & 1.76 & 0.35 & 0.01 & 0.35 & 0.19 & 0.14 & 0.12 & 0.43 & 0.05 & 0.43 \\ 
    \hline
   &\multirow{3}{*}{0.5}&200& 0.98 & 0.93 & 0.97 & 0.97 & 3.00 & 3.45 & 0.69 & 0.05 & 0.69 & 0.35 & 0.28 & 0.21 & 0.83 & 0.05 & 0.83 \\ 
   & & 300&1.00 & 1.00 & 0.97 & 0.97 & 2.17 & 2.74 & 0.51 & 0.09 & 0.50 & 0.29 & 0.23 & 0.17 & 0.63 & 0.03 & 0.63 \\ 
   & &400& 1.00 & 1.00 & 0.95 & 0.94 & 1.94 & 2.33 & 0.50 & 0.05 & 0.50 & 0.25 & 0.20 & 0.15 & 0.61 & 0.00 & 0.61 \\ 
    \hline \hline
 \multirow{9}{*}{0.1}  &\multirow{3}{*}{0.25}&200& 0.97 & 0.96 & 0.95 & 0.95 & 1.02 & 1.09 & 0.25 & 0.04 & 0.25 & 0.18 & 0.13 & 0.12 & 0.27 & 0.00 & 0.27 \\ 
   & &300&1.00 & 0.99 & 0.95 & 0.96 & 0.70 & 0.87 & 0.18 & 0.03 & 0.18 & 0.15 & 0.10 & 0.10 & 0.21 & 0.00 & 0.21 \\ 
   & &400&1.00 & 1.00 & 0.96 & 0.96 & 0.65 & 0.75 & 0.16 & 0.03 & 0.16 & 0.12 & 0.09 & 0.08 & 0.19 & 0.00 & 0.19 \\ 
    \hline
   &\multirow{3}{*}{0.375}&200& 0.99 &0.99 & 0.95 & 0.97 & 1.43 & 1.59 & 0.36 & 0.06 & 0.35 & 0.25 & 0.20 & 0.16 & 0.39 & 0.00 & 0.39 \\ 
   &&300& 1.00 & 1.00 & 0.96 & 0.98 & 1.01 & 1.26 & 0.24 & 0.05 & 0.23 & 0.20 & 0.17 & 0.12 & 0.29 & 0.00 & 0.29 \\ 
   &&400& 1.00 & 1.00 & 0.95 & 0.96 & 0.91 & 1.08 & 0.23 & 0.02 & 0.22 & 0.17 & 0.14 & 0.10 & 0.26 & 0.02 & 0.26 \\ 
    \hline
   & \multirow{3}{*}{0.5}& 200&1.00 & 1.00 & 0.97 & 0.97 & 1.87 & 2.09 & 0.44 & 0.05 & 0.44 & 0.30 & 0.26 & 0.17 & 0.50 & 0.02 & 0.50 \\ 
   &&300& 1.00 & 1.00 & 0.96 & 0.97 & 1.32 & 1.66 & 0.32 & 0.07 & 0.32 & 0.26 & 0.22 & 0.14 & 0.40 & 0.00 & 0.40 \\ 
   &&400& 1.00 & 1.00 & 0.96 & 0.94 & 1.20 & 1.42 & 0.30 & 0.05 & 0.29 & 0.22 & 0.19 & 0.12 & 0.35 & 0.00 & 0.35 \\ 
    \hline\hline
\multirow{9}{*}{0.25}  &\multirow{3}{*}{0.25}& 200& 0.99 & 1.00 & 0.93 & 0.92 & 0.59 & 0.62 & 0.16 & 0.02 & 0.16 & 0.17 & 0.11 & 0.13 & 0.17 & 0.01 & 0.17 \\ 
  & &300& 1.00 & 1.00 & 0.95 & 0.95 & 0.44 & 0.50 & 0.12 & 0.03 & 0.11 & 0.14 & 0.10 & 0.09 & 0.13 & 0.00 & 0.13 \\ 
  && 400& 1.00 & 1.00 & 0.94 & 0.95 & 0.39 & 0.45 & 0.11 & 0.02 & 0.10 & 0.12 & 0.09 & 0.08 & 0.12 & 0.01 & 0.12 \\ 
    \hline
   &\multirow{3}{*}{0.375}&200& 1.00 & 1.00 & 0.96 & 0.96 & 0.81 & 0.84 & 0.21 & 0.05 & 0.20 & 0.22 & 0.18 & 0.12 & 0.21 & 0.00 & 0.21 \\ 
   &&300& 1.00 & 1.00 & 0.95 & 0.96 & 0.56 & 0.68 & 0.14 & 0.04 & 0.14 & 0.18 & 0.15 & 0.10 & 0.16 & 0.00 & 0.16 \\ 
   &&400& 1.00 & 1.00 & 0.96 & 0.96 & 0.52 & 0.59 & 0.13 & 0.02 & 0.13 & 0.15 & 0.13 & 0.09 & 0.15 & 0.00 & 0.15 \\ 
    \hline
  &\multirow{3}{*}{0.5}&200& 1.00 & 1.00 & 0.96 & 0.96 & 1.01 & 1.07 & 0.25 & 0.05 & 0.25 & 0.27 & 0.24 & 0.13 & 0.27 & 0.02 & 0.27 \\ 
   && 300& 1.00 & 1.00 & 0.91 & 0.94 & 0.69 & 0.86 & 0.19 & 0.05 & 0.19 & 0.22 & 0.19 & 0.12 & 0.22 & 0.01 & 0.22 \\ 
   && 400&1.00 & 1.00 & 0.93 & 0.95 & 0.64 & 0.74 & 0.17 & 0.03 & 0.17 & 0.19 & 0.16 & 0.10 & 0.19 & 0.00 & 0.19 \\ 
    \hline   \hline
\multirow{9}{*}{0.5}  &\multirow{3}{*}{0.25}& 200& 0.96 & 0.99 & 0.94 & 0.91 & 0.48 & 0.41 & 0.13 & 0.02 & 0.13 & 0.15 & 0.10 & 0.11 & 0.12 & 0.00 & 0.12 \\ 
  && 300& 1.00 & 1.00 & 0.89 & 0.91 & 0.34 & 0.35 & 0.10 & 0.02 & 0.10 & 0.13 & 0.09 & 0.09 & 0.10 & 0.00 & 0.10 \\ 
  && 400& 1.00 & 1.00 & 0.93 & 0.92 & 0.31 & 0.32 & 0.09 & 0.01 & 0.09 & 0.11 & 0.07 & 0.08 & 0.09 & 0.00 & 0.09 \\ 
    \hline
  &\multirow{3}{*}{0.375}& 200&1.00 & 1.00 & 0.91 & 0.92 & 0.49 & 0.46 & 0.14 & 0.03 & 0.13 & 0.20 & 0.16 & 0.12 & 0.13 & 0.00 & 0.13 \\ 
  &&300& 1.00 & 1.00 & 0.94 & 0.95 & 0.38 & 0.39 & 0.11 & 0.01 & 0.10 & 0.15 & 0.12 & 0.10 & 0.11 & 0.01 & 0.11 \\ 
  &&400& 1.00 & 1.00 & 0.92 & 0.95 & 0.33 & 0.36 & 0.09 & 0.02 & 0.09 & 0.13 & 0.11 & 0.08 & 0.09 & 0.00 & 0.09 \\ 
    \hline
 &\multirow{3}{*}{0.5} &200& 1.00 & 1.00 & 0.92 & 0.94 & 0.53 & 0.54 & 0.15 & 0.04 & 0.14 & 0.24 & 0.21 & 0.13 & 0.14 & 0.00 & 0.14 \\ 
  &&300 & 1.00 & 1.00 & 0.94 & 0.94 & 0.42 & 0.45 & 0.11 & 0.02 & 0.11 & 0.18 & 0.15 & 0.10 & 0.12 & 0.01 & 0.12 \\ 
  && 400& 1.00 & 1.00 & 0.92 & 0.93 & 0.36 & 0.40 & 0.10 & 0.02 & 0.10 & 0.17 & 0.14 & 0.09 & 0.11 & 0.00 & 0.11 \\ 
   \hline  \hline
\end{tabular}
}
\caption{Performance of the HITS hypothesis testing, in comparison with the plug-in Debiased Estimator (``Deb"), with respect to empirical rejection rate (ERR) as well as the empirical coverage (Coverage) and length (Len) of the CIs under the decaying loading $x_{\rm new, j}=Ratio*j^{-\delta}$. Reported also are the RMSE, bias and the standard error (SE) of the HITS estimator compared to the  plug-in Lasso estimator (``Lasso") and the plug-in Debiased Estimator (``Deb").}
\label{tab: decaying setting complete}
\end{table}

\end{document}